\documentclass[aps,prx,twocolumn,aps,floatfix,showpacs,showkeys,superscriptaddress]{revtex4-2}
\usepackage{amssymb}
\usepackage{lineno}
\usepackage{amsbsy}
\usepackage{amsmath}
\usepackage{epsfig}
\usepackage{color}
\usepackage{float}
\usepackage{xr-hyper}
\usepackage{bm}
\usepackage[colorlinks,linktocpage,bookmarks=false,citecolor=blue,linkcolor=red,urlcolor=blue]{hyperref}

\newcommand{\beg}{\begin{gather}}
\newcommand{\eeg}{\end{gather}}
\newcommand{\beq}{\begin{equation}}
\newcommand{\eeq}{\end{equation}}
\newcommand{\bea}{\begin{eqnarray}}
\newcommand{\eea}{\end{eqnarray}}

\newcommand{\bra}[1]{\ensuremath{\left\langle#1\right|}}
\newcommand{\ket}[1]{\ensuremath{\left|#1\right\rangle}}

\newcommand{\be}{\begin{equation}}
\newcommand{\ee}{\end{equation}}
\def\ba{\begin{aligned}}
\def\ea{\end{aligned}}
\def\bes{\begin{subequations}}
\def\ees{\end{subequations}}
\def\bal{\begin{align}}
\def\eal{\end{align}}

\newcommand{\rg}[1]{{\color{black}#1}}
\newcommand{\roop}[1]{{\color{black}#1}}

\usepackage{soul}
\usepackage{ulem} 
\usepackage{hyperref}
\begin{document}
\title{Harmonic Control of Dynamical Freezing in Programmable Rydberg Atom Arrays}
\author{Madhumita Sarkar}
\thanks{These authors contributed equally to this work}
\affiliation{Department of Physics and Astronomy, University College London, Gower Street, WC1E 6BT, London, United Kingdom.}
\author{Ben Zindorf}
\thanks{These authors contributed equally to this work}
\affiliation{Department of Physics and Astronomy, University College London, Gower Street, WC1E 6BT, London, United Kingdom.}
\author{Bhaskar Mukherjee}
\affiliation{Department of Physics, S.N. Bose Centre for Basic Sciences, Kolkata, 700106, India.}
\affiliation{Department of Physics and Astronomy, University College London, Gower Street, WC1E 6BT, London, United Kingdom.}
\author{Sougato Bose}
\affiliation{Department of Physics and Astronomy, University College London, Gower Street, WC1E 6BT, London, United Kingdom.}
\author{Roopayan Ghosh}
\thanks{Corresponding author: Roopayan Ghosh, e-mail: roopayanghosh@yahoo.co.in}
\affiliation{Department of Physics and Astronomy, University College London, Gower Street, WC1E 6BT, London, United Kingdom.}
\affiliation{School of Basic Sciences, Indian Institute of Technology, Bhubaneswar, 752051, India.}
\begin{abstract}

\rg{Periodic driving enables the engineering of complex quantum matter, yet in interacting systems it generically leads to energy absorption, which limits the lifetime of the engineered states. To address this challenge, dynamical freezing has been proposed as a mechanism for stabilizing non-equilibrium states over parametrically long timescales. While theory predicts robust freezing under simplifying assumptions, realistic platforms inevitably include additional interaction processes that alter its stability. Here, we report the experimental observation of dynamical freezing in programmable Rydberg atom arrays of up to 100 atoms in one and two dimensions. We find that while single-frequency driving produces pronounced suppression of excitation dynamics, the freezing behavior is restricted to a narrow parameter regime due to interaction-induced heating channels present in realistic simulators. Using a perturbative Floquet analysis of the fully interacting atomic system, we identify the dominant microscopic heating processes responsible for this destabilization. Leveraging this understanding, we design a dual-parameter modulation of detuning and Rabi frequency that coherently suppresses these absorption pathways and substantially broadens the freezing regime, making it also robust across different geometries. Our results reveal how heating processes shape the stability of dynamical freezing in interacting Floquet systems and demonstrates a route to control driven many-body dynamics in realistic experimental platforms.}
\end{abstract}

\maketitle

\section*{Introduction}
Controlling quantum matter far from equilibrium enables the realization of phases and dynamical phenomena with no equilibrium analogue. Periodic driving provides one of the most versatile frameworks for achieving this objective by enabling the engineering of effective Hamiltonians and emergent states. 
Such driving has enabled the realization of synthetic gauge fields, tailored interactions, topological phases and dynamically stabilized states~\cite{ doi:10.1126/science.abg2530,Will2025, Zhao2023, Bukov04032015,oka2019floquet,PhysRevResearch.6.L022059,PhysRevB.103.184309,PhysRevB.101.245107,PhysRevB.102.075123,He2025,PhysRevLett.130.120401,ghosh2025destructive}. 
However, interacting systems generically absorb energy from the drive, leading to runaway heating that ultimately destabilizes the engineered dynamics. Suppressing this heating has therefore become a central challenge for Floquet quantum technologies~\cite{PhysRevLett.115.030402, PhysRevLett.114.140401,PhysRevLett.116.250401,PhysRevLett.124.190601,PhysRevLett.115.256803,Peng2021}.
Among the various stabilization mechanisms, \textit {dynamical freezing} has emerged as a promising route to preserve non-equilibrium states over parametrically long timescales, leading to an effective inhibition of ergodic behaviour~\cite{PhysRevB.82.172402,PhysRevX.11.021008,haldar2024dynamicalfreezingthermodynamiclimit,PhysRevB.94.075130,PhysRevB.97.014309,ggk3-6cf8}. 
While theoretical studies predict remarkably long-lived freezing regimes under idealized conditions, its behaviour in realistic interacting many-body systems remains poorly understood.

\begin{figure}[h!]
    \centering
\includegraphics[width=1.\linewidth,height=0.68\linewidth]{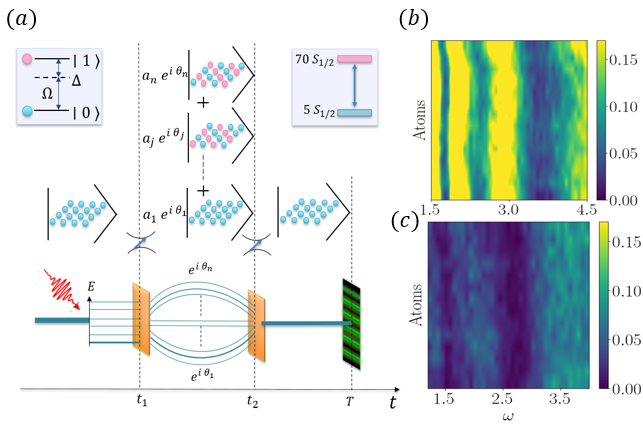}
    \caption{\rg{\textbf{Many-body dynamical freezing from temporal interference.}
\textbf{a}, Schematic illustration of many-body temporal interference in a periodically driven Rydberg array. Experimentally observed freezing fringes in the drive-frequency domain ($\omega$ in rad/$\mu$s) for (b) single-frequency and (c) bi-frequency driving. Dark fringes indicate destructive interference corresponding to strong excitation suppression leading to dynamical freezing, while bright regions indicate enhanced excitation due to heating channels.}}
    \label{fig:schematic}
\end{figure}

\roop{Temporal quantum interference provides a microscopic framework for
understanding dynamical freezing. Such interference naturally arises in
periodically driven systems, where excitation amplitudes accumulated
along distinct time-domain pathways recombine across drive cycles
(see Fig.~\ref{fig:schematic}(a)). In effective two-level systems, this
process gives rise to Landau--Zener--St\"uckelberg interference
~\cite{SHEVCHENKO20101,IVAKHNENKO20231,doi:10.1126/science.1119678,PhysRevLett.97.150502,PhysRevLett.123.240401,PhysRevB.102.075448,PhysRevLett.107.207002,PhysRevLett.112.116803,PhysRevResearch.4.023135,Wang2025,PhysRevA.80.063407}.
Interactions can disrupt the coherent interference underlying
dynamical freezing, and recent theoretical work has begun to explore
this question using simplified interaction models
~\cite{PhysRevX.11.021008,mukherjee2024floquet}. In realistic
many-body systems, however, finite-range interactions, geometry, and
additional interaction pathways can substantially modify the phase
accumulation and excitation processes, and hence the freezing
conditions. Understanding and controlling these interaction-induced
modifications is therefore essential for realizing robust dynamical
freezing in programmable quantum simulators.}

Programmable Rydberg atom arrays provide a powerful platform to systematically investigate these effects. These systems realize strongly interacting quantum spin models with tunable geometry and long coherence times, enabling exploration of many-body dynamics far beyond few-qubit realizations~\cite{bernien2017,ebadi2021}. They have already enabled discoveries ranging from quantum many-body scars~\cite{Turner2018} to exotic quantum phases of matter~\cite{SciPostPhys.14.1.004} including topological spin liquids~\cite{semeghini2021}, quantum critical dynamics and Kibble–Zurek mechanisms~\cite{keesling2019,darbha2025probingemergentprethermaldynamics}, quantum optimization~\cite{ebadi2022}, and logical qubit encoding~\cite{dolev2023}. Here we experimentally probe dynamical freezing in programmable Rydberg arrays of up to $100$ atoms across one- and two-dimensional geometries. We find that single-frequency driving produces excitation suppression at specific frequencies in a one-dimensional atomic chain, yielding frequency-domain interference fringes (see Fig.~\ref{fig:schematic}(b)). Finite-range interaction tails, however, activate interaction-dependent heating channels that already weaken freezing in one dimension and become increasingly detrimental in higher-dimensional geometries.

To overcome this limitation, we introduce dual-parameter modulation of detuning and Rabi frequency that coherently suppresses the dominant heating channels (see Fig.~\ref{fig:schematic}(c)). This dual frequency protocol restores and substantially broadens the freezing regime across geometries, producing near-complete excitation suppression even in regimes where conventional driving would strongly heat the system. Combined with a perturbative Floquet framework that incorporates realistic interaction corrections, our results directly reveal how interaction-induced processes destabilize dynamical freezing and demonstrate how these channels can be systematically eliminated. Beyond providing a microscopic understanding of freezing in realistic platforms, our work exposes the limits of idealized Floquet predictions and establishes interference-engineered suppression of heating as a practical route toward stabilizing interacting non-equilibrium quantum matter.

\section*{Results}
\subsection*{Programmable Rydberg Platform and Drive Protocols}

We employ QuEra’s programmable neutral-atom quantum simulator, Aquila, which can host up to 256 Rydberg atoms with flexible geometry. For this work we restrict to systems of approximately 100 atoms. This scale is set jointly by device geometry and the finite probability of preparing defect-free arrays~\cite{Querawhitepaper}.

Each atom is encoded as a two-level system consisting of a stable electronic ground state $\ket{g}$ and a high-energy Rydberg state $\ket{r}$. The collective vacuum state, $\ket{v}$ corresponds to all atoms in $\ket{g}$. The system dynamics are governed by the Hamiltonian  
\begin{equation}
H(t) = -\sum_i \Delta(t)\, n_i + \sum_i \frac{\Omega(t)}{2}\,\sigma_i^x + V_{\rm int},
\label{eq:rydbergmaster}
\end{equation}
where $n_i = \ket{r}\bra{r}_i$, $\Delta(t)$ is the laser detuning, $V_{\rm int}=\sum_{i<j} \frac{C_6}{r_{ij}^6} n_i n_j$ encodes van der Waals interactions between atoms separated by distance $r_{ij}$, $\Omega(t)$ the Rabi frequency.  

\roop{We probe dynamical freezing under two driving protocols}. In the single-frequency case, the detuning is modulated sinusoidally,
\begin{equation}
\Delta(t) = \Delta_0 \cos(\omega t), \qquad \Omega(t)= \Omega_0.
\end{equation}
In the bi-frequency case, both detuning and Rabi coupling are modulated,
\begin{equation}
\Delta(t) = \Delta_0 \cos(\omega t), \qquad
\Omega(t) = \tfrac{\Omega_0}{2}\bigl[1+\cos(r \omega t)\bigr],
\end{equation}
with $r=0$ recovering the single-frequency protocol and $r=2$ introducing a second harmonic that enhances suppression of excitations and visibility of interference pattern. Higher harmonics can also be implemented which produces alternative interference patterns (see Methods). 
The amplitudes $\Delta_0$ and $\Omega_0$ are chosen such that minima of the measured Rydberg density remain above the \roop{state-preparation and measurement} (SPAM) error bound, with visibility and fringe structure controlled systematically by tuning these values.

\rg{\subsection*{Dynamical freezing in Rydberg chains}}
\begin{figure*}
\hspace{0.02\linewidth}\includegraphics[width=0.02\linewidth,height=0.025\linewidth]{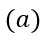}
\hspace{0.22\linewidth}\includegraphics[width=0.02\linewidth,height=0.02\linewidth]{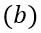}
\hspace{0.25\linewidth}\includegraphics[width=0.02\linewidth,height=0.02\linewidth]{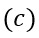}
\hspace{0.22\linewidth}\includegraphics[width=0.02\linewidth,height=0.02\linewidth]{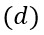}\\
\hspace{0.02\linewidth}\includegraphics[width=0.2 \linewidth,height=0.11\linewidth]{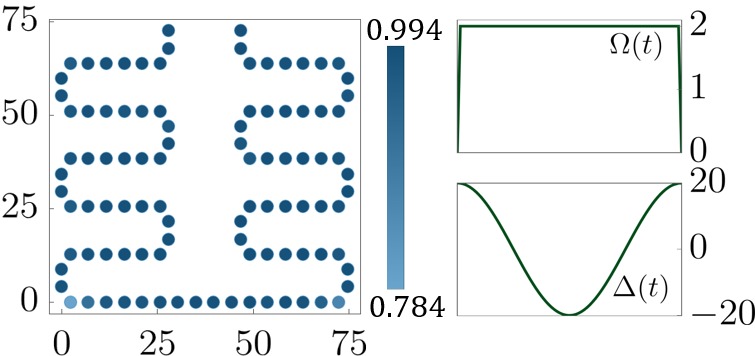}
\hspace{0.03\linewidth} 
\includegraphics[width=0.2 \linewidth,height=0.11\linewidth]{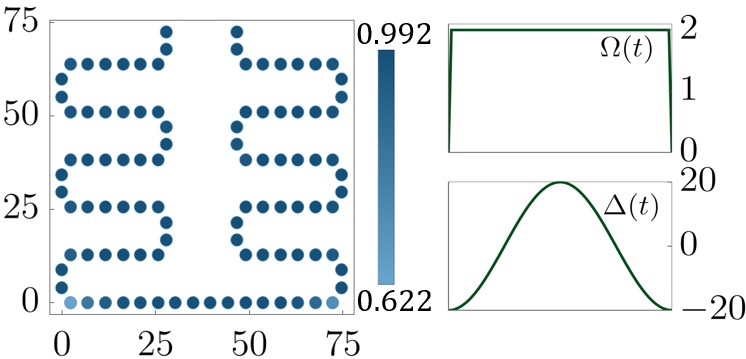}
\hspace{0.01\linewidth} 
\hspace{0.03\linewidth}\includegraphics[width=0.2 \linewidth,height=0.11\linewidth]{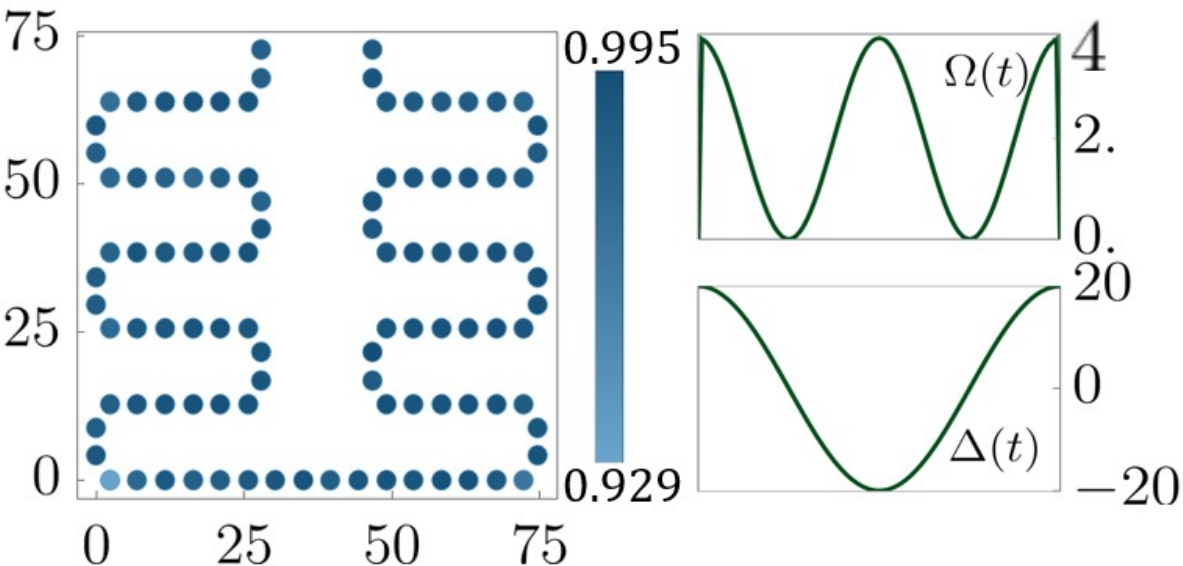}
\hspace{0.025\linewidth} 
\includegraphics[width=0.2 \linewidth,height=0.11\linewidth]{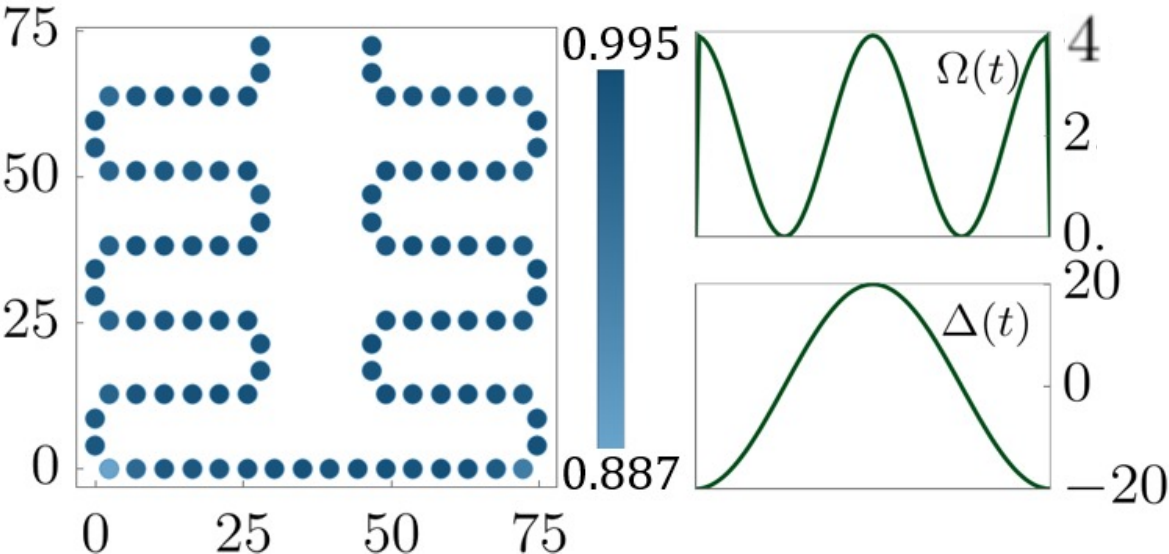}\\
\includegraphics[width=0.24 \linewidth,height=0.17\linewidth]{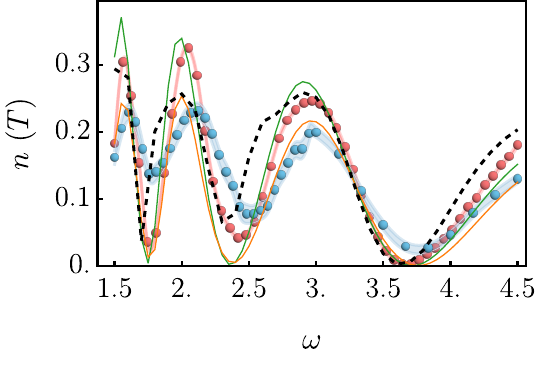}
\includegraphics[width=0.24 \linewidth,height=0.17\linewidth]{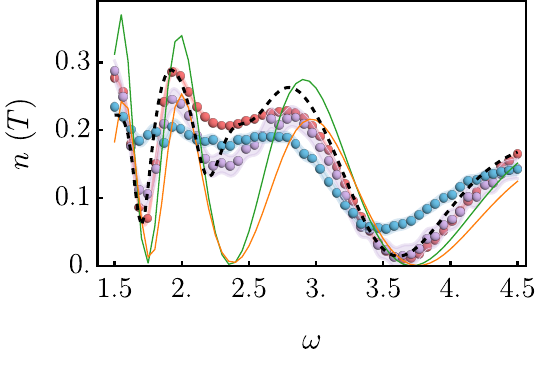}
\includegraphics[width=0.24\linewidth,height=0.17\linewidth]{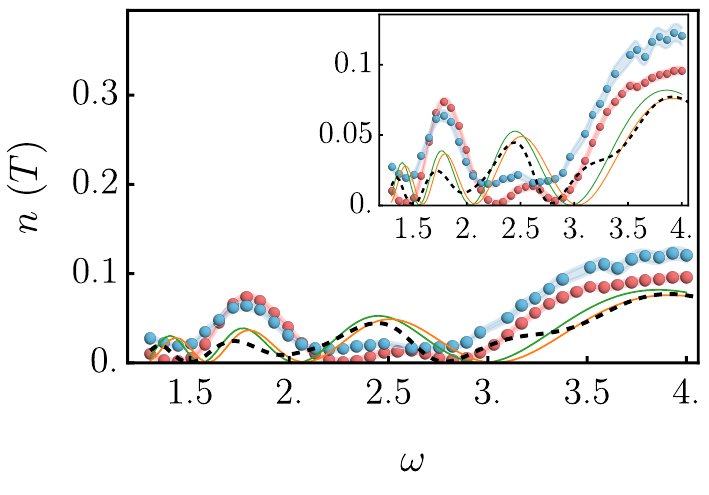}
\includegraphics[width=0.24 \linewidth,height=0.17\linewidth]{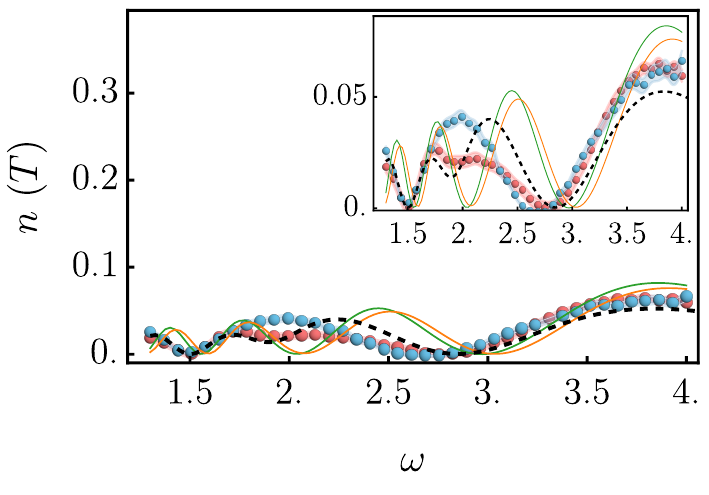}
\hspace{0.02\linewidth}\includegraphics[width=0.92\linewidth,height=0.045\linewidth]{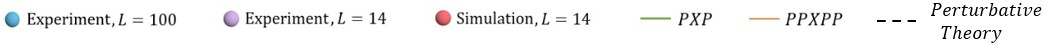}
\caption{\textbf{Dynamical freezing in one-dimensional Rydberg chains.} 
Frequency-dependent average Rydberg excitation density $n(T)$ (drive frequency $\omega$ in rad$/\mu$s) measured after a single drive cycle in $100$-atom chains with spacing $d=4.7~\mu$m. 
\textbf{Top:} Snake geometry of the atomic array (positions in $\mu$m) and corresponding drive protocols (frequencies in rad$/\mu$s); shading indicates atom loading probability. 
\textbf{a}, Single-frequency driving with constant $\Omega$ and $\Delta_0>0$ (vacuum as instantaneous excited state). 
\textbf{b}, Single-frequency driving with constant $\Omega$ and $\Delta_0<0$ (vacuum as instantaneous ground state). 
\textbf{c}, Bi-frequency driving with modulated $\Omega$ and $\Delta_0>0$. 
\textbf{d}, Bi-frequency driving with modulated $\Omega$ and $\Delta_0<0$. 
Blue circles: experimental data from the Aquila quantum processor ($L=100$). 
Purple: experimental data for a smaller system ($L=14$, where shown). 
Black dashed curves: perturbative Floquet theory predictions.
Green/yellow: constrained-model simulations (PXP/PPXPP). 
Red: Bloqade simulations of the full Rydberg Hamiltonian. 
Shaded regions indicate statistical uncertainty. 
Interference fringes appear as oscillations of $n(T)$ with drive frequency $\omega$: bright regions correspond to enhanced excitation, while minima indicate destructive temporal interference and strong dynamical freezing of the vacuum state.}
\label{fig:1dchain}
\end{figure*}

\roop{One-dimensional chains provide an ideal setting to uncover the microscopic
physics of many-body dynamical freezing}. We arrange $100$ Rydberg atoms in a
snake-like configuration with $4.7~\mu$m spacing (Fig.~\ref{fig:1dchain}, top
panels), initialize the system in the vacuum state $\ket{v}$, and measure the
Rydberg excitation density $n(T)$ after a single drive cycle. Using
$\Omega_0=2~\text{rad}/\mu\text{s}$ and $|\Delta_0|=20~\text{rad}/\mu\text{s}$,
we observe pronounced oscillations in $n(T)$ as the drive frequency $\omega$
is varied. \roop{These oscillations manifest alternating regimes of enhanced
excitation and strong dynamical freezing of the vacuum state, arising from
temporal interference between dynamical phases accumulated during successive
passages through the Ising-type quantum critical transition of the Rydberg
chain~\cite{bernien2017}, providing a many-body analogue of LZS interference.
During a drive cycle, the system passes through this transition region twice:
at the first crossing, the wavefunction splits into components evolving along
different energy branches; at the second, these components recombine and
interfere. Their relative phase determines whether the interference is
constructive, yielding enhanced excitation, or destructive, producing
dynamical freezing.}

In our system, enhanced Rydberg excitation corresponds to the ``bright''
fringes, while \rg{the minima represent strong dynamical freezing of the
vacuum}. The resulting interference pattern extends coherently across the
entire $100$-atom array, demonstrating \rg{controllable many-body freezing}.
To characterize the fringes, we analyze their visibility and residual
excitation density at the freezing minima.

\roop{We define the fringe visibility as
$V=(n^{\max}-n^{\min})/(n^{\max}+n^{\min})$, where $n^{\max}$ and $n^{\min}$
are the average excitation densities at a local maximum and its neighboring
freezing minimum. At high frequencies ($\omega\sim3.5~\text{rad}/\mu\text{s}$),
$n^{\max}\simeq0.20$ and $n^{\min}\simeq0.026$, giving $V\sim0.77$ and strong
fringe contrast.} The excitation density at the freezing minimum approaches
the SPAM-error floor. As the frequency decreases, however, the contrast drops
to $V\sim0.35$ at $\omega\sim1.5~\text{rad}/\mu\text{s}$ for $\Delta_0>0$, and
still lower for $\Delta_0<0$, \rg{while the excitation density at the minima
correspondingly increases}. \roop{This degradation reflects the increasing
importance of interaction-induced processes at lower drive frequencies. In
the ideal two-level LZS picture, the two passages through the avoided crossing
lead to coherent phase accumulation and recombination, producing constructive
or destructive interference. As the drive energy scale decreases relative to
the interaction-induced shifts, additional excitation pathways modify the
accumulated phases and distort the ideal interference pattern. The effect
becomes more pronounced in larger systems, where more interaction-assisted
pathways are available. The semi-analytical Floquet predictions reproduce
these frequency-dependent features.} Part of the reduction also arises from
the finite $4~\mu$s coherence time and finite-size and geometry effects, as
seen in the comparison with $L=14$ experimental data. Nevertheless, clear
freezing minima persist in the interacting ensemble.

However, constrained-model simulations capture the qualitative structure of
the freezing fringes but systematically overestimate the degree of excitation
suppression. In particular, when $\ket{v}$ is the ground state at $t=0$, they
predict an additional minimum near $\omega\sim2.2~\text{rad}/\mu\text{s}$
that is absent even in small-size blockade simulations. This discrepancy
indicates that interaction processes beyond the ideal blockade constraint
modify the phase accumulation responsible for dynamical freezing.

We account for these discrepancies using perturbative Floquet theory (discussed later). Within this framework, dynamical freezing occurs when
destructive interference suppresses excitation pathways, which happens at
frequencies where the leading Bessel-function channel
$J_0(\Delta_0/\omega)$ vanishes, leading to an emergent conserved total
$\sigma^z$. Constrained models achieve near-ideal freezing by inhibiting
competing pathways, while finite-range Rydberg interaction tails activate
additional excitation channels, breaking the conservation law strongly or
weakly depending on the strength of interaction and frequency. These
processes reshape the freezing landscape, explaining the observed detuning
asymmetry, fringe shifts, and geometry dependence.

\rg{Motivated by this degradation at lower frequencies, we explore whether
additional control parameters can stabilize the destructive interference
responsible for freezing.} We demonstrate enhanced control through
dual-frequency driving in Fig.~\ref{fig:1dchain}(c,d). Introducing
$\Omega(t)=\tfrac{\Omega_0}{2}[1+\cos(2\omega t)]$ with
\roop{$\Omega_0=4~\text{rad}/\mu\text{s}$}, \rg{provides independent control over
the instantaneous energy gap and mixing at the avoided crossings}.
\rg{This additional handle tunes the relative phases of different dynamical
pathways, suppressing excitation channels that would otherwise lead to
heating.} As a result, excitations are strongly suppressed across the
frequency range. \rg{The suppression brings the excitation density close to
the noise floor; we therefore increase the peak Rabi amplitude from $2$ to
$5~\text{rad}/\mu\text{s}$ to resolve the residual interference structure.}
Even in the low-frequency regime where single-frequency driving exhibits
substantial excitation, bifrequency driving maintains strong freezing,
demonstrating that \rg{multi-parameter control of temporal interference
provides a powerful route to stabilizing dynamical freezing in interacting
many-body systems.}
\begin{figure}
\includegraphics[width=1.\linewidth,height=0.55\linewidth]{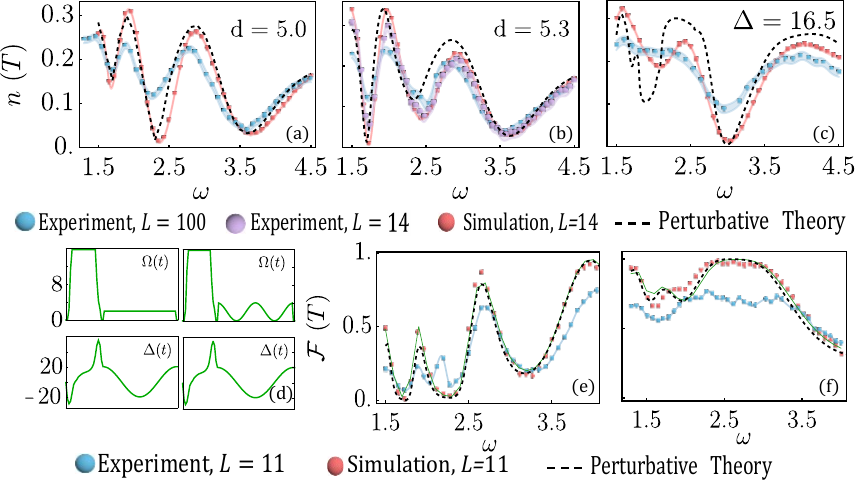}
\caption{\roop{\textbf{Interaction, detuning, and initial-state dependence of dynamical freezing.}
\textbf{a,b,} Frequency-dependent excitation density $n(T)$ for $100$-atom
chains at $d=5.0$ and $5.3~\mu$m, showing interaction-dependent shifts and
changes in freezing-minimum visibility. \textbf{c,} Reduced detuning amplitude
$\Delta_0=16.5~\mathrm{rad}/\mu\mathrm{s}$ shifts the freezing minima to lower
frequencies. \textbf{d,} Preparation of the $Z_2$ state in an $L=11$ chain
($d=6.7~\mu$m) followed by single- (left) and bi-frequency (right) driving.
\textbf{e,f,} Corresponding state fidelity
$\mathcal{F}(T)=|\langle\psi(T)|\psi(0)\rangle|^2$. Blue circles show Aquila
data, purple circles the $L=14$ Aquila data in \textbf{b}, and red circles
Bloqade simulations; system sizes are $L=100$ in \textbf{a,c} and $L=11$ in
\textbf{e,f}. Green curves in \textbf{e,f} show ideal $Z_2$ preparation,
while experimental and Bloqade results include the finite preparation
fidelity (91\% and 99\%, respectively, after SPAM correction). Black dashed
curves show perturbative Floquet predictions. Shaded regions indicate
statistical uncertainty.}}
\label{fig:diffdistance}
\end{figure}

\roop{\subsection*{State and interaction dependent freezing landscape}}

We next demonstrate how interatomic spacing controls \rg{dynamical freezing}, revealing finite-range interactions' decisive role in \rg{the freezing landscape} (Fig.~\ref{fig:diffdistance}). Unlike constrained models where blockade radius changes negligibly affect interference, the full Rydberg Hamiltonian exhibits marked distance sensitivity from van der Waals tails extending beyond nominal blockade radii.

At our spacings ($d=4.7$-$5.3~\mu$m), next-nearest-neighbor interactions remain $O(\Omega_0, \Delta_0)$, which is a regime where interaction and drive energies compete directly, making \rg{freezing patterns} acutely distance-sensitive. At $d=4.7~\mu$m, no \rg{freezing minimum} appears near $\omega \sim 2.4$ rad$/\mu$s, yet this feature emerges at $d=5.0~\mu$m in both Bloqade simulations and experiments, albeit with reduced visibility. At $d=5.3~\mu$m, the strongest \rg{dynamical freezing} shifts to $\omega \sim 1.7$ rad$/\mu$s, qualitatively reproduced across $L=14$ and $L=100$ despite decreasing visibility with system size.

This distance sensitivity reveals how finite-range tails continuously reshape \rg{the conditions for dynamical freezing} as atoms separate, shifting the relative contributions of neighbor couplings and altering phase accumulation. Floquet perturbation theory (see Methods) quantifies this by showing that the interaction corrections to resonances depend explicitly on distance, producing the \rg{observed shifts in freezing minima} that fixed-blockade $PXP$ models miss entirely.

Complementing geometric control, detuning amplitude $\Delta_0$ provides energetic tuning (Fig.~\ref{fig:diffdistance}(c)). First-order theory predicts \rg{freezing} at Bessel-function zeros $J_0(\Delta_0/\omega)=0$, yielding frequencies scaling with $\Delta_0$. Experiments confirm this: \rg{freezing minima} shift systematically to lower frequencies as $\Delta_0$ decreases. \roop{The perturbative Floquet description reproduces the essential features of these shifts and the observed freezing landscape, with quantitative agreement becoming somewhat less accurate at smaller $\Delta_0$ as the relevant perturbative ratios increase.} 

\roop{Freezing also occurs for nontrivial initial states beyond the vacuum.
We prepare the $Z_2$ state $\ket{10101010101}$ experimentally using optimal control and
study its dynamics for $L=11$ (Fig.~\ref{fig:diffdistance}(d--f)), with the
odd system size chosen to remove the degeneracy between the two $Z_2$
configurations. While arbitrary state preparation in an analogue device is
challenging and is best achieved by digital-analogue hybrid control, the $Z_2$ state
provides an experimentally accessible nontrivial many-body initial state in Aquila hardware.
Related freezing behaviour for other initial states is also observed
numerically (Supplementary Information). Because freezing here corresponds
to preserving a specific many-body state, we use its state fidelity ($\mathcal{F}(T)=|\langle \psi(T)|\psi(0)\rangle|^2$) as the
relevant measure. The SPAM-corrected (see Methods) experimental data retain dynamical freezing behaviour (high fidelity points) despite coherence limitations at low frequencies and agree well
with simulations. The high-fidelity freezing points for the $Z_2$ state occur at
frequencies distinct from the vacuum-state minima, with the shift
particularly pronounced for bi-frequency driving, further demonstrating
how interactions and the initial state reshape the freezing landscape.
Single-frequency driving shows weaker freezing at low frequencies, with
high-fidelity freezing emerging only at higher frequencies, consistent with
the vacuum-state results. Bi-frequency driving substantially broadens the
high-fidelity freezing regime towards lower frequencies, demonstrating
enhanced robustness for a nontrivial initial state.}
These results establish distance, geometry, and drive parameters as a multidimensional control space for engineering dynamical freezing for different initial states, providing practical routes to stabilizing quantum states and designing robust Floquet protocols in interacting many-body systems.
\begin{figure*}
\hspace{0.02\linewidth}\includegraphics[width=0.02\linewidth,height=0.025\linewidth]{figa.jpg}
\hspace{0.22\linewidth}\includegraphics[width=0.02\linewidth,height=0.02\linewidth]{figb.jpg}
\hspace{0.22\linewidth}\includegraphics[width=0.02\linewidth,height=0.02\linewidth]{figc.jpg}
\hspace{0.225\linewidth}\includegraphics[width=0.02\linewidth,height=0.02\linewidth]{figd.jpg}\\
\hspace{0.19 in}\includegraphics[width=0.2 \linewidth,height=0.11\linewidth]{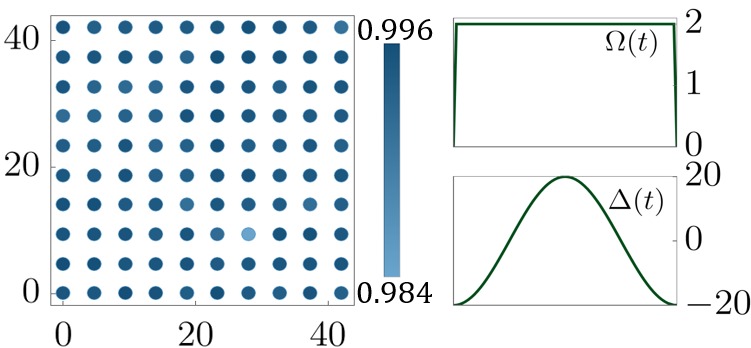}
\hspace{0.045\linewidth} 
\includegraphics[width=0.2 \linewidth,height=0.112\linewidth]{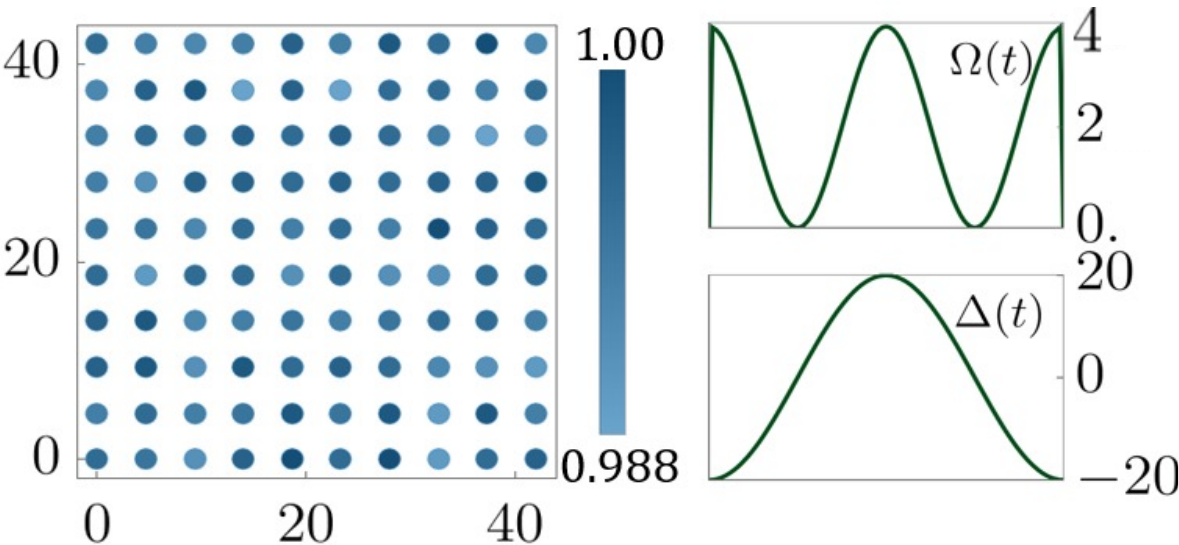}
\hspace{0.04\linewidth}\includegraphics[width=0.2 \linewidth,height=0.11\linewidth]{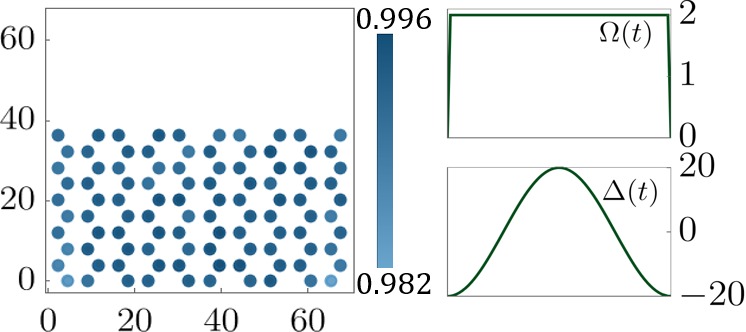}
\hspace{0.04\linewidth} 
\includegraphics[width=0.2 \linewidth,height=0.11\linewidth]{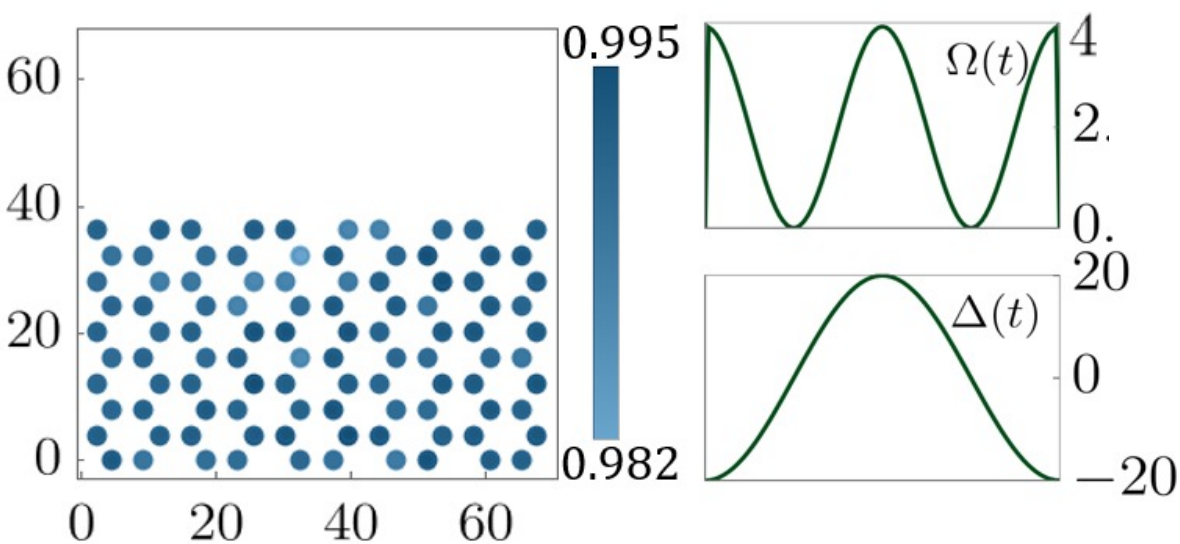}\\
\includegraphics[width=0.24 \linewidth,height=0.16\linewidth]{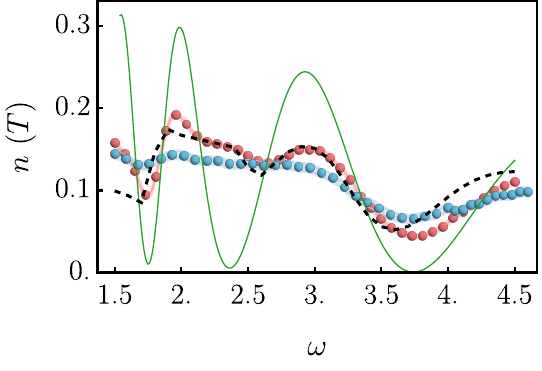}
\includegraphics[width=0.24 \linewidth,height=0.16\linewidth]{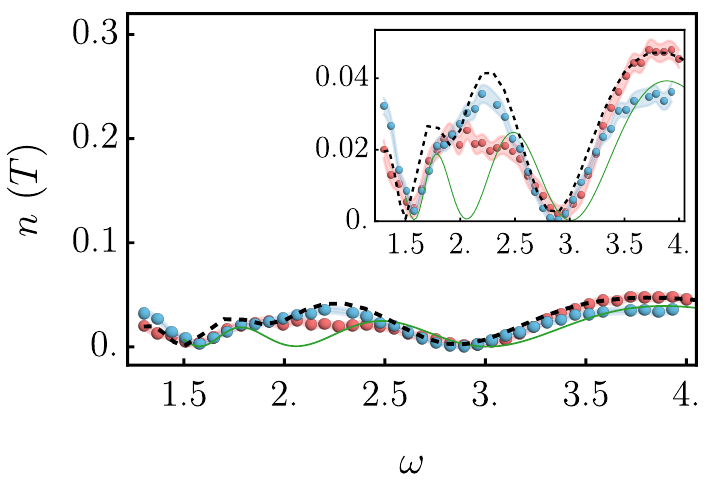}
\includegraphics[width=0.24\linewidth,height=0.16\linewidth]{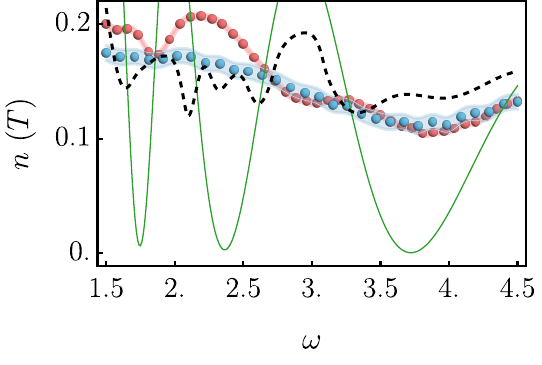}
\includegraphics[width=0.24 \linewidth,height=0.16\linewidth]{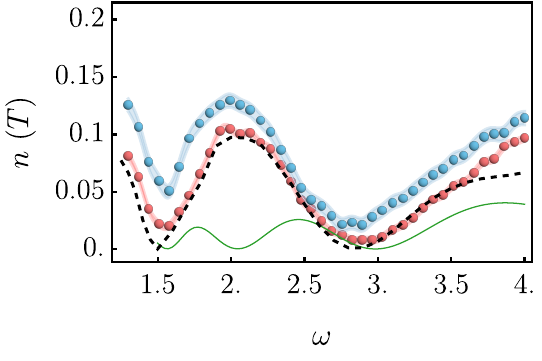}\\
\hspace{0.02\linewidth}\includegraphics[width=0.73\linewidth,height=0.03\linewidth]{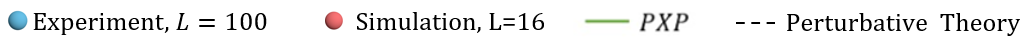}
\caption{\textbf{\rg{Geometry-dependent dynamical freezing in  Rydberg arrays.}} 
\textbf{Top:} Atomic arrangements, drive protocols, and loading probabilities. 
\rg{a,c,} Frequency-dependent Rydberg excitation density $n(T)$ (drive frequency $\omega$ in rad$/\mu$s) measured after one drive cycle in two-dimensional arrays under single-frequency driving. \rg{Square (a) and honeycomb (c) geometries exhibit strongly reduced freezing contrast compared to one-dimensional chains, with only weak minima visible.} 
\rg{b,d,} Dual-frequency modulation restores high-contrast freezing patterns. \rg{In square lattices the visibility reaches $V\sim0.9$, while honeycomb geometries exhibit substantial recovery despite stronger interaction-induced heating channels.} 
\textbf{a,b:} Square lattice. \textbf{c,d:} Honeycomb lattice. 
Blue circles: experiment (Aquila, $L=100$); red: Bloqade simulations ($L=16$); Black dashed curves: perturbative Floquet predictions; \rg{green curves: constrained-model simulations (PXP).} 
\rg{Minima correspond to dynamical freezing of the vacuum state.}}
\label{fig:2dgeom}
\end{figure*}
\rg{\subsection*{Freezing beyond one-dimension}}

We next extend dynamical freezing to two-dimensional lattices, where atoms
interact with multiple neighbours and acquire geometry-dependent phases.
These additional pathways provide a stringent test of whether the coherent
interference and associated vacuum freezing survive increased many-body
complexity. Moreover, entanglement growth and long-range couplings make
tensor-network simulations inefficient in two dimensions, highlighting the
value of experimental access. Figure~\ref{fig:2dgeom} compares square and
honeycomb geometries, with \textit{Bloqade} simulations for $L=16$ benchmarked
against \textit{Aquila} experiments at $L=100$ and constrained $PXP$
simulations. All data are initialized from the vacuum state with
$\Delta_0=-20$~rad/$\mu$s.

Compared to one-dimensional chains, single-frequency driving shows strongly
reduced freezing in two dimensions. At $L=100$, both geometries
retain only a single weak freezing minimum, with substantial finite-size
effects relative to the $L=16$ simulations. The suppression in freezing behavior is stronger in
the honeycomb geometry, particularly at low frequencies where finite
coherence further limits contrast. This geometry dependence reflects the
increased role of next-nearest-neighbour interactions and competing
interference channels, consistent with our analytical framework (black dashed lines). In contrast,
constrained $PXP$ simulations remain insensitive to geometry and dimensionality,
reproducing the one-dimensional behaviour.

Dual-frequency modulation,
$\Omega(t)=\tfrac{\Omega_0}{2}[1+\cos(2\omega t)]$ with
\roop{$\Omega_0=4$~rad/$\mu$s} (Fig.~\ref{fig:2dgeom}), restores high-contrast
interference. In the square lattice, the excitation suppression gives
$V\sim(0.055-0.004)/(0.055+0.004)\sim0.86$, corresponding to strong vacuum
freezing with $\langle n\rangle\sim0.01$, exceeding the one-dimensional
performance. In the honeycomb lattice, bi-frequency driving stabilizes
interference from $L=16$ to $L=100$, where single-frequency driving fails.
The additional modulation provides a secondary interference channel that
mitigates the competing phase contributions from the interaction network,
with visibility further enhanced by reducing $\Omega_0$. The resulting
stabilization depends on the harmonic choice, with only selected
commensurate frequencies producing strong cancellation.

These results demonstrate that multi-frequency driving provides an effective
route to \roop{robust dynamical freezing} in two dimensions, where complex
interaction networks make coherent cancellation substantially more fragile.

\begin{figure}
    \centering
    \includegraphics[width=1.\linewidth]{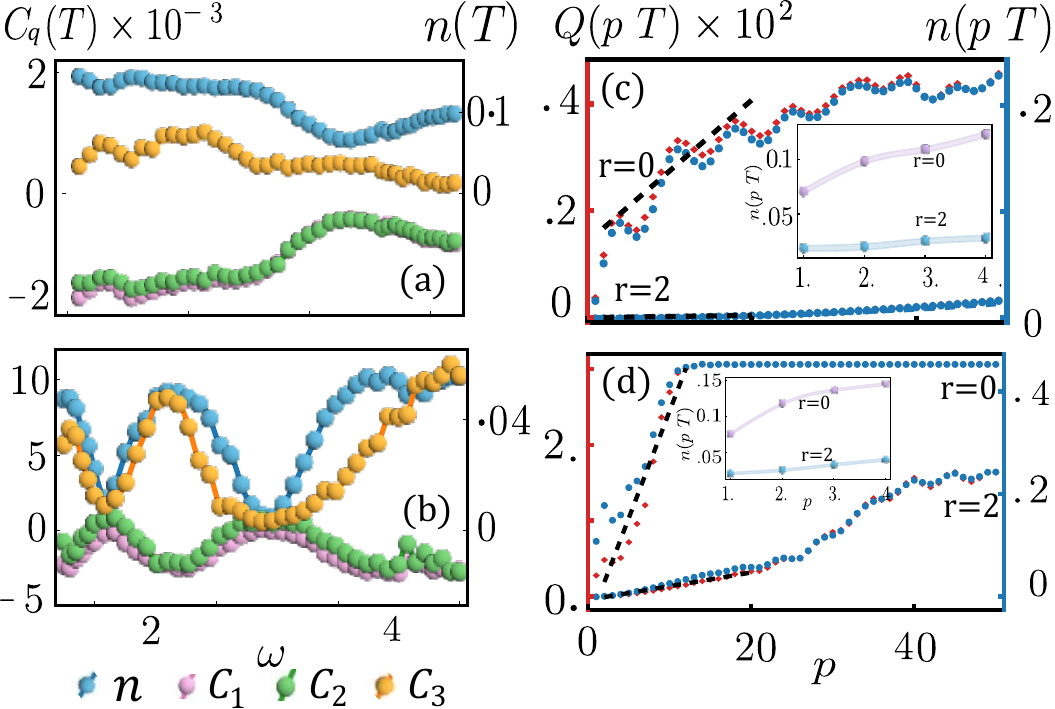}
    \caption{\roop{\textbf{Interaction-induced correlations and heating under
single- and dual-frequency driving.}
\textbf{a,b,} Experimental excitation density $n(T)$ and connected two-point correlations
$C_q(T)=\langle n_i n_{i+q}\rangle-\langle n_i\rangle\langle n_{i+q}\rangle$
for the first three neighbour shells ($q=1,2,3$) after one drive cycle in a
square lattice. Here, $q$ labels the neighbour shell of site $i$. \textbf{a,}
Single-frequency driving, for which substantial interaction-induced
correlations persist at the freezing minimum. \textbf{b,} Dual frequency driving,
showing strong suppression of both the excitation density and spatial
correlations. The left and right axes denote $C_q(T)$ and $n(T)$,
respectively. The parameters are same as Fig.~\ref{fig:2dgeom}.
\textbf{c,d,} Heating $Q$ and excitation density $n$ as a function of the
number of drive cycles $p$ for \textbf{c,} an $L=14$ linear chain and
\textbf{d,} an $L=16$ square lattice, evaluated at the corresponding
single- and dual- frequency freezing frequencies. For the linear chain,
$\omega=3.73$ and $2.89~\mathrm{rad}/\mu\mathrm{s}$, and for the square
lattice, $\omega=3.55$ and $3.00~\mathrm{rad}/\mu\mathrm{s}$, for
single- and dual-frequency driving, respectively. Dashed lines show linear fits
to $Q$ over $2\leq p\leq20$, yielding heating rates of $1.33$ and $0.023$ for
single- and dual-frequency driving in the linear chain, corresponding to a
$>50\times$ suppression, and $28.20$ and $1.79$ in the square lattice,
corresponding to a $>15\times$ suppression. Insets show the corresponding
experimental excitation density over four drive cycles for $100$-atom linear
chains and square arrays, respectively, exhibiting the same contrast between
single- and dual-frequency driving. The left and right axes in \textbf{c,d}
denote $Q$ and $n$, respectively.}}
    \label{fig:diff_corr}
\end{figure}

\roop{\subsection*{Spatial correlations and heating}
Dynamical freezing suppresses not only excitation density but also the buildup
of spatial correlations. Figure~\ref{fig:diff_corr}(a,b) shows the connected
correlations $C_q(T)$ for the first three neighbor shells together with the
excitation density after one drive cycle on a square lattice. The correlations
extend beyond nearest neighbors, reflecting the finite-range interaction tails
and demonstrating that the dynamics cannot be characterized solely by
independent local excitations. For single-frequency driving [Fig.~\ref{fig:diff_corr}(a)],
the different correlation shells remain substantial and exhibit a
reduced suppression at the excitation-density minimum. In contrast,
bi-frequency driving [Fig.~\ref{fig:diff_corr}(b)] produces pronounced
frequency-dependent oscillations in the correlations, with enhanced suppression
coinciding with the minima of the excitation density. This shows that the
engineered interference suppresses the buildup of many-body correlations
together with the excitation density, rather than merely reducing the
single-site excitation probability. Additional measurements and simulations
for different interatomic spacings and lattice geometries are presented in the
Supplementary Information, where the corresponding collective $Z$-spin
fluctuations further quantify the many-body correlations and provide a quantum
Fisher information-based entanglement witness.

The suppression persists over multiple drive cycles. Fig.~\ref{fig:diff_corr}(c,d) shows the absorbed energy 
$Q(pT)=\langle\psi(pT)|H(0)|\psi(pT)\rangle
- \langle\psi(0)|H(0)|\psi(0)\rangle$, and excitation density $n$ during repeated driving at representative freezing frequencies. The heating rate extracted from $Q$ is reduced by more than an order of magnitude under bi-frequency driving in both one- and two-dimensional geometries, consistent with the simultaneous suppression of excitation and the persistence of spatial correlations. Experimental measurements over four cycles on $100$-atom arrays show the same trend despite the finite coherence time of the hardware. Together, the suppression of excitation and heating, accompanied by spatially extended correlations, provides strong evidence for robust many-body dynamical freezing.}

\roop{\subsection*{Microscopic origin of Dynamical Freezing}}

\roop{Finally, to understand the rich structures observed in the interference patterns,
including their sensitivity to the drive protocol, detuning, interaction
range, and geometry, we develop an analytical description based on
Floquet perturbation theory over a single drive cycle (see Methods and
Supplementary Material). The stroboscopic evolution can be written as
\begin{equation}
U(T)=e^{-iT H_F^{(0)}}
\exp\left[-iT\left(H_F^{(1)}+H_F^{(2)}\right)\right]
+\mathcal{O}(\Omega^3),
\end{equation}
where
$H_F^{(0)}=\sum_{i,j}V_{ij}n_i n_j$ and the leading interaction-picture contribution is
\begin{equation}
H_F^{(1)}
=
\Omega\sum_i
\left[
\tilde{\sigma}_i^+\mathcal{I}(\alpha_i)
+\tilde{\sigma}_i^-\mathcal{I}^*(\alpha_i)
\right],
\end{equation}
with
\begin{equation}
\mathcal{I}(\alpha)=
\left(e^{i2\pi\alpha/\omega}-1\right)
\sum_{n=-\infty}^{\infty}
\frac{J_n(\Delta_0/\omega)}
{iT(\alpha+n\omega)} .
\label{eq:maintextI}
\end{equation}
Here $\alpha_i=\sum_{j\in{\rm LR}(i)}V_{ij}n_j$ is the
interaction-induced energy shift at site $i$, ${\rm LR}(i)$ denotes atoms
outside the blockade radius, $J_n$ is the $n$th Bessel function,
$T=2\pi/\omega$, and
$\tilde{\sigma}_i^\pm=\mathcal{P}\sigma_i^\pm\mathcal{P}$ are the
spin-flip operators projected into the blockade-constrained subspace.

For the vacuum state, the relevant first-order matrix elements are
independent of $\alpha_i$, so that the effect of finite-range
interactions first enters through the second-order contribution. We
therefore retain
\begin{equation}
H_F^{(2)}
=
-\frac{i\Omega^2}{T}
\sum_{i,j}\sum_{s,s'=\pm}
\left[
K_{ij}^{(s,s')}
-\frac{T^2}{2}\mathcal{I}_i^{(s)}\mathcal{I}_j^{(s')}
\right]
\tilde{\sigma}_i^s\tilde{\sigma}_j^{s'},
\label{eq:HF2_main}
\end{equation}
where $\mathcal{I}_i^{(+)}=\mathcal{I}(\alpha_i),\qquad
\mathcal{I}_i^{(-)}=\mathcal{I}^*(\alpha_i)$, and the second-order kernel $K_{ij}^{(s,s')}$ is given explicitly in
the Methods. This expression is independent of the underlying lattice
geometry: geometry enters only through the interaction matrix $V_{ij}$
and the corresponding local shifts $\alpha_i$. We use the same
stroboscopic hamiltonian, with terms retained through second order, to
obtain the semi-analytical predictions shown in Figs.~\ref{fig:1dchain}-\ref{fig:2dgeom}, without
adjustable fitting parameters.}

\roop{This expression provides the microscopic interpretation of the conditions for dynamical freezing}. In the non-interacting or perfectly
constrained limit ($\alpha_i=0$), till the second order, harmonic contributions recombine destructively
over a full cycle, leaving only the $J_0(\Delta_0/\omega)$ channel active.
Frequencies satisfying $J_0(\Delta_0/\omega)=0$ eliminate this channel entirely,
producing dynamical freezing. This explains both the emergence of interference
minima and why $PXP$ models practically mimic single-particle LZS interference
fringes: at leading order, the same single-site excitation amplitudes govern the
dynamics, with the blockade constraint restricting the accessible excitation
pathways. Crucially, reversing detuning sign preserves $|\mathcal{I}(0)|$,
consistent with the symmetry exhibited by constrained models.

\roop{Finite-range interactions open interaction-assisted excitation channels that
qualitatively modify the dynamics. When $\alpha_i\neq0$, excitation amplitudes
become conditional on surrounding occupations, allowing interactions with nearby
Rydberg atoms to enable or suppress spin flips. Thus, for states containing
Rydberg excitations, interaction-dependent corrections already appear at first
order through the $\alpha_i$ dependence of Eq.~(\ref{eq:maintextI}). This also
shows that the stability of dynamical freezing depends on the initial excitation
density: states with more pre-existing Rydberg excitations activate more
interaction-assisted channels and therefore can exhibit a different freezing
landscape. For the vacuum state, however, the relevant first-order matrix
elements are evaluated with $\alpha_i=0$, so the interaction-induced correction
to freezing first appears at second order through virtual excitation processes,
as described by $H_F^{(2)}$. The shifted denominators
$(\alpha_i+n\omega)$ redistribute spectral weight across harmonics, breaking
perfect destructive interference and generating the detuning asymmetry absent
in constrained $PXP$ models. Because $\alpha_i$ depends on interatomic distance, resonance frequencies shift and fringe
visibility degrades when low harmonics approach resonance, activating heating
channels and producing the spacing-dependent suppression of freezing observed
in Fig.~\ref{fig:diffdistance}. This state dependence is observed experimentally for the $Z_2$ and other initial states,
for which the freezing minima are shifted relative to the vacuum case (See Supplementary material for more examples).}

This framework also clarifies the dimensional dependence of freezing. In two
dimensions, each atom couples to multiple neighbors at different distances,
broadening the distribution of $\alpha_i$ and introducing competing phase
contributions. In square lattices, nearest and next-nearest neighbors remain
blockaded, but the next interaction shell at $2d$, corresponding to the
one-dimensional next-nearest-neighbor distance, falls within the tail regime,
with four such neighbors contributing (compared to two in one dimension),
thereby increasing the number of competing excitation channels and reducing
fringe visibility. Honeycomb lattices are more fragile: six next-nearest
neighbors at distance $\sqrt{3}d$ produce interaction shifts
$V_{\rm NNN}\sim\Delta_0$, placing them directly in resonance with the drive
and opening strong heating channels that severely suppress interference
and hence the freezing observed in experiments. \roop{The semi-analytical
predictions obtained from the effective Floquet Hamiltonian reproduce the
main geometry-dependent features, while becoming less regular for the
honeycomb lattice, where the larger number of interaction-induced
channels enhances the importance of higher-order processes. This highlights
the role of these channels in shaping the observed freezing landscapes.}

\roop{To mitigate the interaction-induced degradation of freezing, we therefore
seek to engineer additional interference pathways through the drive
itself. Bi-frequency driving provides a direct route to this control. For
$\Omega(t)=\Omega_0[1+\cos(r\omega t)]/2$, the leading excitation
amplitude contains the harmonic combination
\begin{equation}
J_0(x)+\frac{J_r(x)+J_{-r}(x)}{2},
\qquad x=\frac{\Delta_0}{\omega}.
\end{equation}
For odd $r$, the additional harmonic cancels by symmetry, leaving the
leading $J_0$ channel unchanged, whereas even $r$ introduces an
additional harmonic. The case $r=2$ is particularly effective because
the Bessel recurrence relation gives
\begin{equation}
J_0(x)+J_2(x)=\frac{2J_1(x)}{x}.
\end{equation}
Thus, at large $x$, the leading excitation amplitude is suppressed from
$J_0(x)\sim x^{-1/2}$ to
$J_0(x)+J_2(x)\sim x^{-3/2}$. The additional harmonic suppresses the excitation amplitude through
destructive interference, leading to an overall suppression of
excitations across the low-frequency regime and substantially deeper
freezing minima. The corresponding freezing points are shifted from the
zeros of $J_0$ to those of $J_1$ as a consequence of the modified
interference condition.

More generally, the same leading-order destructive interference occurs
for harmonics $r=4m+2$. Their large-$x$ asymptotic form is
\begin{equation}
J_0(x)+J_r(x)
\sim
\frac{r^2}{2}
\sqrt{\frac{2}{\pi}}\,x^{-3/2}
\sin\left(x-\frac{\pi}{4}\right),
\end{equation}
showing that the residual amplitude grows as $r^2$. Consequently,
higher harmonics require progressively larger $x=\Delta_0/\omega$,
or equivalently lower drive frequencies, for the destructive
interference to become effective. This makes $r=2$ the most effective
harmonic in the experimentally accessible regime, while establishing
that the underlying mechanism is not specific to a single harmonic.

The interference-engineering principle is also not restricted to
sinusoidal dual frequency driving. We find numerically that other harmonic
choices produce distinct freezing landscapes, while a square-pulse
protocol, with a different Fourier spectrum and hence without the same
Bessel-function structure, also produces well-defined freezing regimes
at shifted frequencies (Supplementary Information). More general
waveform engineering could therefore provide additional control over
the relative amplitudes and phases of excitation channels, potentially
allowing still more effective stabilization through optimized drive
protocols. 

Together, these results show that the stability of dynamical freezing is
controlled by the interplay between interaction-induced phase
accumulation and the harmonic content of the drive. Interaction range,
geometry, and drive waveform therefore provide complementary routes for
engineering robust dynamical freezing in interacting Floquet systems.}

\section*{Discussion}

\roop{Our demonstration of controllable many-body dynamical freezing establishes an
experimentally demonstrated strategy for stabilizing non-equilibrium quantum
matter in the presence of realistic interactions.} By strongly suppressing
heating and delaying thermalization, dynamical freezing can dramatically
extend coherent lifetimes and the retention of many-body memory, attributes
valuable for quantum control and information processing in driven systems.
Moreover, the dual frequency driving protocols we establish provide practical
design principles for stabilizing driven many-body systems beyond what can
be achieved using single-frequency protocols. The strong sensitivity of
freezing points to sub-micron variations in interatomic spacing also suggests
potential applications in precision sensing, where collective Rydberg
responses could transduce small interaction changes into measurable shifts
of the freezing condition.

\roop{The interference mechanism behind this dynamical freezing also governs
non-adiabatic quantum annealing}, where systems traverse avoided crossings at
finite speed. By engineering when dynamical phases accumulate or cancel, one
can accelerate quantum state preparation while preserving high fidelity,
offering advantages beyond conventional adiabatic schemes. In this context,
dynamical freezing provides a mechanism for suppressing unwanted excitations
during optimization protocols, identifying interference-based control of
non-equilibrium dynamics as \roop{a potential resource for quantum
optimization and algorithm design}. Recent demonstrations of gate-engineered
interference in digital quantum circuits~\cite{Abanin2025} \roop{further
highlight the connection between interference-based control in analog and
digital architectures.}

\roop{More broadly, the interference-engineering approach developed here can
provide a route to controlling dynamical freezing in other forms of
nonequilibrium quantum matter, where related freezing phenomena have also
been recently explored~\cite{PhysRevX.11.021008,mukherjee2024floquet,ggk3-6cf8}.
Our large-scale experiments demonstrate the robustness of this approach by
systematically tuning the interaction strength, lattice geometry and initial
state, while observing controlled dynamical freezing across one- and
two-dimensional arrays and extending to systems containing up to 100 atoms.
The persistence and enhancement of freezing across these distinct
microscopic conditions show that the phenomenon is not tied to the
idealized PXP limit or to a particular geometry or initial state. Numerical
results further show that the interference can also be engineered using
other drive waveforms (see Supplementary Material), indicating that the
control strategy is not restricted to a specific sinusoidal protocol.
While the optimal waveform and harmonic content will depend on the
microscopic Hamiltonian of the setup, our results demonstrate that harmonic engineering
provides a robust and experimentally accessible strategy for stabilizing
driven many-body dynamics in realistic experimental platforms.}

Finally, our results place the striking dynamical freezing predicted in constrained models into a realistic many-body setting. Dynamical freezing arises from emergent conservation laws that strongly suppress thermalization, producing long-lived non-ergodic dynamics with persistent many-body memory. In idealized PXP descriptions, where the Rydberg blockade is treated as perfect, the Hilbert-space constraint prevents processes that would break these conservation laws, allowing freezing to persist over extremely long timescales. Our experiments show how realistic Rydberg interactions modify this picture: finite-range interaction tails reshape the interference conditions underlying freezing, producing patterns that depart from constrained descriptions and shifting the regimes where stabilization occurs. \roop{Rather than representing these interaction effects solely as a limitation, our results show that they can be characterized and actively controlled through the interaction geometry and driving protocol.} In this way, our results clarify both the limits of idealized blockade models and \roop{practical routes for engineering robust dynamical freezing in interacting Floquet quantum matter.}

\section*{Methods}

\subsection{Experimental platform}

Experiments were performed on QuEra’s  programmable quantum processor, \textit{Aquila}~\cite{Querawhitepaper}, which traps up to 256 individual $^{87}$Rb atoms in reconfigurable optical tweezer arrays. Each atom encodes a qubit between the ground state $\ket{g}=\ket{5S_{1/2}}$ and a Rydberg state $\ket{r}=\ket{70S_{1/2}}$, coupled by a far-detuned two-photon transition (420~nm + 1013~nm). Pairs of Rydberg excitations interact via a van der Waals potential
\[
V_{ij}=\frac{C_6}{r_{ij}^6},\qquad C_6 =5,420,503~\mu\mathrm{m}^6~\mathrm{rad}/\mu\mathrm{s},
\]
(calibrated for $\ket{70S_{1/2}}$). The maximum Rabi frequency which can be used is $\Omega_0^{\mathrm{max}} = 15.6~\mathrm{rad}/\mu\mathrm{s}$ (corresponding to $\approx 2.48$~MHz), which sets a characteristic static blockade radius for zero detuning, i.e. $\Delta=0$,
\[
R_b = \left(\frac{C_6}{|\Omega_0^{\mathrm{max}}|}\right)^{1/6} \approx 8.4~\mu\mathrm{m}.
\]
In our driven protocols $\Omega(t)$ and $\Delta(t)$ vary periodically, so the instantaneous blockade radius $R_b(t)=(C_6/\sqrt{\Omega(t)^2+\Delta(t)^2})^{1/6}$ is itself time-dependent over the cycle. Unless noted otherwise, quoted blockade values refer to $t{=}0$.

Atoms were optically pumped to the $\ket{g}$ state with fidelity exceeding $99\%$ before each experimental run. Finite site-loading imperfections introduce an average defect probability of $p \approx 0.007$ per site. Thus it is expected that we obtain defect-free configurations in approximately $58\%$ of attempts for $\sim100$-atom arrays~\cite{Querawhitepaper}, subject to some corrections based on geometry. All reported data were obtained exclusively from verified defect-free realizations, identified via fluorescence imaging. To visualize geometry-dependent trapping reliability, the probability of successful atom loading is indicated by graded colour shading in the geometry figures shown in the main text. To ensure uniformity while maintaining experimental efficiency, we restrict all measurements to arrays containing $84–100$ atoms across one- and two-dimensional geometries.

One-dimensional “snake” chains were limited by the $75\times75~\mu\mathrm{m}^2$ active region of the device. At an interatomic spacing of $d=4.7~\mu$m, up to $100$ atoms could be accommodated; increasing $d$ to $5.0$ and $5.3~\mu$m reduced the usable system size to 88 and 84, respectively. Two-dimensional square and honeycomb arrays of comparable scale were realized under interatomic spacing of $4.7 \mu m$ as well.

\subsection{Drive protocols}

Time-dependent control of the detuning $\Delta(t)$ and Rabi frequency $\Omega(t)$ was implemented using piecewise-linear waveforms with 50~ns temporal resolution. We employed two primary protocols:
\begin{equation}
\Delta(t)=\Delta_0\cos(\omega t), \qquad
\Omega(t)=\frac{\Omega_0}{2}\big[1+\cos(r\,\omega t)\big],
\end{equation}
where $r=0$ (single-frequency) modulates only detuning with constant Rabi amplitude $\Omega_0$, and $r=2$ (dual-frequency) adds a commensurate Rabi modulation.

Drive angular frequencies spanned $\omega = 1.2$–$4.5~\mathrm{rad}/\mu\mathrm{s}$, with most data taken between $1.5$ and $4.5~\mathrm{rad}/\mu\mathrm{s}$ so that single-cycle durations $T=2\pi/\omega \lesssim 4~\mu\mathrm{s}$ remain within the system coherence window reported in Aquila whitepaper~\cite{Querawhitepaper}. Each cycle began and ended at $\Omega=0$, with $\sim 250$~ns rise and fall times set by hardware constraints.

A practical limitation at higher $\omega$ is the fixed rise and fall overhead in $\Omega(t)$. Each cycle includes $\sim50$~ns ramps at both the beginning and end, amounting to a total overhead of $0.1~\mu$s per period. The fractional time spent ramping,
\[
f_{\mathrm{ramp}} = \frac{2\tau_{\mathrm{ramp}}}{T} = \frac{\tau_{\mathrm{ramp}}\omega}{\pi},
\]
increases linearly with drive frequency. Consequently, around  $10\%$ of each cycle is consumed by ramps at $\omega=4.5~\mathrm{rad}/\mu\mathrm{s}$, rendering still higher frequencies inefficient. Most experiments were therefore restricted to $\omega \leq 4.5~\mathrm{rad}/\mu\mathrm{s}$, where the effective portion of the cycle used for coherent evolution exceeds $90\%$.

Typical control amplitudes were $\Omega_0 = 2$–$4~\mathrm{rad}/\mu\mathrm{s}$ and $\Delta_0 = 20~\mathrm{rad}/\mu\mathrm{s}$, unless otherwise stated. All measurements report the Rydberg population after exactly one complete drive cycle ($N=1$).

\subsection{Measurement and error correction}
Rydberg state populations were measured via fluorescence detection after each drive sequence, averaging at least $400$ defect-free shots per data point (corresponding to $900-1500$ total attempts depending on loading probability). State-preparation and measurement (SPAM) errors were corrected using calibrated values $\epsilon_{\text{det},g} = 0.01$ (false ground detection) and $\epsilon_{\text{det},r} = 0.08$ (false Rydberg detection)~\cite{Querawhitepaper}. For local observables, the measured Rydberg population was corrected according to
\begin{equation}
n(T) = \frac{n_{\text{meas}} - \epsilon_{\text{det},g}}
{1 - \epsilon_{\text{det},g} - \epsilon_{\text{det},r}}.
\end{equation}
\roop{For the probability of an exact many-body configuration, such as the $Z_2$ state $|10101010101\rangle$, this single-site correction is insufficient, since detection errors can transfer probability between different many-body bitstrings. We therefore additionally perform a full SPAM correction of the measured bitstring distribution. Denoting by $\mathbf{p}_{\mathrm{meas}}$ and $\mathbf{p}_{\mathrm{true}}$ the measured and underlying probability distributions over the $2^{11}$ possible configurations, respectively, we use
\begin{equation}
\mathbf{p}_{\mathrm{meas}} = \mathcal{M}\mathbf{p}_{\mathrm{true}},
\end{equation}
where $\mathcal{M}$ is the $2^{11}\times2^{11}$ confusion matrix constructed from the calibrated single-atom detection errors. Assuming independent detection errors between atoms, its elements are given by
\begin{equation}
\mathcal{M}_{\mathbf{m},\mathbf{n}}
=
\prod_{i=1}^{11}
P(m_i|n_i),
\end{equation}
with $P(1|0)=\epsilon_{\text{det},g}$ and $P(0|1)=\epsilon_{\text{det},r}$. The underlying many-body distribution is then obtained by solving
\begin{equation}
\mathbf{p}_{\mathrm{true}}
=
\mathcal{M}^{-1}\mathbf{p}_{\mathrm{meas}},
\end{equation}
and the $Z_2$ fidelity is identified with the corrected probability $p_{\mathrm{true}}(10101010101)$. This full-distribution correction is required for the many-body fidelity, whereas the single-site correction above is sufficient for local quantities such as $\langle n\rangle$.}

Error bars in figures represent standard error of the mean computed from shot-to-shot fluctuations. Nearest-neighbour interatomic distances were uniform within each geometry to within $\pm 50$~nm, except at snake-pattern bends where natural geometric variations occur.

\subsection{Analytical understanding}
\label{app:analytical}
\roop{Relegating the detailed derivations to the Supplementary Material, we
summarize here the effective Floquet Hamiltonian used to obtain the
semi-analytical results shown in the main text. For single-frequency
driving, the stroboscopic evolution over one period is written as~\cite{PhysRevB.102.235114}
\begin{equation}
U(T)
=
e^{-iT H_F^{(0)}}
\exp\left[-iT\left(H_F^{(1)}+H_F^{(2)}\right)\right]
+\mathcal{O}(\Omega_0^3),
\end{equation}
where the zeroth-order contribution is
\begin{equation}
H_F^{(0)}
=
\sum_i\sum_{j\in\mathrm{LR}(i)}
V_{ij}\,n_i n_j ,
\label{eq:HF0}
\end{equation}
and as discussed in the main text, the first-order interaction-picture contribution is
\begin{equation}
H_F^{(1)}
=
\Omega_0 \sum_i
\left[
\tilde{\sigma}_i^+\,\mathcal{I}(\alpha_i)
+
\tilde{\sigma}_i^-\,\mathcal{I}^\ast(\alpha_i)
\right],
\label{eq:HF1}
\end{equation}
with
\begin{equation}
\mathcal{I}(\alpha)
=
\left(e^{i2\pi\alpha/\omega}-1\right)
\sum_{n=-\infty}^{\infty}
\frac{J_n(\Delta_0/\omega)}
{iT(\alpha+n\omega)}.
\label{eq:Ialpha_closed}
\end{equation}
Here
$\alpha_i=\sum_{j\in\mathrm{LR}(i)}V_{ij}n_j$ is the interaction-induced energy shift at site $i$.

For the vacuum initial state, the first-order matrix elements relevant
to vacuum-to-one-excitation transitions are independent of the
interaction shifts $\alpha_i$. Consequently, the leading
interaction-induced modification of the freezing condition enters
through the second-order contribution, which we retain below:
\begin{equation}
H_F^{(2)}
=
-\frac{i\Omega_0^2}{T}
\sum_{i,j}\sum_{s,s'=\pm}
\left[
K_{ij}^{(s,s')}
-\frac{T^2}{2}
\mathcal{I}_i^{(s)}
\mathcal{I}_j^{(s')}
\right]
\tilde{\sigma}_i^s\tilde{\sigma}_j^{s'},
\label{eq:HF2}
\end{equation}
where
\begin{equation}
\mathcal{I}_i^{(+)}=\mathcal{I}(\alpha_i),
\qquad
\mathcal{I}_i^{(-)}=\mathcal{I}^{\ast}(\alpha_i),
\end{equation}
and
\begin{equation}
K_{ij}^{(s,s')}
=
\int_0^Tdt_1\int_0^{t_1}dt_2\,
e^{is\phi_i(t_1)}
e^{is'\phi_j(t_2)}
e^{-iss'V_{ij}t_2},
\label{eq:Kij}
\end{equation}
with
\begin{equation}
\phi_i(t)
=
\alpha_i t+
\frac{\Delta_0}{\omega}\sin(\omega t).
\end{equation}

To evaluate the second-order kernel, we expand the oscillatory phase
factors using the Jacobi--Anger identity. This reduces the kernel to a
sum over products of Bessel-function amplitudes and elementary time
integrals,
\begin{equation}
K_{ij}^{(s,s')}
=
\sum_{m,n=-\infty}^{\infty}
J_m\!\left(\frac{\Delta_0}{\omega}\right)
J_n\!\left(\frac{\Delta_0}{\omega}\right)
\mathcal{J}_{mn}^{(s,s')},
\label{eq:Kij_sum}
\end{equation}
where
\begin{equation}
\mathcal{J}_{mn}^{(s,s')}
=
\int_0^Tdt_1\,e^{iAt_1}
\int_0^{t_1}dt_2\,e^{iBt_2},
\end{equation}
with
\begin{equation}
A=s(\alpha_i+m\omega),\qquad B=s'(\alpha_j-sV_{ij}+n\omega).
\end{equation}
For $A\neq0$, $B\neq0$, and $A+B\neq0$, the integral evaluates to
\begin{equation}
\mathcal{J}_{mn}^{(s,s')}
=
-\left[
\frac{e^{i(A+B)T}-1}{B(A+B)}
-
\frac{e^{iAT}-1}{AB}
\right].
\label{eq:Jkernel}
\end{equation}
The cases in which any of the denominators vanish are obtained from
the corresponding continuous limits of Eq.~(\ref{eq:Jkernel}); these
are removable singularities of the integrated expression. Using
$T=2\pi/\omega$, the integer-harmonic factors further simplify since
$e^{im\omega T}=1$. The complete evaluation of the resulting
second-order kernel is given in the Supplementary Material.

The semi-analytical results are obtained by constructing the effective
Floquet propagator from $H_F^{(0)}$, $H_F^{(1)}$, and $H_F^{(2)}$ and
evolving the same initial state as in the corresponding full numerical
calculation. The resulting excitation densities give the
semi-analytical curves shown in Figs.~2--4. For numerical evaluation,
the Bessel-function sums are truncated at $|n|\leq20$ for
one-dimensional systems and $|n|\leq16$ for two-dimensional systems.
These cutoffs retain the relevant harmonic contributions while keeping
the evaluation of the effective Hamiltonian numerically tractable. No
parameters are fitted to the experimental data; all interaction
strengths, drive parameters, and geometrical parameters are taken
directly from the microscopic Hamiltonian and experimental
calibration.

The Floquet analysis generalizes directly to an amplitude-modulated
transverse drive
\begin{equation}
\Omega(t)=\frac{\Omega_0}{2}\left[1+\cos(r\omega t)\right],
\end{equation}
where $r$ is an integer. To first order in the interaction-picture
expansion, the effective Floquet Hamiltonian is
\begin{equation}
H_F^{(1)}
=
\frac{\Omega_0}{2}
\sum_i
\left[
\tilde{\sigma}_i^+\mathcal{I}_r(\alpha_i)
+
\tilde{\sigma}_i^-\mathcal{I}_r^\ast(\alpha_i)
\right],
\label{eq:HFr1}
\end{equation}
where $\alpha_i=\sum_{j\in\mathrm{LR}(i)}V_{ij}n_j$ and
\begin{widetext}
\begin{equation}
\mathcal{I}_r(\alpha)
=
\left(e^{i2\pi\alpha/\omega}-1\right)
\sum_{n=-\infty}^{\infty}
\frac{J_n(\Delta_0/\omega)}{iT}
\left[
\frac{1}{\alpha+n\omega}
+
\frac{1}{2}
\left(
\frac{1}{\alpha+(n+r)\omega}
+
\frac{1}{\alpha+(n-r)\omega}
\right)
\right],
\qquad
T=\frac{2\pi}{\omega}.
\label{eq:Ir}
\end{equation}
\end{widetext}

It is convenient to introduce the effective harmonic coefficient
\begin{equation}
\mathcal{B}_n^{(r)}(x)
=
J_n(x)
+\frac{1}{2}J_{n-r}(x)
+\frac{1}{2}J_{n+r}(x),
\qquad
x=\frac{\Delta_0}{\omega},
\label{eq:Br}
\end{equation}
in terms of which Eq.~(\ref{eq:Ir}) can equivalently be written as
\begin{equation}
\mathcal{I}_r(\alpha)
=
\sum_{n=-\infty}^{\infty}
\mathcal{B}_n^{(r)}(x)
\frac{e^{i(\alpha+n\omega)T}-1}
{iT(\alpha+n\omega)}.
\label{eq:Ir_compact}
\end{equation}
For the vacuum state, as in the single-frequency case, the
interaction-induced energy shift satisfies
$\alpha_i|\mathrm{vac}\rangle=0$. The first-order Floquet amplitude then
reduces to
\begin{equation}
\mathcal{I}_r(0)
=
J_0(x)
+
\frac{1+(-1)^r}{2}J_r(x),
\qquad
x=\frac{\Delta_0}{\omega}.
\label{eq:Ir_zero}
\end{equation}
Interaction-induced higher-order corrections can nevertheless shift and
reshape the freezing minima, so that the interacting dynamics need not
reproduce all interference minima of the ideal $PXP$ limit
(see Figs.~\ref{fig:1dchain} and~\ref{fig:2dgeom}).

For even $r$, the additional harmonic interferes with the $J_0$
contribution. In particular, for $r=2$,
\begin{equation}
\mathcal{B}_0^{(2)}(x)
=
J_0(x)+J_2(x)
=
\frac{2J_1(x)}{x},
\end{equation}
as discussed above. The corresponding modification
of the leading harmonic amplitude for representative values of $r$ is
shown in Fig.~\ref{fig:besselcheck}.

\begin{figure}
    \centering
    \includegraphics[width=0.98\linewidth]{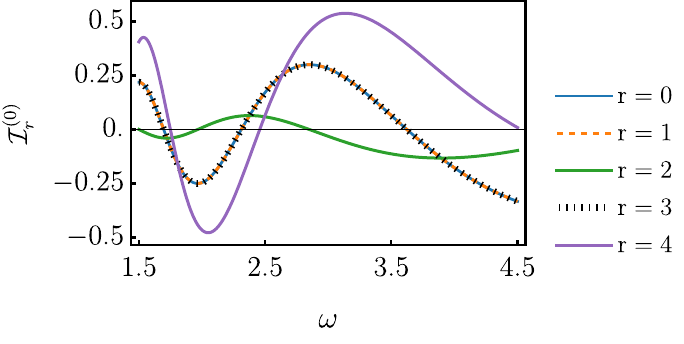}
    \caption{Modification of the leading Floquet amplitude for
    amplitude-modulated driving with different values of $r$, as
    determined by Eq.~\eqref{eq:Ir_zero}.}
    \label{fig:besselcheck}
\end{figure}

The same modified harmonic coefficients enter the higher-order Floquet
contributions. In particular, the second-order kernel can be written
compactly as
\begin{equation}
K_{ij,r}^{(s,s')}
=
\sum_{m,n=-\infty}^{\infty}
\mathcal{B}_m^{(r)}(x)
\mathcal{B}_n^{(r)}(x)
\mathcal{J}_{mn}^{(s,s')},
\label{eq:Kij_r}
\end{equation}
where $\mathcal{J}_{mn}^{(s,s')}$ is the elementary time integral
defined above for the single-frequency case. Thus, the single-frequency
Bessel products
$J_m(x)J_n(x)$ are replaced by the corresponding products
$\mathcal{B}_m^{(r)}(x)\mathcal{B}_n^{(r)}(x)$. Expanding these products
gives nine Bessel-function combinations; their full expression and
contribution to the second-order Floquet Hamiltonian are given in the
Supplementary Material. Equation~(\ref{eq:Kij_r}) is the compact form
used in the numerical evaluation.

For $r=2$, the vanishing of
$\mathcal{B}_0^{(2)}=2J_1(x)/x$ suppresses the dominant harmonic channel
not only at first order but also in second-order pathways containing
this channel. Residual contributions from nonzero harmonic indices
remain and account for the finite interaction-induced dynamics near
the freezing points.

For the numerical evaluation of the analytical expressions, we construct
the effective Floquet Hamiltonian for $L=12$ sites in one dimension and
$L=16$ sites for the two-dimensional geometries. The Bessel-function sums
are truncated at $|n|\leq20$ and $|n|\leq16$ for the one- and
two-dimensional cases, respectively. The analytical expressions
themselves are not restricted to these system sizes and can be evaluated
for larger systems; the finite sizes used here provide a computationally
tractable evaluation of the effective Floquet Hamiltonian. The resulting
semi-analytical predictions are shown without any adjustable fitting
parameters.}
\\
\paragraph{\textbf{Data availability:}} The data that support the findings of this study are available at https://doi.org/10.5281/zenodo.22677714
\paragraph{\textbf{Code availability:}} The codes used for the simulations and analysis in this study are available at https://doi.org/10.5281/zenodo.22677714.
\begin{acknowledgements}
This research used resources of the National Energy Research Scientific Computing Center (NERSC), a Department of Energy Office of Science User Facility under Contract No. DE-AC02-05CH11231 using NERSC award for QCAN Project DDR-ERCAP0033861. This project was also funded and supported by the UK National Quantum Computer Centre [NQCC200921], which is a UKRI Centre and part of the UK National Quantum Technologies Programme (NQTP). SB, MS and RG acknowledge EPSRC-SFI funded project EP/X039889/1 (GeQuantumBus). BZ was supported by the Engineering and Physical Sciences Research Council [grant numbers EP/R513143/1, EP/T517793/1]. B. M. was funded by Department of Science \& Technology, Government of India via the INSPIRE Faculty programme. B.M was also funded by the European Research Council (ERC)
under the European Union’s Horizon 2020 research and
innovation programme (Grant Agreement No. 853368), near the beginning of this project. The authors would like to thank Pedro Lopez, Milan Kornja\v{c}a, Katherine Klymko and other members of the QuEra-NERSC team for discussions. RG and MS would like to thank Krishnendu Sengupta, Arnab Sen and Alexander Nico-Katz for discussions. BM would like to thank Krishnendu Sengupta and Dolev Bluvstein for discussions. 
\end{acknowledgements}
\bibliography{ref}

\paragraph{\textbf{Author contributions:}} B.Z. prepared the experiments in QuEra \textit{Aquila}, which were jointly performed by R.G., M.S. and B.Z. M.S. performed the data analysis and generated the figures. M.S. and R.G. carried out the analytical calculations and numerical simulations for the long-range Rydberg systems. B.M. performed the PXP numerics. The project was conceived during discussions among S.B., R.G., M.S. and B.M. R.G. co-ordinated the experimental, numerical and theoretical components of the work. All authors contributed to writing the manuscript.
\\
\paragraph{\textbf{Competing Interests:}} The authors declare no competing interests.

\appendix
\onecolumngrid

\section{The Floquet Perturbation expansion}

In the main text we define,
For our driven Rydberg system, we define,
$H(t)=H_0(t)+H_1$, with
\begin{align}
H_0(t)
&=
\Delta_0 \cos(\omega t)\sum_i n_i
+
\sum_{j\in\mathrm{LR}(i)} V_{ij}\, n_i n_j ,
\label{eq:H0}\\
H_1
&=
\Omega \sum_i \tilde{\sigma}_i^x .
\label{eq:H1maintext}
\end{align}
The Floquet perturbation expansion proceeds analogously to a Dyson series formulated in the interaction picture~\cite{PhysRevB.102.235114}. The zeroth-order term defines the unperturbed evolution operator
\begin{equation}
    U_0(T,0) = e^{-i H_F^{(0)} T} = \mathcal{T} \exp\left[-i \int_0^T H_0(t)\, dt \right],
\end{equation}
where $\mathcal{T}$ denotes time ordering. Since $H_0(t)$ is diagonal at all times, the corresponding zeroth-order evolution operator is obtained exactly as
\begin{equation}
U_0(t)
=
\exp\!\left[
-i\frac{\Delta_0}{\omega}\sin(\omega t)\sum_i n_i
\right]
\exp\!\left[
-i t\sum_{j\in\mathrm{LR}(i)} V_{ij}\, n_i n_j
\right].
\label{eq:U0}
\end{equation}

The zeroth-order contribution to the Floquet Hamiltonian arises from the evolution operator $U_0(T)$ over one driving period. Since $\sin(\omega T)=0$ for $T=2\pi/\omega$, the contribution from the time-dependent detuning vanishes over a full period, and one finds
\begin{equation}
U_0(T)
=
\exp\!\left[
-i T \sum_{j\in\mathrm{LR}(i)} V_{ij}\, n_i n_j
\right].
\label{eq:U0T}
\end{equation}
This defines the zeroth-order Floquet Hamiltonian
\begin{equation}
H_F^{(0)}
=
\sum_{j\in\mathrm{LR}(i)} V_{ij}\, n_i n_j ,
\label{eq:HF0}
\end{equation}
which corresponds to the static interaction energy within the blockade-constrained Hilbert space.

To calculate the first order term, we start from the interaction-picture Hamiltonian
\begin{equation}
H_I(t)=U_0^\dagger(t) H_1 U_0(t),
\end{equation}
which takes the form
\begin{equation}
H_I(t)
=
\Omega \sum_i
\left[
\tilde{\sigma}_i^+ e^{+i\phi_i(t)}
+
\tilde{\sigma}_i^- e^{-i\phi_i(t)}
\right],
\label{eq:HI_supp}
\end{equation}
with
\begin{equation}
\phi_i(t)
=
\alpha_i t + \frac{\Delta_0}{\omega}\sin(\omega t),
\qquad
\alpha_i = \sum_{j\in\mathrm{LR}(i)} V_{ij}\, n_j .
\label{eq:phi_supp}
\end{equation}
Here $\tilde{\sigma}_i^\pm$ are projected spin-flip operators acting within the blockade subspace. The quantity $\alpha_i$ is diagonal in the occupation-number basis and therefore commutes with itself at different times, allowing it to be treated as a c-number within the time integral.
The effective Floquet Hamiltonian $H_F$ is then defined in the interaction frame generated by $U_0(t)$.
Transforming back to the laboratory frame yields $U(T)=U_0(T)e^{-iH_F T}$, where $U_0(T)$ is diagonal in the occupation basis and contributes only configuration-dependent phases, leaving the effective coupling structure unchanged.

\subsection*{First-order Dyson expansion}

The time-evolution operator in the interaction picture is
\begin{equation}
U_I(T)
=
\mathcal{T}
\exp\!\left(
-i\int_0^T H_I(t)\,dt
\right),
\end{equation}
where $\mathcal{T}$ denotes time ordering. To first order in the Dyson expansion, this reduces to
\begin{equation}
U_I(T)
\simeq
\mathbb{I}
-i\int_0^T H_I(t)\,dt .
\end{equation}
The effective Floquet Hamiltonian at this order is defined through
\begin{equation}
U_I(T) \equiv e^{-i H_F^{(1)} T},
\end{equation}
which yields
\begin{equation}
H_F^{(1)}
=
\frac{1}{T}\int_0^T H_I(t)\,dt,
\qquad
T=\frac{2\pi}{\omega}.
\label{eq:HF1_def_supp}
\end{equation}

Substituting Eq.~\eqref{eq:HI_supp} into Eq.~\eqref{eq:HF1_def_supp} gives
\begin{equation}
H_F^{(1)}
=
\Omega \sum_i
\left[
\tilde{\sigma}_i^+\,\mathcal{I}(\alpha_i)
+
\tilde{\sigma}_i^-\,\mathcal{I}^\ast(\alpha_i)
\right],
\label{eq:HF1}
\end{equation}
where
\begin{equation}
\mathcal{I}(\alpha)
=
\frac{1}{T}
\int_0^T
dt\;
e^{i\left[\alpha t + (\Delta_0/\omega)\sin(\omega t)\right]}.
\label{eq:Ialpha_supp}
\end{equation}

\subsection*{Evaluation of the time integral}

To evaluate Eq.~\eqref{eq:Ialpha_supp}, we employ the Jacobi--Anger expansion
\begin{equation}
e^{iz\sin\theta}
=
\sum_{n=-\infty}^{\infty}
J_n(z)\,e^{in\theta}.
\label{eq:jacobi_anger}
\end{equation}
This allows us to write
\begin{equation}
e^{i(\Delta_0/\omega)\sin(\omega t)}
=
\sum_{n=-\infty}^{\infty}
J_n(\Delta_0/\omega)\,e^{in\omega t}.
\end{equation}
Substituting into Eq.~\eqref{eq:Ialpha_supp}, we obtain
\begin{align}
\mathcal{I}(\alpha)
&=
\frac{1}{T}
\sum_{n=-\infty}^{\infty}
J_n(\Delta_0/\omega)
\int_0^T
dt\;
e^{i(\alpha+n\omega)t}.
\end{align}

The remaining time integral can be evaluated explicitly as
\begin{equation}
\int_0^T dt\, e^{i(\alpha+n\omega)t}
=
\frac{e^{i(\alpha+n\omega)T}-1}{i(\alpha+n\omega)}.
\end{equation}
Using $T=2\pi/\omega$ and the identity $e^{in\omega T}=e^{i2\pi n}=1$, this simplifies to
\begin{equation}
\int_0^T dt\, e^{i(\alpha+n\omega)t}
=
\frac{e^{i2\pi\alpha/\omega}-1}{i(\alpha+n\omega)}.
\end{equation}

Combining terms yields the closed-form expression
\begin{equation}
\mathcal{I}(\alpha)
=
\big(e^{i2\pi\alpha/\omega}-1\big)
\sum_{n=-\infty}^{\infty}
\frac{J_n(\Delta_0/\omega)}{iT(\alpha+n\omega)},
\label{eq:Ialpha_closed_supp}
\end{equation}
which corresponds to Eq.~(12) in the main text.

\subsection*{Physical interpretation}

Since $\alpha_i$ depends only on occupations outside the blockade radius, the coefficient $\mathcal{I}(\alpha_i)$ conditions the spin-flip dynamics at site $i$ on the surrounding Rydberg configuration. The resulting effective Hamiltonian therefore describes density-assisted spin-flip processes arising from interaction tails beyond the blockade radius.

In the absence of interactions beyond the blockade radius ($V_{ij}=0$), 
Eqs.~\eqref{eq:HF0} and~\eqref{eq:HF1} reduce to
$H_F^{(0)}=0$ and
$H_F^{(1)}=\Omega J_0(\Delta_0/\omega)\sum_i \tilde{\sigma}_i^x$.
This limit corresponds to the PXP model, which retains only constrained coherent spin flips and neglects interaction-induced corrections arising beyond the blockade radius.

\rg{\subsection*{Breakdown of the first-order description}

The first-order Floquet Hamiltonian captures density-assisted single spin-flip processes conditioned on the interaction-induced energy shifts $\alpha_i$. Consequently, all matrix elements connecting configurations containing one or more Rydberg excitations are described through the coefficient $\mathcal{I}(\alpha_i)$.

However, the vacuum sector requires special consideration. Since the vacuum state contains no Rydberg excitations, one has
$\alpha_i=0$,
for every site. The corresponding first-order coefficient therefore reduces to $\mathcal{I}(0)=J_0\!\left(\frac{\Delta_0}{\omega}\right)$.
Consequently, near the dynamical freezing points determined by the zeros of $J_0$, the first-order Floquet Hamiltonian predicts a complete suppression of transitions out of the vacuum.

While this reproduces the freezing expected within the PXP description, it fails to capture the interaction-induced corrections observed in the main text. In particular, the first-order theory cannot account for the finite dynamics that persist in the vicinity of the freezing points, nor for the interaction-induced shifts and changes of the effective coupling across successive Bessel zeros. These effects originate from higher-order virtual processes and therefore require extending the Floquet perturbation expansion beyond first order.

\subsection*{Second-order Dyson expansion}

To capture the leading coupling of the vacuum sector to states containing Rydberg excitations, we extend the Floquet perturbation expansion to second order. The Dyson expansion of the interaction-picture evolution operator till second order is

\begin{equation}
U_I(T)
=
\mathbb{I}
-i\int_0^Tdt\,H_I(t)
\nonumber
-\int_0^Tdt_1
\int_0^{t_1}dt_2\,
H_I(t_1)H_I(t_2)
+O(\Omega^3).
\label{eq:DysonSecond}
\end{equation}

The effective Floquet Hamiltonian is obtained from the logarithm of the evolution operator,
$U_I(T)=e^{-iH_FT}$,
which yields, up to second order,

\begin{equation}
H_F
=
H_F^{(1)}
+
H_F^{(2)}
+
O(\Omega^3).
\end{equation}

Thus one can expand the interaction-picture propagator as $U_I(T,0)
=
1
+
U_1(T,0)
+
U_2(T,0)
+
O(\Omega^3)$,
and define, 

\begin{align}
U_1(T,0)
&=
-i\int_0^Tdt\,H_I(t),\\
U_2(T,0)
&=
(-i)^2
\int_0^Tdt_1
\int_0^{t_1}dt_2\,
H_I(t_1)H_I(t_2).
\end{align}
Comparing term by term one can show that $H_F^{(2)}=\frac{i}{T}(U_2-\frac{U_1^2}{2})$.
The effective Floquet Hamiltonian till the desired order is obtained by combining the interaction-picture expansion with the exactly known zeroth-order evolution operator $U_0(T)$, as discussed in the previous section.

For the second-order calculation, it is convenient to rewrite
Eq.~\eqref{eq:HI_supp} in a compact form by introducing the sign
variable $s=\pm1$, with
$
\tilde{\sigma}_i^{(+1)}\equiv\tilde{\sigma}_i^{+},
\tilde{\sigma}_i^{(-1)}\equiv\tilde{\sigma}_i^{-}$.

Equation~\eqref{eq:HI_supp} can then be written as
\begin{equation}
H_I(t)
=
\Omega
\sum_i
\sum_{s=\pm1}
e^{is\phi_i(t)}
\tilde{\sigma}_i^{\,s},
\label{eq:HIcompact}
\end{equation}
where $\phi_i(t)$ is defined in Eq.~\eqref{eq:phi_supp}.

Substituting Eq.~\eqref{eq:HIcompact} into the second-order Dyson term gives

\begin{align}
U_2(T,0)
&=
(-i)^2
\Omega^2
\sum_{ij}
\sum_{s_1,s_2=\pm1}
\int_0^Tdt_1
\int_0^{t_1}dt_2
\,
e^{is_1\phi_i(t_1)}
\tilde{\sigma}_i^{\,s_1}
e^{is_2\phi_j(t_2)}
\tilde{\sigma}_j^{\,s_2}.
\end{align}

Using the commutation relation

\begin{equation}
\tilde{\sigma}_i^{\,s_1}
e^{is_2\alpha_jt}
=
e^{-is_1s_2V_{ji}t}
e^{is_2\alpha_jt}
\tilde{\sigma}_i^{\,s_1},
\label{eq:commutation}
\end{equation}

together with the fact that all operators $\alpha_i$ commute, the second-order contribution can be written as

\begin{equation}
U_2(T,0)
=
(-i)^2
\Omega^2
\sum_{ij}
\sum_{s_1,s_2=\pm1}
K_{ij}^{(s_1,s_2)}
\,
\tilde{\sigma}_i^{\,s_1}
\tilde{\sigma}_j^{\,s_2},
\label{eq:U2kernel}
\end{equation}

where

\begin{equation}
K_{ij}^{(s_1,s_2)}=\int_0^T dt_1\int_0^{t_1}dt_2\,
e^{is_1\phi_i(t_1)}\,e^{is_2\phi_j(t_2)}\,e^{-is_1s_2V_{ij}t_2}.
\label{eq:KernelDefinition}
\end{equation}

Substituting the explicit expression $\phi_i(t)
=
\alpha_i t
+
\frac{\Delta_0}{\omega}\sin(\omega t)$,
into Eq.~\eqref{eq:KernelDefinition} gives

\begin{equation}
K_{ij}^{(s_1,s_2)}=\int_0^T dt_1\int_0^{t_1}dt_2\,
e^{is_1\frac{\Delta_0}{\omega}\sin(\omega t_1)}e^{is_2\frac{\Delta_0}{\omega}\sin(\omega t_2)}
e^{it_1s_1\alpha_i}\,e^{it_2s_2(\alpha_j-s_1V_{ij})}.
\label{eq:KernelExpanded1}
\end{equation}

Expanding the oscillatory phase factors using the Jacobi-Anger identity, 
$e^{is_1\frac{\Delta_0}{\omega}\sin(\omega t)}
=
\sum_{n=-\infty}^{\infty}
J_n\!\left(\frac{\Delta_0}{\omega}\right)
e^{is_1n\omega t}$, 
one finds, $K_{ij}^{(s_1,s_2)}
= \sum_{n,m=-\infty}^{\infty}
J_n\!\left(\frac{\Delta_0}{\omega}\right)
J_m\!\left(\frac{\Delta_0}{\omega}\right)
\,I_{nm}^{(s_1,s_2)}$, where 
\begin{equation}
I_{nm}^{(s_1,s_2)}=\int_0^T dt_1\,e^{it_1s_1(\alpha_i+n\omega)}\int_0^{t_1}dt_2\,e^{it_2s_2(\alpha_j-s_1V_{ij}+m\omega)}.
\label{eq:Inm}
\end{equation}
The expressions above determine the second-order contribution to the
interaction-picture propagator. 
Accordingly, the full Floquet propagator over one driving period is

\begin{equation}
\boxed{
U(T)
=
U_0(T)
\left(
1+U_1+U_2
\right)
=
U_0(T)
\exp\!\left[
-iT(H_F^{(1)}+H_F^{(2)})
\right]
+
O(\Omega^3).
}
\label{eq:FullPropagator}
\end{equation}
\subsection*{Evaluation of the second-order kernel}

The time integral appearing in Eq.~\eqref{eq:Inm} can be written in the compact form

\begin{equation}
I_{nm}^{(s_1,s_2)}
=
\int_0^Tdt_1\,
e^{iAt_1}
\int_0^{t_1}dt_2\,
e^{iBt_2},
\label{eq:InmCompact}
\end{equation}

where we have introduced the shorthand notation $A=s_1(\alpha_i+n\omega),\qquad B=s_2(\alpha_j-s_1V_{ij}+m\omega).$. The inner integral evaluates straightforwardly to
$\int_0^{t_1}dt_2\,e^{iBt_2}
=
\frac{e^{iBt_1}-1}{iB}$, 
provided $B\neq0$. 
Performing the remaining time integration yields

\begin{equation}
\boxed{
I_{nm}^{(s_1,s_2)}
=
-
\left[
\frac{e^{i(A+B)T}-1}{B(A+B)}
-
\frac{e^{iAT}-1}{AB}
\right],
}
\label{eq:GeneralKernel}
\end{equation}
which is valid whenever $A\neq0,
B\neq0$ and
$A+B\neq0$.

Using $T=2\pi/\omega$, the exponentials involving integer multiples of the drive frequency simplify according to $e^{in\omega T}=1$,

such that Eq.~\eqref{eq:GeneralKernel} may be written explicitly as

\begin{equation}
\boxed{
I_{nm}^{(s_1,s_2)}=-\left[\frac{e^{iT[s_1\alpha_i+s_2(\alpha_j-s_1V_{ij})]}-1}
{[s_2(\alpha_j-s_1V_{ij}+m\omega)]\,[s_1(\alpha_i+n\omega)+s_2(\alpha_j-s_1V_{ij}+m\omega)]}
-\frac{e^{iTs_1\alpha_i}-1}{[s_1(\alpha_i+n\omega)]\,[s_2(\alpha_j-s_1V_{ij}+m\omega)]}\right].}
\label{eq:GeneralKernelExplicit}
\end{equation}

For notational convenience, we continue to denote the two energy denominators by
$A=s_1(\alpha_i+n\omega),
\qquad
B=s_2(\alpha_j-rV_{ij}+m\omega)$, 
since the resonant limits are then naturally classified according to the conditions
$A=0$, $B=0$, and $A+B=0$.

The expression in Eq.~\eqref{eq:GeneralKernel} possesses removable
singularities whenever one or more of the energy denominators vanish.
These correspond to resonant virtual processes. While the results below
may be obtained by taking the appropriate limits of
Eq.~\eqref{eq:GeneralKernel}, we also list their explicit forms since
these are employed directly in our numerical implementation.
\paragraph*{\textbf{Case I: }}

When $B=0$, the inner integral becomes 
$\int_0^{t_1}dt_2=t_1$, 
and therefore $I_{nm}^{(s_1,s_2)}
= \int_0^Tdt_1\,t_1e^{iAt_1}
= \frac{-1+e^{iAT}(1-iAT)}{A^2}$.

\paragraph*{\textbf{Case II: }}
When $A=0$, the outer exponential disappears and 
$I_{nm}^{(s_1,s_2)}=
\int_0^Tdt_1
\int_0^{t_1}dt_2\,e^{iBt_2}
=
\frac{1-e^{iBT}+iBT}{B^2}$.

\paragraph*{\textbf{Case III:}}
When $A+B=0$,
the first exponential in Eq.~\eqref{eq:GeneralKernel} is replaced by
unity, giving 
$I_{nm}^{(s_1,s_2)}
= \frac{1-e^{iAT}+iAT}{A^2}$.

\paragraph*{\textbf{Case IV:}}

Finally, when both resonance conditions are simultaneously satisfied, $A=B=0$, 
the double integral evaluates to $I_{nm}^{(s_1,s_2)}
=
\frac{T^2}{2}$.
The doubly resonant contribution proportional to $T^2/2$ corresponds to
disconnected first-order processes and is exactly cancelled by the
$\frac{1}{2}U_1^2$ term when constructing the second-order Floquet
Hamiltonian from the Dyson series. Consequently, only connected virtual
processes can contribute to the effective Hamiltonian at second order.
\subsection*{Small-interaction expansion of the Floquet Hamiltonian}

The expressions derived above are exact within Floquet perturbation theory and remain valid for arbitrary interaction strengths beyond the blockade radius. However, the dependence of the effective Hamiltonian on the local interaction energies $\alpha_i$ obscures the microscopic origin of the interaction-assisted processes. To expose these effects explicitly, we now consider the regime
$\Omega \sim V_{ij} \sim \omega \ll \Delta_0$, 
where the long-range interaction tails may be treated perturbatively while remaining controlled by the large detuning scale. This expansion provides transparent analytical expressions for the effective Floquet Hamiltonian, allowing the influence of finite-range interactions to be identified term by term.

Since the local interaction energies $\alpha_i=\sum_jV_{ij}n_j$
depend explicitly on the occupations of neighbouring atoms, the resulting effective couplings become configuration dependent, providing a direct analytical description of how finite-range interactions generate genuine many-body corrections to the ideal PXP dynamics. Consequently, the expansion reveals how different excitation configurations reshape the Floquet landscape and give rise to interaction-assisted processes that are absent in the non-interacting PXP limit.}

Up to first order in the interaction-picture Dyson expansion, the effective Floquet Hamiltonian can be written as
\begin{equation}
H_F
=
H_F^{(0)} + H_F^{(1)} + O\!\left(\frac{\Omega V^2}{\Delta_0^2}\right),
\end{equation}
where the zeroth-order contribution originates entirely from the diagonal part of the Hamiltonian retained in $U_0(t)$,
\begin{equation}
H_F^{(0)}
=
\sum_{i}\sum_{j\in\mathrm{LR}(i)}
V_{ij}\, n_i n_j .
\end{equation}

The first-order contribution $H_F^{(1)}$ arises from the time average of the interaction-picture Hamiltonian and is controlled by the coefficient
\begin{equation}
\mathcal{I}(\alpha)
=
\frac{1}{T}
\int_0^T
dt\;
e^{i\left[\alpha t + (\Delta_0/\omega)\sin(\omega t)\right]} .
\end{equation}
In the present regime, the interaction-induced shift
\begin{equation}
\alpha_i = \sum_{j\in\mathrm{LR}(i)} V_{ij}\, n_j
\end{equation}
satisfies $|\alpha_i| \ll \Delta_0$, allowing $\mathcal{I}(\alpha_i)$ to be expanded perturbatively in $\alpha_i$.

Using the Jacobi--Anger expansion and expanding the integrand to linear order in $\alpha_i$, we obtain
\begin{align}
\mathcal{I}(\alpha_i)
&=
\frac{1}{T}
\int_0^T dt\;
e^{i(\Delta_0/\omega)\sin(\omega t)}
\left(
1 + i\alpha_i t + O(\alpha_i^2)
\right) \nonumber\\
&=
\frac{1}{T}
\int_0^T dt\;
e^{i(\Delta_0/\omega)\sin(\omega t)}
+
\frac{i\alpha_i}{T}
\int_0^T dt\;
t\,e^{i(\Delta_0/\omega)\sin(\omega t)}
+ O(\alpha_i^2).
\end{align}

The first term evaluates to the zeroth-order Bessel function,
\begin{equation}
\frac{1}{T}
\int_0^T dt\;
e^{i(\Delta_0/\omega)\sin(\omega t)}
=
J_0(\Delta_0/\omega),
\end{equation}
while the second term can be evaluated by inserting the Jacobi--Anger expansion and performing the time integral term by term, yielding
\begin{equation}
\frac{i}{T}
\int_0^T dt\;
t\,e^{i(\Delta_0/\omega)\sin(\omega t)}
=
\sum_{n\neq 0}
\frac{J_n(\Delta_0/\omega)}{n\omega}.
\end{equation}
Combining these results, the coefficient $\mathcal{I}(\alpha_i)$ admits the expansion
\begin{equation}
\mathcal{I}(\alpha_i)
=
J_0(\Delta_0/\omega)
+
\alpha_i
\sum_{n\neq 0}
\frac{J_n(\Delta_0/\omega)}{n\omega}
+
O\!\left(\frac{V^2}{\Delta_0^2}\right).
\end{equation}

Substituting this expression into the first-order Floquet Hamiltonian yields
\begin{equation}
\begin{aligned}
H_F^{(1)}
&=
\Omega J_0(\Delta_0/\omega)
\sum_i \tilde{\sigma}_i^x \quad+
\Omega
\sum_i
\left(
\sum_{j\in\mathrm{LR}(i)}
\lambda_{ij}\, n_j
\right)
\tilde{\sigma}_i^x ,
\end{aligned}
\end{equation}

where the interaction-induced coupling is given by
\begin{equation}
\lambda_{ij}
=
\frac{V_{ij}}{\omega}
\sum_{n\neq 0}
\frac{J_n(\Delta_0/\omega)}{n}.
\end{equation}

This expression makes explicit that, even for weak interaction tails, the effective Floquet Hamiltonian acquires density-assisted spin-flip terms that are absent in the PXP model and arise solely from interactions beyond the blockade radius.

The leading interaction-induced correction therefore corresponds to a density-assisted spin-flip process: a spin at site $i$ can flip only in the presence of a finite interaction-induced energy shift generated by Rydberg excitations at neighbouring sites beyond the blockade radius. In this regime, interactions do not merely renormalize the global flip amplitude, but condition local spin dynamics on the surrounding Rydberg configuration.
Such density-assisted processes are absent in the PXP model and provide the leading correction required to capture the experimentally observed dynamics when interactions beyond the blockade radius are retained.
\subsection*{Dual-frequency drive}

We now generalize the above analysis to an rabi-frequency modulated transverse drive of the form
\begin{equation}
\Omega(t)=\frac{\Omega_0}{2}\left[1+\cos(r\omega t)\right],
\qquad r\in\mathbb{Z}.
\end{equation}
The interaction-picture Hamiltonian becomes
\begin{equation}
H_I(t)
=
\frac{\Omega_0}{2}
\sum_i
\left[1+\cos(r\omega t)\right]
\left[
\tilde{\sigma}_i^+ e^{+i\phi_i(t)}
+
\tilde{\sigma}_i^- e^{-i\phi_i(t)}
\right],
\end{equation}
where $\phi_i(t)$ is defined as in Eq.~(S\ref{eq:phi_supp}),
\begin{equation}
\phi_i(t)
=
\alpha_i t + \frac{\Delta_0}{\omega}\sin(\omega t),
\qquad
\alpha_i=\sum_{j\in\mathrm{LR}(i)}V_{ij}n_j.
\end{equation}

To first order in the interaction-picture Dyson expansion, the effective Floquet Hamiltonian is given by the time average
$H_F^{(1)}
=
\frac{1}{T}\int_0^T H_I(t)\,dt$ with  
$T=\frac{2\pi}{\omega}$. 
Substituting the expression for $H_I(t)$ and performing the time average yields
\begin{equation}
H_F^{(1)}
=
\frac{\Omega_0}{2}
\sum_i
\left[
\tilde{\sigma}_i^+\,\mathcal{I}_r(\alpha_i)
+
\tilde{\sigma}_i^-\,\mathcal{I}_r^\ast(\alpha_i)
\right],
\end{equation}
with
\begin{equation}
\mathcal{I}_r(\alpha)
=
\frac{1}{T}
\int_0^T dt\;
\left[1+\cos(r\omega t)\right]
e^{i\left[\alpha t + (\Delta_0/\omega)\sin(\omega t)\right]}.
\label{eq:Ir_def_supp}
\end{equation}

To evaluate this integral, we again employ the Jacobi--Anger expansion
\begin{equation}
e^{i(\Delta_0/\omega)\sin(\omega t)}
=
\sum_{n=-\infty}^{\infty}
J_n(\Delta_0/\omega)\,e^{in\omega t}.
\end{equation}
Using $\cos(r\omega t)=(e^{ir\omega t}+e^{-ir\omega t})/2$, Eq.~\eqref{eq:Ir_def_supp} becomes
\begin{align}
\mathcal{I}_r(\alpha)
&=
\frac{1}{T}
\sum_{n=-\infty}^{\infty}
J_n(\Delta_0/\omega)
\int_0^T dt\;
e^{i(\alpha+n\omega)t}
+
\frac{1}{2T}
\sum_{n=-\infty}^{\infty}
J_n(\Delta_0/\omega)
\int_0^T dt\;
\left[
e^{i(\alpha+(n+r)\omega)t}
+
e^{i(\alpha+(n-r)\omega)t}
\right].
\end{align}

Each time integral can be evaluated explicitly as
\begin{equation}
\int_0^T dt\; e^{i(\alpha+m\omega)t}
=
\frac{e^{i(\alpha+m\omega)T}-1}{i(\alpha+m\omega)},
\end{equation}
which, using $T=2\pi/\omega$ and $e^{im\omega T}=1$ for integer $m$, simplifies to
\begin{equation}
\int_0^T dt\; e^{i(\alpha+m\omega)t}
=
\frac{e^{i2\pi\alpha/\omega}-1}{i(\alpha+m\omega)}.
\end{equation}
Collecting terms, we arrive at the closed-form expression

\begin{equation}
\mathcal{I}_r(\alpha)
=
\big(e^{i2\pi\alpha/\omega}-1\big)
\sum_{n=-\infty}^{\infty}
\frac{J_n(\Delta_0/\omega)}{iT}
\left[
\frac{1}{\alpha+n\omega}
+
\frac{1}{2}
\left(
\frac{1}{\alpha+(n+r)\omega}
+
\frac{1}{\alpha+(n-r)\omega}
\right)
\right],
\end{equation}

which reduces to the single-frequency result for $r=0$.
\rg{
The second-order Dyson expansion proceeds identically to the
single-frequency case and therefore will not be repeated here. The only
modification arises from the product of the two time-dependent Rabi
amplitudes,
\begin{equation}
\Omega(t_1)\Omega(t_2)
=
\frac{\Omega_0^2}{4}
\left[1+\cos(r\omega t_1)\right]
\left[1+\cos(r\omega t_2)\right],
\end{equation}
which introduces additional Floquet sidebands associated with each of the
two time integrals. Expanding the product gives
\begin{align}
\Omega(t_1)\Omega(t_2)
=
\frac{\Omega_0^2}{4}
\Big[
1
+\cos(r\omega t_1)
+\cos(r\omega t_2)
+\cos(r\omega t_1)\cos(r\omega t_2)
\Big].
\end{align}
Each cosine shifts the corresponding Floquet harmonic by
$\pm r$, so that every second-order virtual process acquires independent
sideband contributions from the two integration times. Consequently, the
second-order kernel becomes a weighted superposition of the
single-frequency kernels evaluated at the shifted harmonic indices,
\begin{align}
K_{ij}^{(s_1,s_2)}
=
\sum_{n,m=-\infty}^{\infty}
J_nJ_m
\sum_{\delta_n,\delta_m=0,\pm r}
c_{\delta_n}c_{\delta_m}\,
K_{ij}^{(s_1,s_2)}
(n+\delta_n,m+\delta_m),
\label{eq:KernelDual}
\end{align}
where
$c_0=1$,
$c_{+r}=c_{-r}=1/2$,
and
$J_n\equiv J_n(\Delta_0/\omega)$.
Explicitly, this corresponds to the original kernel together with eight
additional sideband contributions,
\begin{align}
K_{ij}^{(s_1,s_2)}(n,m)
&+\frac12K_{ij}^{(s_1,s_2)}(n+r,m)
+\frac12K_{ij}^{(s_1,s_2)}(n-r,m)
+\frac12K_{ij}^{(s_1,s_2)}(n,m+r)
+\frac12K_{ij}^{(s_1,s_2)}(n,m-r)
\nonumber\\
&+\frac14K_{ij}^{(s_1,s_2)}(n+r,m+r)
+\frac14K_{ij}^{(s_1,s_2)}(n+r,m-r)
+\frac14K_{ij}^{(s_1,s_2)}(n-r,m+r)
+\frac14K_{ij}^{(s_1,s_2)}(n-r,m-r),
\end{align}
which is precisely the form employed in the numerical implementation. Thus,
the bifrequency modulation preserves the structure of the second-order
Floquet expansion while introducing additional virtual processes through
independent sideband shifts of the two Floquet harmonics.

\subsection*{Remarks}

The derivation above shows that the bifrequency modulation preserves the
structure of the Floquet perturbation theory. Rather than introducing
new classes of virtual processes, the additional modulation frequency
dresses every first- and second-order process through Floquet sidebands
shifted by $\pm r$. Consequently, both the direct constrained spin-flip
processes and the interaction-assisted virtual processes acquire the same
sideband structure.

In the experimentally relevant regime $|\alpha_i|\ll\omega$, the
first-order coefficient admits the expansion
\begin{equation}
\mathcal I_r(\alpha_i)
=
\mathcal I_r^{(0)}
+
\alpha_i\,\mathcal I_r^{(1)}
+
O(\alpha_i^2),
\end{equation}
where
\begin{equation}
\mathcal I_r^{(0)}
=
J_0\!\left(\frac{\Delta_0}{\omega}\right)
+
\frac{1+(-1)^r}{2}
J_r\!\left(\frac{\Delta_0}{\omega}\right),
\end{equation}
and
\begin{equation}
\mathcal I_r^{(1)}
=
\sum_{n\neq0,\pm r}
J_n\!\left(\frac{\Delta_0}{\omega}\right)
\left[
\frac{1}{n\omega}
+
\frac12
\left(
\frac{1}{(n+r)\omega}
+
\frac{1}{(n-r)\omega}
\right)
\right].
\end{equation}

For the experimentally relevant choice $r=2$, the resonant constrained
spin-flip amplitude is therefore governed by the combination
$J_0(\Delta_0/\omega)+J_2(\Delta_0/\omega)$ rather than $J_0(\Delta_0/\omega)$
alone. Furthermore, the same sideband dressing enters the second-order
Floquet kernels, implying that the amplitudes of the interaction-assisted
virtual processes are renormalized by the same bifrequency mechanism.
Consequently, near parameter values for which
$J_0(\Delta_0/\omega)+J_2(\Delta_0/\omega)$ is minimized, the dominant
interaction-assisted excitation pathways are likewise substantially
suppressed. This provides an analytical explanation for why bifrequency
driving produces more robust dynamical freezing than the single-frequency
protocol, even in the presence of finite-range interaction tails.}

\roop{\section{Higher order connected correlators and Entanglement witness}

To characterize the spatial correlations generated during the Floquet dynamics, we first compute the connected density-density correlation functions. The local Rydberg excitation on site $i$ is described by the number operator $n_i$  The connected correlator between two sites separated by a distance $q$ is defined as
\begin{equation}
C_q=\frac{1}{L-q}\sum_{i=1}^{L-q}
\left(
\langle n_i n_{i+q}\rangle
-
\langle n_i\rangle\langle n_{i+q}\rangle
\right),
\label{eq:connectedcorr}
\end{equation}
where $L$ is the number of atoms in the chain. By subtracting the product of the local expectation values, the disconnected contribution arising from statistically independent excitations is removed, ensuring that $C_q$ captures only genuine many-body correlations.

The sign of the connected correlators provides direct insight into the underlying spatial ordering. A positive value indicates that two atoms are more likely to be excited simultaneously than expected for statistically independent particles, corresponding to correlated or cooperative excitation dynamics. Conversely, a negative value signifies anti-correlations, implying that the excitation of one atom suppresses the excitation probability of the other. In Rydberg systems, such negative nearest-neighbour correlations arise naturally from the Rydberg blockade, which inhibits the simultaneous excitation of nearby atoms.

Throughout this section, we examine the connected correlators at different neighbour separations, for example, $C_1$, $C_2$, and $C_3$, corresponding to the nearest-neighbour, next-nearest neighbor and next to next nearest-neighbour and so on. While $C_1$ directly probes the blockade-induced suppression of neighbouring excitations, the higher-order correlators $C_2$, $C_3$, $C_4$ and $C_5$ characterize how correlations extend beyond the nearest-neighbour blockade, thereby providing a more complete picture of the spatial structure of the Floquet many-body state.

To directly probe the entanglement generated by the Floquet dynamics, we use the quantum Fisher information (QFI) associated with collective $Z$-spin operators~\cite{PhysRevA.85.022322, PhysRevA.85.022321}. For a pure state, the QFI of an observable $A$ is $F_Q[A]=4\mathrm{Var}(A)$. We therefore consider collective operators of the form $J_{\mathbf{s}}^z=\frac{1}{2}\sum_i s_i\sigma_i^z$, with $s_i=\pm1$, for which
\begin{equation}
F_Z(\mathbf{s})\equiv 4\mathrm{Var}(J_{\mathbf{s}}^z)
=\sum_{i,j}s_i s_j C_{ij}^{ZZ},
\label{eq:FZ}
\end{equation}
where $C_{ij}^{ZZ}=\langle\sigma_i^z\sigma_j^z\rangle-\langle\sigma_i^z\rangle\langle\sigma_j^z\rangle$. Since $n_i=(1+\sigma_i^z)/2$, these correlations are directly related to the connected density-density correlations above by $C_{ij}^{ZZ}=4[\langle n_i n_j\rangle-\langle n_i\rangle\langle n_j\rangle]$.

For a fully separable pure state of $L$ spin-$1/2$ particles, $F_Q[J_{\mathbf{s}}^z]\leq L$ for any choice of ${s_i}$. Thus, $F_Z>L$ rules out full separability and provides a entanglement witness~\cite{PhysRevA.85.022322, PhysRevA.85.022321}. This motivates searching for the collective mode that gives the largest value of $F_Z$. A uniform choice, $s_i=1$, is not particularly suited to the Rydberg-blockaded system because the predominantly negative nearest-neighbour correlations generated by blockade cancel in the uniform sum. A staggered choice $s_i=(-1)^i$, is necessary to compensate for these nearest-neighbour anti-correlations. However, longer-range correlations need not follow the same staggered pattern, especially in the geometries considered here. We therefore optimize over all binary sign configurations,
\begin{equation}
F_Z^{\mathrm{opt}}=\max_{{s_i=\pm1}}F_Z(\mathbf{s}).
\label{eq:FZopt}
\end{equation}
The optimal sign configuration is determined independently at each driving frequency, allowing the entanglement witness to adapt to the spatial structure generated by the Floquet dynamics.

\subsection{Connected correlators as a signature of many-body dynamical freezing}

The connected density-density correlations provide a complementary probe of dynamical freezing and, importantly, of the many-body character of the driven dynamics. The system is initially prepared in the vacuum state, $\ket{\psi_0}=\ket{0}^{\otimes L}$,
for which the Rydberg excitation density vanishes. At a dynamical freezing point, the Floquet evolution suppresses the generation of excitations and the system approximately returns to its initial state. Thus, for an ideal freezing point,
   $ \langle n_i\rangle = 0,
    \langle n_i n_j\rangle = 0$, for all sites $i,j$, and consequently
\begin{equation}
    C_q =
    \langle n_i n_{i+q}\rangle
    -
    \langle n_i\rangle
    \langle n_{i+q}\rangle
    =0,
    \qquad \forall q.
\end{equation}

Figure~\ref{fig:corr} shows the experimentally measured excitation density together with the connected correlators, allowing the two quantities to be examined simultaneously. The left vertical axes show the connected correlators $C_q(T)$, while the right vertical axes show the Rydberg excitation density $n(T)$. Panels (a)--(c) correspond to single-frequency driving, whereas panels (d)--(f) show the corresponding results for bifrequency driving. The one-dimensional chain geometry with positive detuning $\Delta>0$ is shown in panels (a) and (d), the one-dimensional chain geometry with negative detuning $\Delta<0$ in panels (b) and (e), and the hexagonal geometry in panels (c) and (f).

\begin{figure}[h!]
    \centering
    \includegraphics[width=1.\linewidth,height=0.4\linewidth]{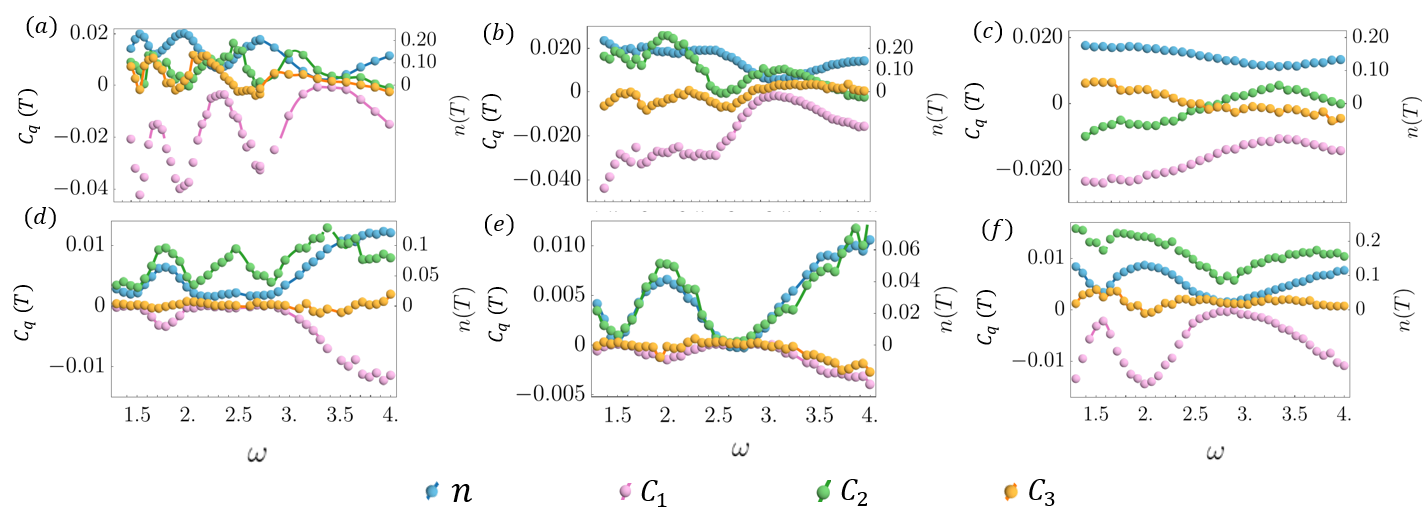}
    \caption{\textbf{Experimental connected two-point correlations under single- and bifrequency driving.}
    Experimentally measured connected two-point correlation functions $C_q(T)$ (left vertical axes) and Rydberg excitation density $n(T)$ (right vertical axes) as functions of the drive frequency $\omega$. Panels (a) and (d) correspond to the one-dimensional snake geometry with positive detuning $\Delta>0$, panels (b) and (e) to the one-dimensional snake geometry with negative detuning $\Delta<0$, and panels (c) and (f) to the hexagonal geometry. The top row, panels (a)--(c), shows single-frequency driving, while the bottom row, panels (d)--(f), shows bifrequency driving. For the one-dimensional geometries, the connected correlators exhibit frequencies at which the different correlation functions simultaneously approach zero, accompanied by a strong suppression of the excitation density, consistent with dynamical freezing. In the hexagonal geometry, the correlators are suppressed but do not simultaneously reach zero at any frequency, indicating weaker freezing. The bifrequency measurements exhibit substantially smaller correlation amplitudes and excitation densities than the corresponding single-frequency results, demonstrating stronger suppression of the driven many-body response even away from the exact freezing frequencies.}
    \label{fig:corr}
\end{figure}

For the one-dimensional geometries, the single-frequency results exhibit pronounced frequencies at which the connected correlators simultaneously collapse towards zero. In the positive-detuning chain shown in Fig.~\ref{fig:corr}(a), the pronounced structures in $C_1$, $C_2$, and $C_3$ are strongly suppressed at the frequencies where the excitation density is minimized. A similar behaviour is observed for the negative-detuning chain in Fig.~\ref{fig:corr}(b), where the different correlators approach zero together at the freezing frequencies. The simultaneous suppression of the different spatial correlators indicates that the freezing is not restricted to a particular correlation length, but involves the correlated excitation dynamics across the chain.

However, for the hexagonal geometry in Fig.~\ref{fig:corr}(c), the connected correlators approach zero over portions of the frequency range, but do not simultaneously reach zero at any particular frequency. Thus, although the correlations are strongly suppressed at certain frequencies, the freezing is not as complete as in the one-dimensional geometries. This difference indicates that the geometry of the array influences the efficiency with which the periodically driven system returns towards its initial state. 

The bifrequency results in the lower panels, Figs.~\ref{fig:corr}(d)--(f), also show an important reduction in the magnitude of the connected correlators compared with the single-frequency results. This is evident directly from the scales of the left vertical axes. Even away from the frequencies where the correlators are minimized, their magnitudes in the bifrequency case are generally smaller than those observed under single-frequency driving. Since the connected correlations are generated by the presence of Rydberg excitations and their interaction-mediated spatial correlations, their reduced magnitude is consistent with the substantially lower excitation density produced by the bifrequency drive. Thus, bifrequency driving suppresses the correlated response not only at the freezing points but also over a broader range of frequencies.

A vanishing connected correlator by itself, however, does not necessarily imply dynamical freezing. An uncorrelated state with a finite excitation density can also satisfy,$\langle n_i n_j\rangle
    =
    \langle n_i\rangle\langle n_j\rangle$,
and therefore have $C_q=0$. The distinction is particularly important in the large-distance, weakly interacting regime, where the atoms may remain excited but evolve approximately independently. In the present measurements, the excitation density and connected correlators are therefore considered together. The simultaneous suppression of $n(T)$ and $C_q(T)$ at the freezing frequencies provides a much stronger indication that the system is returning towards the initial vacuum state.

Away from the freezing points, the situation is qualitatively different. The drive produces a finite density of Rydberg excitations, while the interactions generate nonzero connected correlations between different sites. For example, negative values of the nearest-neighbour correlator,
    $C_1<0$,
indicate anti-correlations arising from Rydberg blockade, where excitation of one atom suppresses excitation of its neighbour. More generally, the observation of nonzero $C_1$, $C_2$, and $C_3$ at finite excitation density demonstrates that the excitation dynamics at different sites are not statistically independent. The simultaneous presence of correlations provides evidence for an interaction-induced collective response extending beyond an isolated pair of atoms.

Consequently, the frequency-dependent behaviour of the connected correlators provides a many-body characterization of the dynamical freezing. Away from the freezing points, finite excitation density together with nonzero $C_q$ reflects the development of spatially correlated many-body dynamics. At the freezing frequencies, the excitation density and the correlators simultaneously collapse towards zero, consistent with the reconstruction of the initial vacuum state. In contrast, the hexagonal geometry exhibits only an approach towards zero, with residual correlations remaining at all frequencies, indicating weaker freezing. The reduction of both the excitation density and the correlation amplitudes under bifrequency driving further demonstrates the enhanced suppression of the driven many-body response. Thus, the combined behaviour of $n(T)$ and $C_q(T)$ distinguishes genuine dynamical freezing from simple absence of spatial correlations and shows that the observed Floquet response involves the collective dynamics of the interacting Rydberg array.

\subsection{Distance dependence of connected correlations showing genuine many-body effects}

To elucidate the role of the Rydberg interaction in the periodically driven dynamics, we investigate the connected density-density correlators as a function of the lattice spacing $d$, while keeping all other driving parameters fixed. Varying the lattice spacing continuously modifies the interaction strength according to  $V_{ij}=\frac{C_6}{r_{ij}^{6}}$,
and, in particular, the nearest-neighbour interaction is
$
    V_{\mathrm{NN}}=\frac{C_6}{d^6}$.
Thus, changing $d$ provides a direct means of tuning the system from an effectively non-interacting regime to a strongly interacting regime in which Rydberg blockade becomes important. To characterize the resulting spatial correlations, we consider the connected correlators $C_q$ between atoms separated by $q$ lattice sites, corresponding to a physical separation $qd$.

For sufficiently large lattice spacings, the interatomic interactions become negligible compared with the driving scale and the atoms evolve approximately independently. This behaviour is clearly visible for $d=18~\mu\mathrm{m}$ in Fig.~\ref{fig:corr_dist}(d,h), where all correlators $C_q$ remain close to zero over the range of frequencies considered, both in the simulations and in the experiment. A vanishing connected correlator in this regime reflects the factorization of the joint excitation probability,
\begin{equation}
    \langle n_i n_{i+q}\rangle
    \simeq
    \langle n_i\rangle\langle n_{i+q}\rangle,
\end{equation}
rather than the absence of Rydberg excitations. Thus, the near-zero values of $C_q$ at large $d$ indicate that the excitation dynamics of spatially separated atoms are essentially independent.

As the lattice spacing is reduced, the interaction strength increases rapidly as $d^{-6}$, and deviations from this factorized behaviour become increasingly pronounced. This can already be seen at $d=12~\mu\mathrm{m}$ in Fig.~\ref{fig:corr_dist}(c,g), where the nearest-neighbour correlator $C_1$ develops appreciable positive values over a range of driving frequencies. A similar behaviour is observed more strongly at $d=8~\mu\mathrm{m}$ in Fig.~\ref{fig:corr_dist}(b,f). In these regimes, the atoms are sufficiently close for the interactions to correlate their excitation dynamics, while the nearest-neighbour interaction is not yet strong enough to completely suppress simultaneous excitations. A positive value of $C_1$ corresponds to
\begin{equation}
    \langle n_i n_{i+1}\rangle
    >
    \langle n_i\rangle\langle n_{i+1}\rangle,
\end{equation}
indicating an enhanced probability of simultaneous nearest-neighbour excitation relative to statistically independent atoms. The emergence of such finite connected correlations therefore signals the onset of interaction-induced collective dynamics.

\begin{figure}[h!]
    \centering
    \includegraphics[width=0.99\linewidth,height=0.4\linewidth]{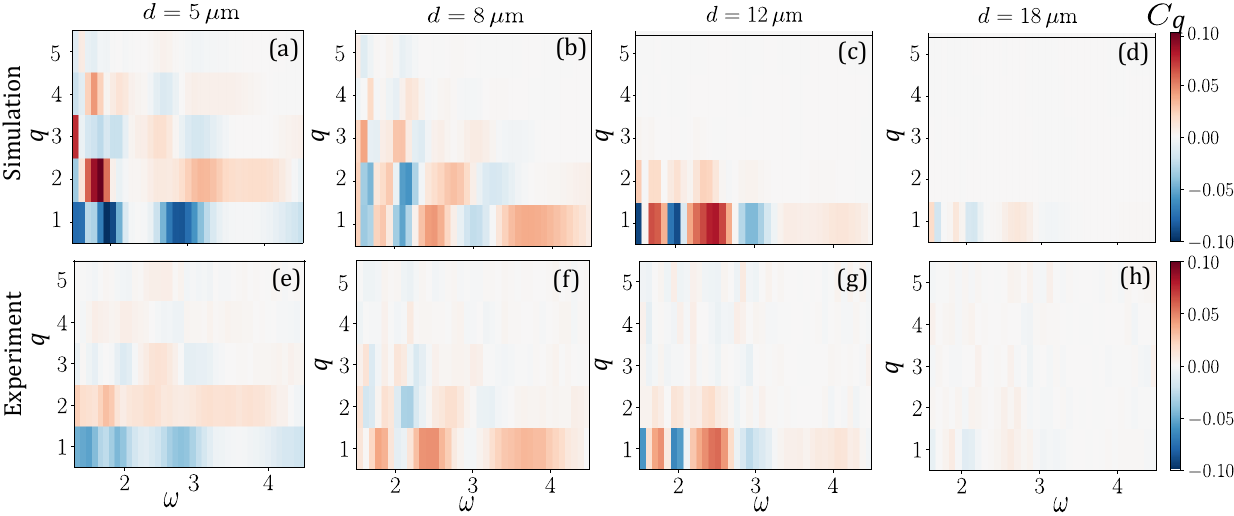}
    \caption{\textbf{Distance dependence of connected correlations: simulations and experiment.}
    Connected two-point correlation functions $C_q(T)$ as functions of the drive frequency $\omega$ for different lattice spacings $d$. The correlators $C_1$--$C_5$ probe pairs of atoms separated by $q$ lattice sites, corresponding to physical separations $qd$. The top row, panels (a)--(d), shows the simulation results, while the bottom row, panels (e)--(h), shows the corresponding experimental measurements. The columns correspond to $d=5~\mu\mathrm{m}$ [(a),(e)], $d=8~\mu\mathrm{m}$ [(b),(f)], $d=12~\mu\mathrm{m}$ [(c),(g)], and $d=18~\mu\mathrm{m}$ [(d),(h)]. At the largest lattice spacing, $d=18~\mu\mathrm{m}$, all connected correlators remain close to zero, consistent with effectively independent atomic dynamics. As the atoms are brought closer together, finite correlations develop at progressively larger spatial separations. At $d=12~\mu\mathrm{m}$ and $8~\mu\mathrm{m}$, the nearest-neighbour correlator $C_1$ becomes positive over significant frequency ranges, while at $d=5~\mu\mathrm{m}$ it becomes predominantly negative as a consequence of Rydberg-blockade-induced anti-correlations. The corresponding experimental results reproduce the qualitative distance-dependent evolution observed in the simulations, demonstrating the emergence and spatial extension of interaction-induced many-body correlations.}
    \label{fig:corr_dist}
\end{figure}

A qualitatively different regime is reached at the smallest lattice spacing considered, $d=5~\mu\mathrm{m}$, shown in Fig.~\ref{fig:corr_dist}(a,e). Here, the nearest-neighbour interaction becomes sufficiently strong to impose the Rydberg blockade, resulting in negative values of $C_1$ over pronounced frequency ranges. In this case,
\begin{equation}
    \langle n_i n_{i+1}\rangle
    <
    \langle n_i\rangle\langle n_{i+1}\rangle,
\end{equation}
so that excitation of one atom suppresses the probability of excitation of its nearest neighbour. The negative $C_1$ therefore directly reflects the anti-correlations generated by the blockade. The evolution from positive $C_1$ at $d=12~\mu\mathrm{m}$ and $8~\mu\mathrm{m}$ to negative $C_1$ at $d=5~\mu\mathrm{m}$ demonstrates that reducing the lattice spacing not only increases the magnitude of the correlations but can also change their character, from enhanced joint excitation to blockade-induced anti-correlation.

The higher-order correlators $C_2$, $C_3$, $C_4$, and $C_5$ provide information about the spatial extent of these interaction-induced correlations. At large separation, all of these correlators remain close to zero, as illustrated most clearly for $d=18~\mu\mathrm{m}$ in Fig.~\ref{fig:corr_dist}(d,h). Upon reducing the lattice spacing, finite correlations progressively appear at larger values of $q$. In particular, at $d=8~\mu\mathrm{m}$ and $d=5~\mu\mathrm{m}$, appreciable correlations are observed not only in $C_1$ but also in higher-order correlators, demonstrating that the interaction-induced correlations extend beyond nearest-neighbour pairs. Since $C_q$ probes atoms separated by the physical distance $qd$, the appearance of nonzero higher-order correlators indicates that the spatially correlated response extends across multiple sites of the array rather than being confined to an isolated pair.

It should be noted that the results shown in Fig.~\ref{fig:corr_dist} correspond to the results after a single drive cycle. As the number of drive cycles is increased, the periodically driven dynamics have more time to build up interaction-induced correlations, leading to an enhancement of the magnitude of the connected correlators. In particular, correlations at larger spatial separations can progressively develop, resulting in appreciable contributions from $C_4$ and $C_5$ and other higher-order correlators. Thus, the correlations shown here after a single cycle already reveal the onset of the many-body response, while evolution over multiple cycles is expected to further enhance and extend these correlations across the array.

The comparison between simulations and experimental measurements in Fig.~\ref{fig:corr_dist} further shows that this distance-dependent evolution of the correlations is consistently captured in both descriptions. The qualitative progression from nearly vanishing correlations at large $d$, to finite correlations at intermediate separations, and finally to pronounced nearest-neighbour anti-correlations in the strongly interacting regime is reproduced experimentally. This agreement demonstrates that the observed spatial correlations originate from the interaction-driven many-body dynamics rather than from an isolated two-atom effect.

The distance dependence of the connected correlators therefore provides direct evidence for the many-body nature of the observed dynamics. At large lattice spacing, the vanishing of all $C_q$ is consistent with effectively independent atomic evolution. As the atoms are brought closer together, the emergence of finite $C_q$ demonstrates that their excitation dynamics become correlated through the interatomic interactions, with correlations extending over increasingly larger spatial separations. In the strongly interacting regime, the nearest-neighbour correlations additionally acquire a negative sign due to Rydberg blockade. Thus, the continuous evolution of $C_1$--$C_5$ with lattice spacing establishes a direct connection between interaction strength and the spatial structure of the driven many-body state, supporting the interpretation of the observed dynamical freezing as a genuinely many-body phenomenon rather than a collection of independent single-particle responses.

\subsection{Entanglement witness via optimized Fisher information}
\begin{figure}
    \centering
\includegraphics[width=0.7\linewidth]{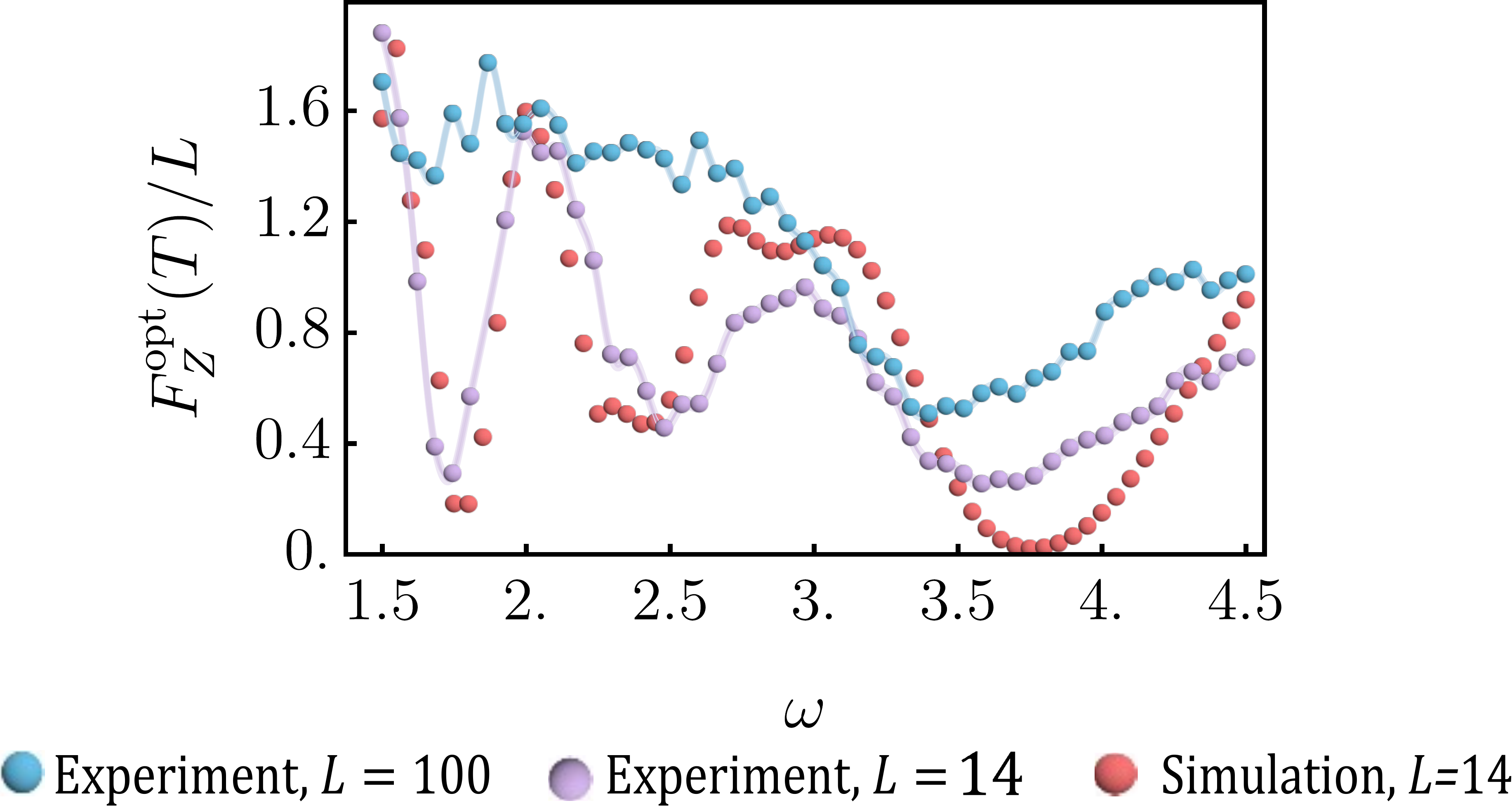}
\caption{\textbf{Optimized Fisher information after one drive cycle.}
Optimized Fisher information per atom, $F_Z^{\mathrm{opt}}(T)/L$, after one drive period $T$ as a function of $\omega$ for an interatomic spacing of $d=4.7~\mu\mathrm{m}$ and driving parameters as in Fig.~2(b) of the main text. Blue: experiment, $L=100$; purple: experiment, $L=14$; red: simulation, $L=14$.}
    \label{fig:FZ}
\end{figure}

To further characterize the many-body nature of the Floquet states, we evaluate the optimized Fisher information $F_Z^{\mathrm{opt}}(T)$ defined above after one driving cycle. Figure~\ref{fig:FZ} shows $F_Z^{\mathrm{opt}}/L$ for the $L=100$ experimental system (blue), the $L=14$ experimental system (purple), and the corresponding $L=14$ numerical results (red). For $L=14$, the optimization is performed by an exhaustive scan over all $2^{14}$ binary sign configurations, while for $L=100$ we determine the optimum using multiple random initial configurations followed by greedy coordinate optimization of the signs.

The optimized Fisher information follows the overall structure of the Floquet dynamics. In particular, in frequency regimes where the dynamics exhibits strong freezing, $F_Z^{\mathrm{opt}}/L$ is suppressed, whereas in the more dynamically active regimes it can exceed unity. Most prominently, at low driving frequencies, $F_Z^{\mathrm{opt}}/L>1$ after a single drive cycle, which rules out a fully separable description of the state. The corresponding behaviour is consistently observed in the $L=100$ experiment and in both the experimental and numerical results for $L=14$.

Combined with the distance-dependent connected correlations discussed above, this provides complementary evidence for the many-body character of the dynamics. The finite values of $C_q$ over multiple spatial separations demonstrate that the interaction-induced correlations extend beyond isolated nearest-neighbour pairs, while $F_Z^{\mathrm{opt}}/L>1$ establishes that the resulting state cannot be described as a product of independent local states. Together, these observations show that the Floquet dynamics generates collective correlations extending across multiple sites, beyond what can be captured by independent-atom or simple mean-field descriptions, and therefore support the interpretation of the observed freezing as a genuinely many-body phenomenon.}

\roop{\section{Initial state dependence of freezing landscape}

We investigate here the dependence of dynamical freezing landscape on the choice of initial many-body state. This provides a way to distinguish effects arising from the driving protocol from those associated with the spatial structure of the initial state and the underlying interaction range. We characterize freezing through the fidelity after one driving period,
\begin{equation}
    \mathcal{F}(T)=|\langle\psi_0|\psi(T)\rangle|^2,
\end{equation}
and plot $\log[1-\mathcal{F}(T)]$ as a function of the drive frequency $\omega$. Minima of this quantity correspond to enhanced fidelity and hence to the strongest dynamical freezing.

\begin{figure}
    \centering
    \includegraphics[width=\linewidth]{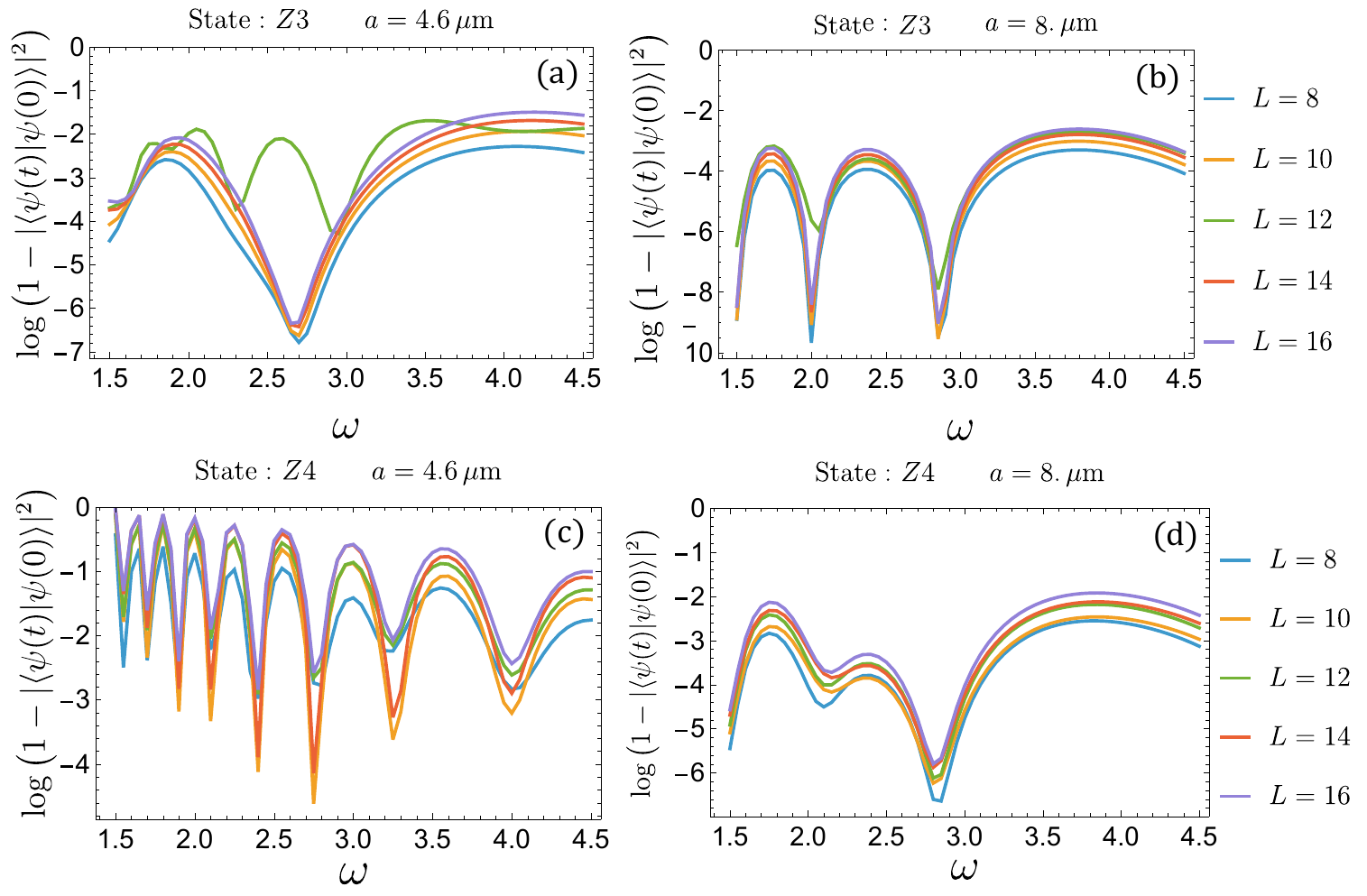}
    \caption{\textbf{Initial-state dependence of dynamical freezing in the full long-range Rydberg model.}
    The quantity $\log[1-\mathcal{F}(T)]$, with $\mathcal{F}(T)=|\langle\psi_0|\psi(T)\rangle|^2$, is shown as a function of the drive frequency $\omega$ for the $\ket{\mathbb{Z}_3}$ initial state at $d=4.6~\mu{\rm m}$ (a) and $d=8~\mu{\rm m}$ (b), and for the $\ket{\mathbb{Z}_4}$ initial state at $d=4.6~\mu{\rm m}$ (c) and $d=8~\mu{\rm m}$ (d). Results are shown for system sizes $L=8,10,12,14,16$. At $d=4.6~\mu{\rm m}$, where longer-range interactions contribute appreciably, the freezing landscape depends on the initial state, with both the locations and number of pronounced freezing points differing between the $\ket{\mathbb{Z}_3}$ and $\ket{\mathbb{Z}_4}$ states. Increasing the atom separation to $d=8~\mu{\rm m}$ suppresses this dependence as the system approaches the PXP limit, with the dominant freezing features occurring at increasingly similar frequencies for the two initial states.}
    \label{fig:diffinit_ryd}
\end{figure}

The vacuum and $\ket{\mathbb{Z}_2}$ initial states were considered in the main text. Here, we extend the analysis to two further commensurate configurations,
\begin{align}
    \ket{\mathbb{Z}_3} &\equiv \ket{100100\cdots}, \nonumber\\
    \ket{\mathbb{Z}_4} &\equiv \ket{10001000\cdots}.
\end{align}
These states contain different spatial arrangements of Rydberg excitations and therefore probe distinct interaction environments in the full long-range model.

Figure~\ref{fig:diffinit_ryd} shows the resulting fidelity landscape for the $\ket{\mathbb{Z}_3}$ and $\ket{\mathbb{Z}_4}$ states at two different atom separations. For $d=4.6~\mu{\rm m}$, where the longer-range interactions are appreciable, the freezing landscape depends strongly on the initial state. The $\ket{\mathbb{Z}_3}$ state exhibits the freezing structure shown in Fig.~\ref{fig:diffinit_ryd}(a), while the $\ket{\mathbb{Z}_4}$ state gives a qualitatively different pattern in Fig.~\ref{fig:diffinit_ryd}(c). In particular, both the locations and the number of pronounced freezing points change between the two initial states. Thus, the freezing frequencies are determined not only by the temporal structure of the drive, but also by how the initial excitation pattern samples the distance-dependent interaction landscape.

This dependence is substantially reduced upon increasing the atom separation to $d=8~\mu{\rm m}$. In this regime, shown in Fig.~\ref{fig:diffinit_ryd}(b,d), the system approaches the PXP limit and the freezing features for the $\ket{\mathbb{Z}_3}$ and $\ket{\mathbb{Z}_4}$ states become increasingly similar. The dominant freezing points occur at approximately the same frequencies, while the differences in their number and strength are considerably suppressed. This shows that the initial-state dependence observed at smaller separation is closely tied to the longer-range components of the Rydberg interactions rather than arising solely from the blockade constraint.

The PXP model provides a useful limiting case in which to examine this behavior directly. In this model, the longer-range interaction terms are absent, leaving the nearest-neighbor blockade constraint as the relevant interaction structure. As shown in Fig.~\ref{fig:diffinit_pxp}, the freezing landscape is then independent of the choice of initial state: the freezing points occur at the same drive frequencies for the different initial configurations. This contrasts with the state-dependent response of the full long-range model at $d=4.6~\mu{\rm m}$ and confirms that the differences observed there originate from the additional spatial dependence introduced by longer-range interactions.

\begin{figure}
    \centering
    \includegraphics[width=0.99\linewidth,height=0.28\linewidth]{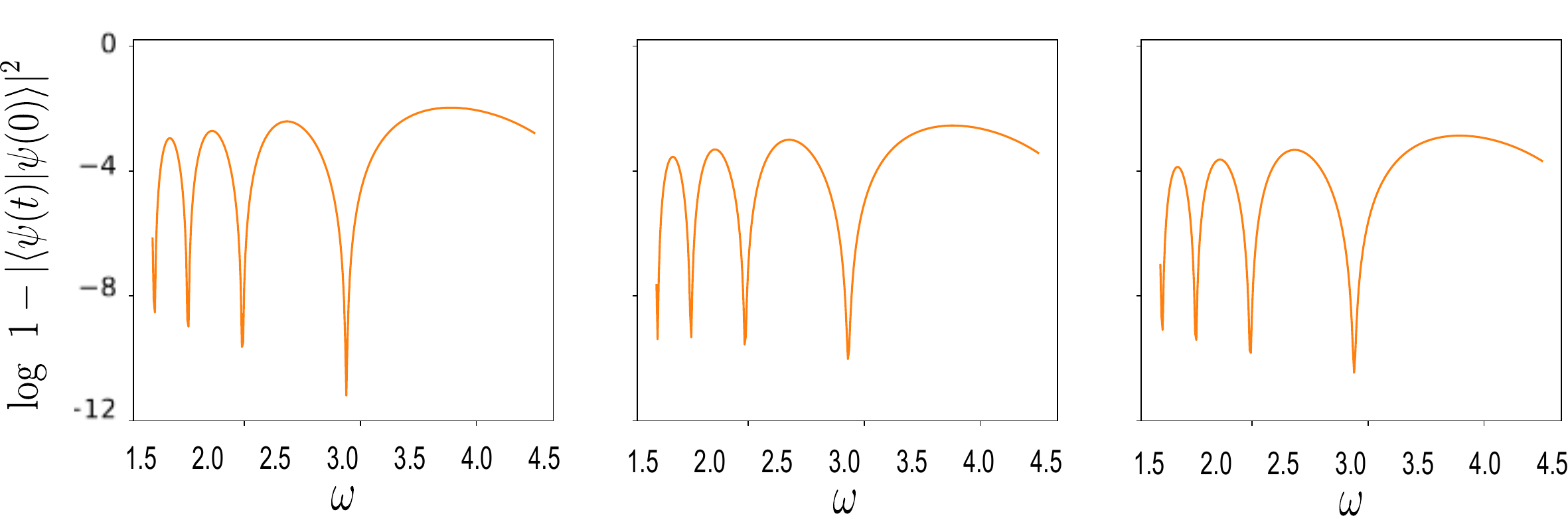}
    \caption{\textbf{Initial-state independence of dynamical freezing in the PXP model.}
The quantity $\log[1-\mathcal{F}(T)]$ is shown as a function of the drive frequency $\omega$ for three different initial states: the single-excitation state $\ket{1}$ (left), the $\ket{\mathbb{Z}_2}$ state (middle), and the $\ket{\mathbb{Z}_3}$ state (right). The freezing points occur at the same frequencies for all three initial states, demonstrating that the freezing landscape in the PXP model is independent of the initial many-body configuration.}
    \label{fig:diffinit_pxp}
\end{figure}

Taken together, these results show that the initial state and interaction range play complementary roles in determining dynamical freezing. In the full long-range Rydberg model, different initial configurations experience different effective interaction landscapes and consequently produce distinct freezing patterns. As the interaction range is reduced, this dependence progressively disappears, reaching an initial-state-independent freezing landscape in the PXP limit. Dynamical freezing therefore provides a sensitive probe of the interplay between the spatial structure of the initial many-body state and the microscopic interaction range of the system.}

\roop{\section{Heating and Stability of Dynamical Freezing}

We next study the heating dynamics of the periodically driven system by monitoring the energy absorbed from the drive over successive driving cycles. This provides a complementary characterization of dynamical freezing: while the excitation density probes the state after a given number of driving periods, the heating dynamics reveal how stable this state remains under prolonged Floquet evolution. In particular, a slower growth of the absorbed energy indicates that the system remains close to its initial low-energy state for a longer time.

To quantify the heating, we evaluate the energy at stroboscopic times $t=nT$ according to
\begin{equation}
    Q(nT)=\Delta E(nT)
    =
    \bra{\psi(nT)}H(t=0)\ket{\psi(nT)}
    -E_0,
\end{equation}
where
\begin{equation}
    E_0=
    \bra{\psi_0}H(t=0)\ket{\psi_0}.
\end{equation}
Since the system is initially prepared in the vacuum state, $E_0=0$ for the parameters considered here. Thus, $\Delta E$ directly measures the energy absorbed from the periodic drive after $n$ driving cycles. Continued growth of $\Delta E$ corresponds to the progressive population of higher-energy states and hence to Floquet heating.

For comparison, we also consider the infinite-temperature value of the energy. At infinite temperature, the corresponding thermal density matrix is
\begin{equation}
    \rho_{\beta=0}
    =
    \frac{e^{-\beta H}}{\mathrm{Tr}(e^{-\beta H})}
    \bigg|_{\beta=0}
    =
    \frac{\mathbb{I}}{\mathrm{HSD}},
\end{equation}
where $\mathrm{HSD}$ is the dimension of the relevant Hilbert space. The corresponding infinite-temperature expectation value of an observable $O$ is therefore
\begin{equation}
    \langle O\rangle_{\beta=0}
    =
    \frac{\mathrm{Tr}[O]}{\mathrm{HSD}}.
\end{equation}
When Floquet ETH applies, local observables can approach these infinite-temperature values at sufficiently long times. For the PXP model, the trace is evaluated within the constrained Hilbert space, whose dimension for a chain of length $L$ is
\begin{equation}
    \mathrm{HSD}=F_{L+2},
\end{equation}
where $F_n$ denotes the $n$th Fibonacci number. The red dashed line in Fig.~\ref{fig:heating} therefore indicates the corresponding infinite-temperature energy scale.

\begin{figure}[H]
    \centering
    \includegraphics[width=0.49\linewidth]{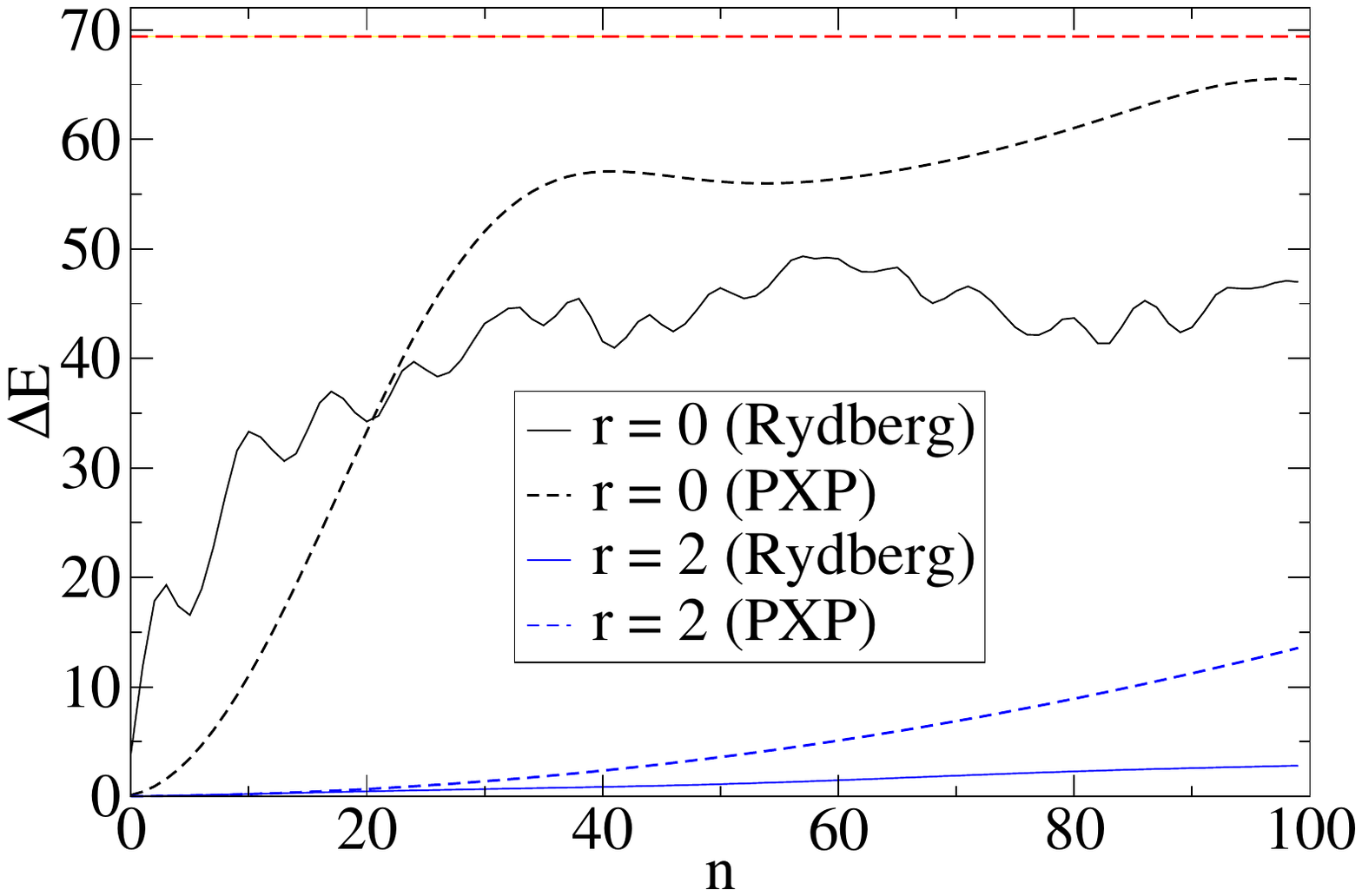}
    \includegraphics[width=0.49\linewidth]{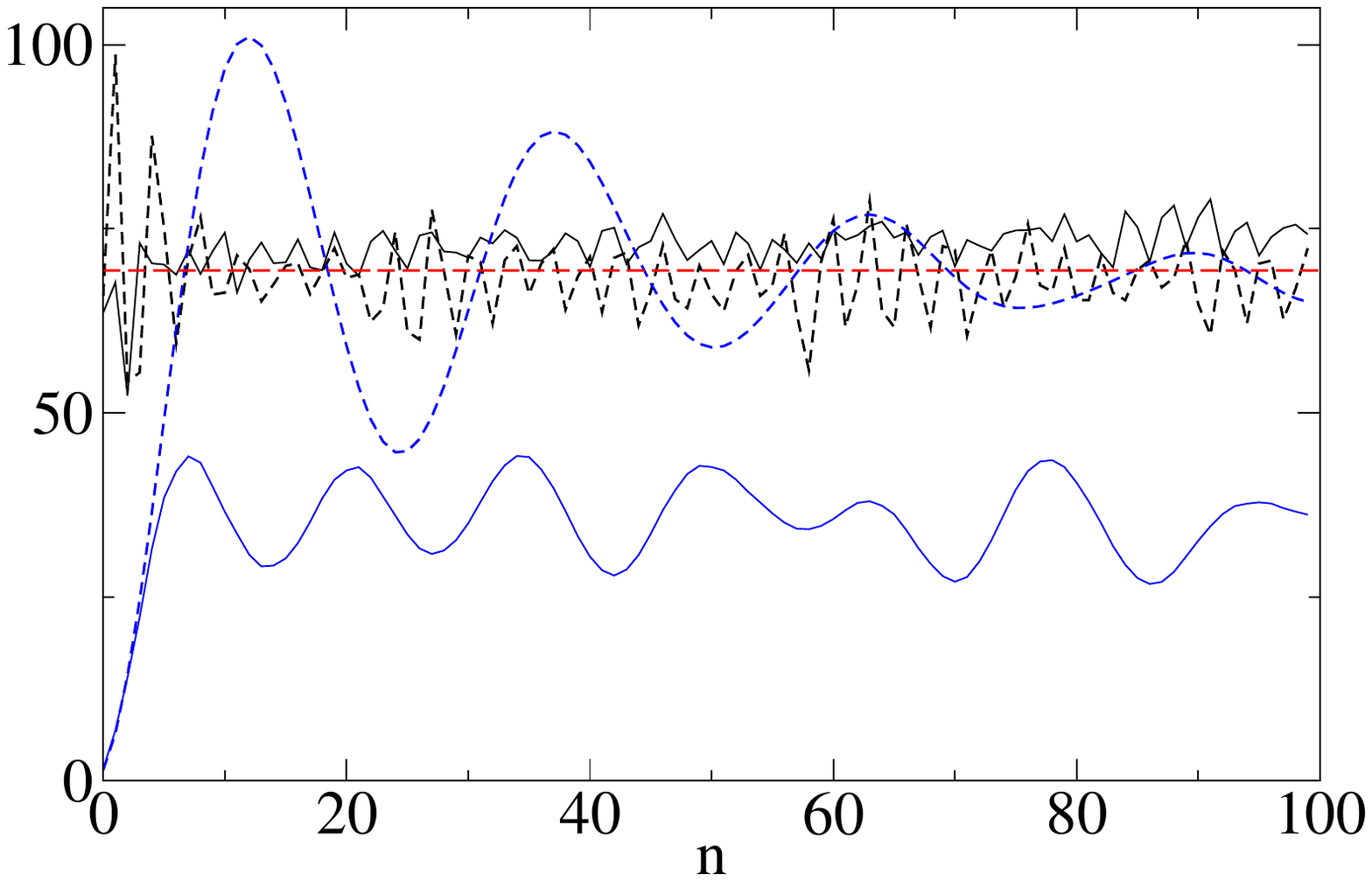}
    \caption{\textbf{Heating dynamics under single- and bifrequency driving.}
    Absorbed energy $\Delta E$ at stroboscopic times $t=nT$ for the full long-range Rydberg Hamiltonian and the PXP model. The left panel corresponds to frequencies near a pronounced freezing minimum, with $\omega=3.73$ for $r=0$ and $\omega=2.89$ for $r=2$, while the right panel corresponds to frequencies near a peak between two freezing minima, with $\omega=2.95$ for $r=0$ and $\omega=2.37$ for $r=2$. Solid and dashed curves correspond to the full Rydberg and PXP models, respectively, while black/gray and blue/light-blue curves denote the single-frequency ($r=0$) and bifrequency ($r=2$) protocols. The red dashed lines indicate the corresponding infinite-temperature energy scale for the PXP model. The results show that the full long-range Rydberg model initially heats more rapidly than the PXP model, but its heating slows at later times due to the additional long-range interactions. In contrast, the bifrequency protocol with $r=2$ exhibits strongly suppressed heating over many driving cycles, demonstrating a substantially more stable and longer-lived freezing response. Parameters: $N=12$ (OBC), $\Omega_0=2.0$, $\Delta_0=20.0$, $200$ Trotter steps, $C_6=5420503$, and $d=4.7$.}
    \label{fig:heating}
\end{figure}

Figure~\ref{fig:heating} compares the heating dynamics at two representative choices of driving frequency. The left panel corresponds to frequencies located near a freezing minimum, where the generation of Rydberg excitations is strongly suppressed. In contrast, the right panel corresponds to frequencies near a peak between two freezing minima and therefore represents a regime in which substantially more excitations are generated. Comparing these two cases allows us to distinguish the heating behaviour close to a frozen state from that in a more strongly excited regime.

The full Rydberg Hamiltonian initially absorbs energy more rapidly than the
PXP model. This is particularly apparent in the left panel, where the energy
of the Rydberg model rises rapidly during the first several driving cycles.
At later times, however, the growth of $\Delta E$ becomes considerably
slower, and the energy remains largely bounded by oscillations. This
crossover reflects the role of the long-range interaction tails beyond the
nearest-neighbour blockade. While these interactions initially introduce
additional interaction-dependent excitation pathways, their associated
energy shifts subsequently modify the resonance conditions and suppress
further energy absorption. This effect does not require a large buildup of
the excitation density and is particularly pronounced under bifrequency
driving, where the dominant excitation channels are already strongly
suppressed. Consequently, although the PXP model heats more slowly at early
times, its residual heating eventually exceeds that of the full Rydberg
model, leading to the observed reversal in the long-time growth of
$\Delta E$.

The same qualitative behaviour is visible in the right panel, although the overall heating is stronger because the chosen frequencies lie away from a freezing minimum. The difference between the two models demonstrates that the long-range interaction modifies not only the excitation dynamics at a single driving period but also the subsequent heating process over many cycles.

A much stronger suppression of heating is obtained for the bifrequency protocol with $r=2$. In both the near-freezing regime shown in the left panel and the more strongly excited regime shown in the right panel, the $r=2$ curves remain substantially lower than their corresponding $r=0$ curves as the number of driving cycles increases. The absorbed energy therefore grows much more slowly under bifrequency driving, indicating that the suppression is not restricted to a particular stroboscopic time or to an isolated freezing minimum.

The reduced long-time heating rate demonstrates that the enhanced freezing produced by the $r=2$ protocol corresponds to a more stable Floquet state. In particular, the persistence of the suppression over many driving cycles shows that bifrequency driving does not merely reduce the excitation density after one period, but also substantially delays the subsequent absorption of energy. Thus, within the parameter range studied, the $r=2$ protocol produces a longer-lived dynamical freezing regime with significantly enhanced resistance to Floquet heating.}

\section{Interference after multiple drive cycles}
\begin{figure}[h!]
    \centering
\includegraphics[width=0.49\linewidth]{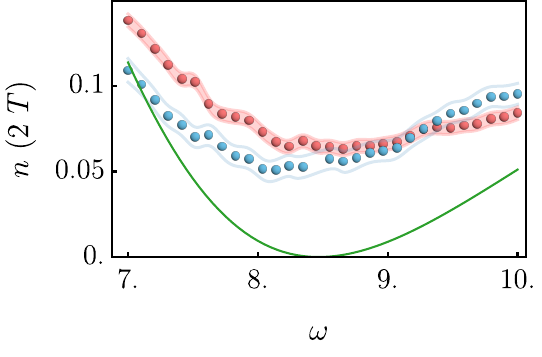}
\includegraphics[width=0.49\linewidth]{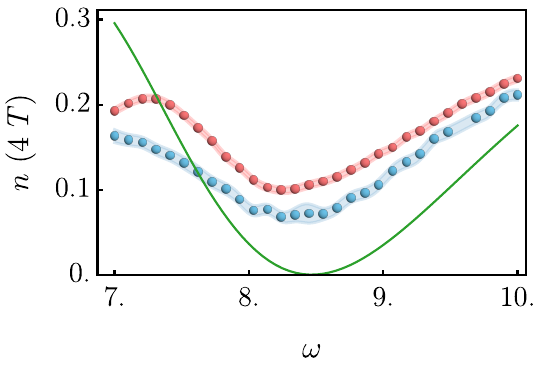}
\hspace{0.02\linewidth}\includegraphics[width=0.58\linewidth,height=0.03\linewidth]{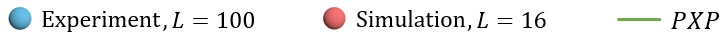}
    \caption{ Average Rydberg excitation density $n$ vs frequency $\omega$ after Left: two drive cycles, Right:  four drive cycles}
    \label{fig:multidrive}
\end{figure}

In this section we report the excitation density under repeated periodic driving. To remain within the experimentally accessible coherence time, we consider higher drive frequencies than in the main text and focus on parameters near the first (smallest) zero of the Bessel function $J_0$. Since the argument of $J_0$ scales as $\Delta_0/\omega$, this corresponds to the highest drive frequency at which the characteristic interference minima remain observable in our experiment. We experimentally measure the excitation density for $L=100$ atoms in a snake geometry after two and four drive cycles, shown in Fig.~\ref{fig:multidrive} (left and right panels, respectively), and compare these results with numerical simulations obtained using Bloqade for a system of size $L=14$, as well as with the constrained PXP model at the same size.

While temporal interference persists over multiple cycles, its visibility is significantly reduced with increasing drive duration. In particular, the excitation density at points of constructive interference is markedly higher after four cycles than after two, indicating the onset of Floquet heating. This behavior is captured both in experiment and in the full Rydberg simulations, consistent with the understanding that dynamical freezing in interacting Floquet systems is generally a transient prethermal regime rather than a truly stable phase, in line with recent theoretical studies attributing such heating processes to instanton-mediated mechanisms~\cite{mukherjee2024floquet}.

By contrast, the $PXP$ model exhibits substantially weaker degradation of interference visibility over the same timescales, reflecting its effectively constrained and non-generic interaction structure. This comparison highlights that finite-range interaction tails play a central role in enabling genuine many-body dynamics and in determining the long-time stability of driven interference phenomena. In this sense, the extended interactions present in realistic Rydberg systems are not merely a perturbation to constrained models but are essential for capturing the experimentally observed heating and coherence properties.

\roop{\section{Dependence of Dynamical Freezing on the drive protocol and harmonics}

To examine the robustness of dynamical freezing beyond the driving protocol considered in the main text, we first consider a square-pulse drive. In this case, the Rabi frequency is kept constant, $\Omega(t)=\Omega_0$, while the detuning is varied using a square-pulse protocol. The resulting drive is shown in the right-hand panels of Fig.~\ref{fig:sqprot}, with $\Omega(t)$ displayed in the upper panel and $\Delta(t)$ in the lower panel. The corresponding excitation probability after one driving period is shown in the left panel as a function of the drive frequency $\omega$.

\begin{figure}[h!]
    \centering
    \includegraphics[width=0.85\linewidth,height=0.32\linewidth]{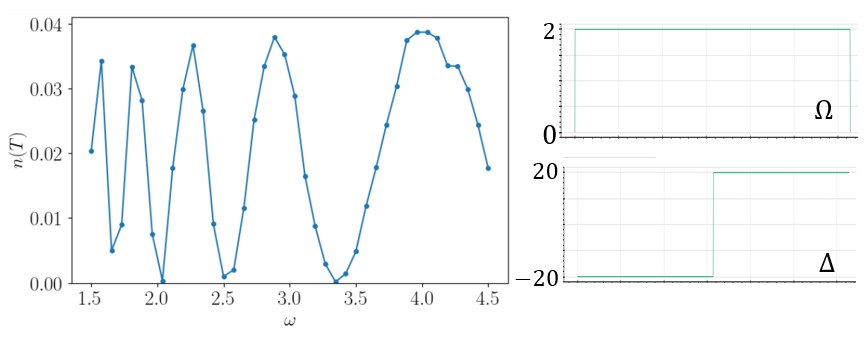}
    \caption{\textbf{Dynamical freezing under a square-pulse driving protocol.}
    The left panel shows the excitation probability as a function of drive frequency $\omega$ for the square-pulse protocol with $r=0$. Pronounced minima in the excitation probability indicate strong dynamical freezing at the corresponding frequencies. The right panels show the corresponding driving protocol: $\Omega(t)$ is shown in the upper panel and $\Delta(t)$ in the lower panel. In this protocol, the Rabi frequency is kept constant while the detuning is modulated as a square pulse.}
    \label{fig:sqprot}
\end{figure}

As shown in Fig.~\ref{fig:sqprot}, pronounced minima in the excitation
probability persist under the square-pulse drive, demonstrating that
dynamical freezing does not rely on the specific smooth sinusoidal
waveform used in the main text. For a symmetric square pulse,
$\Delta(t)=-\Delta_0$ for $0<t<T/2$ and $\Delta(t)=+\Delta_0$ for
$T/2<t<T$, with $T=2\pi/\omega$, we consider the transverse drive
$H_x=(\Omega/2)\sum_i\sigma_i^x$, where
$\sigma_i^x=\sigma_i^++\sigma_i^-$. 

For the vacuum-to-one-particle transition, the interaction energy
difference vanishes. The first-order
Floquet Hamiltonian is therefore
\begin{equation}
\left(H_F^{(1)}\right)
=
\sum_i \frac{\Omega\omega}{\pi\Delta_0}
e^{-i\pi\Delta_0/(2\omega)}
\sin\left(\frac{\pi\Delta_0}{2\omega}\right)  \sigma^x_i
\label{eq:square_FPT_first}
\end{equation}
It consequently vanishes at $\omega_m=\Delta_0/(2m)$, with
$m=1,2,3,\ldots$. For $\Delta_0=20$, the zeros in the frequency range
of Fig.~\ref{fig:sqprot} are $\omega=1.667$, $2.000$, $2.500$, and
$3.333$, in close agreement with the locations of the observed
freezing minima.

The interaction does, however, enter at second order. In the
vacuum-to-two-particle process
$|0\rangle\rightarrow|i\rangle\rightarrow|ij\rangle$, the second flip
acquires the interaction energy
$E_V(|ij\rangle)-E_V(|i\rangle)=V_{ij}$, and hence the second-order FPT
term contains interaction-dependent time integrals. These corrections
set the residual excitation at the first-order freezing points. They
appear comparatively weaker for the square-pulse protocol than for the
sinusoidal drive, although a detailed comparison of higher-order
corrections is beyond the scope of this analysis. As in the sinusoidal
case, the freezing becomes progressively less pronounced at lower
frequencies, consistent with the increasing importance of higher-order
corrections. Thus, the square-pulse protocol provides a simple example
in which the locations of the freezing points are already captured by
the leading-order FPT result, while the robustness of freezing itself
persists despite substantial changes in the waveform.

We next investigate how the different harmonic drive protocols affects the dynamical freezing response. For this purpose, we vary the harmonic parameter $r$ while keeping the remaining parameters of the full long-range Rydberg Hamiltonian fixed. Figure~\ref{fig:diffr} compares the excitation dynamics obtained for different values of $r$. The excitation density $n(T)$ after one driving period is shown as a function of the drive frequency $\omega$, providing a direct measure of the extent to which Rydberg excitations are suppressed by the periodic drive.

\begin{figure}[h!]
    \centering
    \includegraphics[width=0.48\linewidth,height=0.32\linewidth]{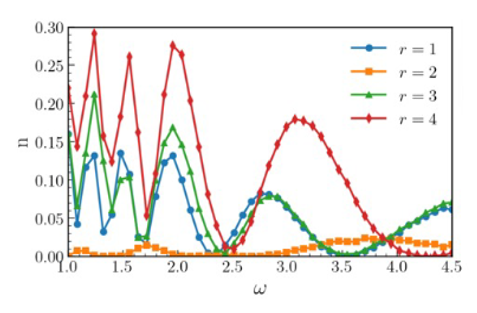}
    \caption{\textbf{Dependence of dynamical freezing on the harmonic parameter $r$.}
    Excitation density $n(T)$ after one driving period as a function of the drive frequency $\omega$ for different values of the harmonic parameter $r$, calculated using the full long-range Rydberg Hamiltonian. The comparison shows that the suppression of the excitation density depends sensitively on the harmonic content of the drive. Within the frequency range considered, the $r=2$ protocol exhibits the strongest suppression of $n(T)$, indicating the most pronounced dynamical freezing among the harmonic configurations studied.}
    \label{fig:diffr}
\end{figure}

The dependence of the freezing response on the harmonic parameter is evident from Fig.~\ref{fig:diffr}. Within the frequency range investigated, the theoretical analysis predicts the strongest suppression of the excitation density for $r=2$, and this behaviour is reflected in the numerical results. In particular, the $r=2$ protocol produces the smallest excitation density over the relevant frequency range, indicating that the periodic drive most effectively suppresses the generation of Rydberg excitations for this choice of harmonic.

The comparison between different values of $r$ demonstrates that the occurrence and strength of dynamical freezing are not determined solely by the overall driving amplitude, but also depend on the temporal structure of the drive. Changing $r$ modifies the relative harmonic content of the driving protocol and consequently changes the frequency-dependent response of the system. Among the harmonic configurations considered here, $r=2$ therefore provides the strongest suppression of the excitation density and is consequently the harmonic configuration adopted for the results presented in the main text.

These results further demonstrate that dynamical freezing is sensitive to the detailed form of the driving protocol. While freezing persists for different harmonic choices, its strength varies systematically with $r$, with $r=2$ yielding the most pronounced suppression within the parameter range studied.}

\section{Robustness of Dynamical Freezing in PXP Model}

We use exact numerical simulations of the PXP model to examine the robustness of the dynamical freezing phenomenon with respect to system size, boundary conditions, and the ratio of the two driving frequencies. Time evolution is performed using second-order Trotterization with 400 time steps per driving period. Unless otherwise specified, open boundary conditions are used.

\begin{figure}[H]
    \centering
    \includegraphics[width=0.32\linewidth]{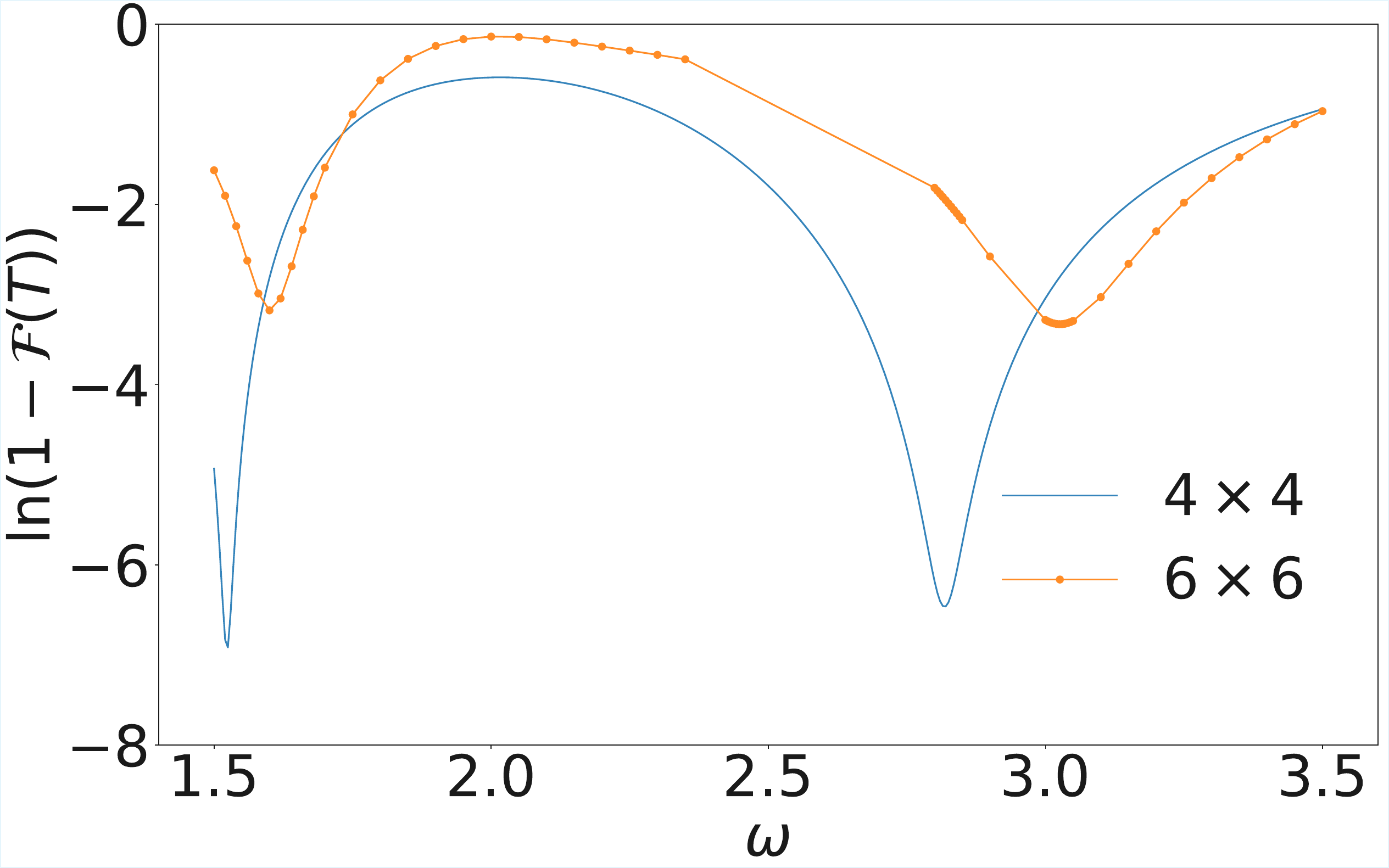}
    \includegraphics[width=0.32\linewidth]{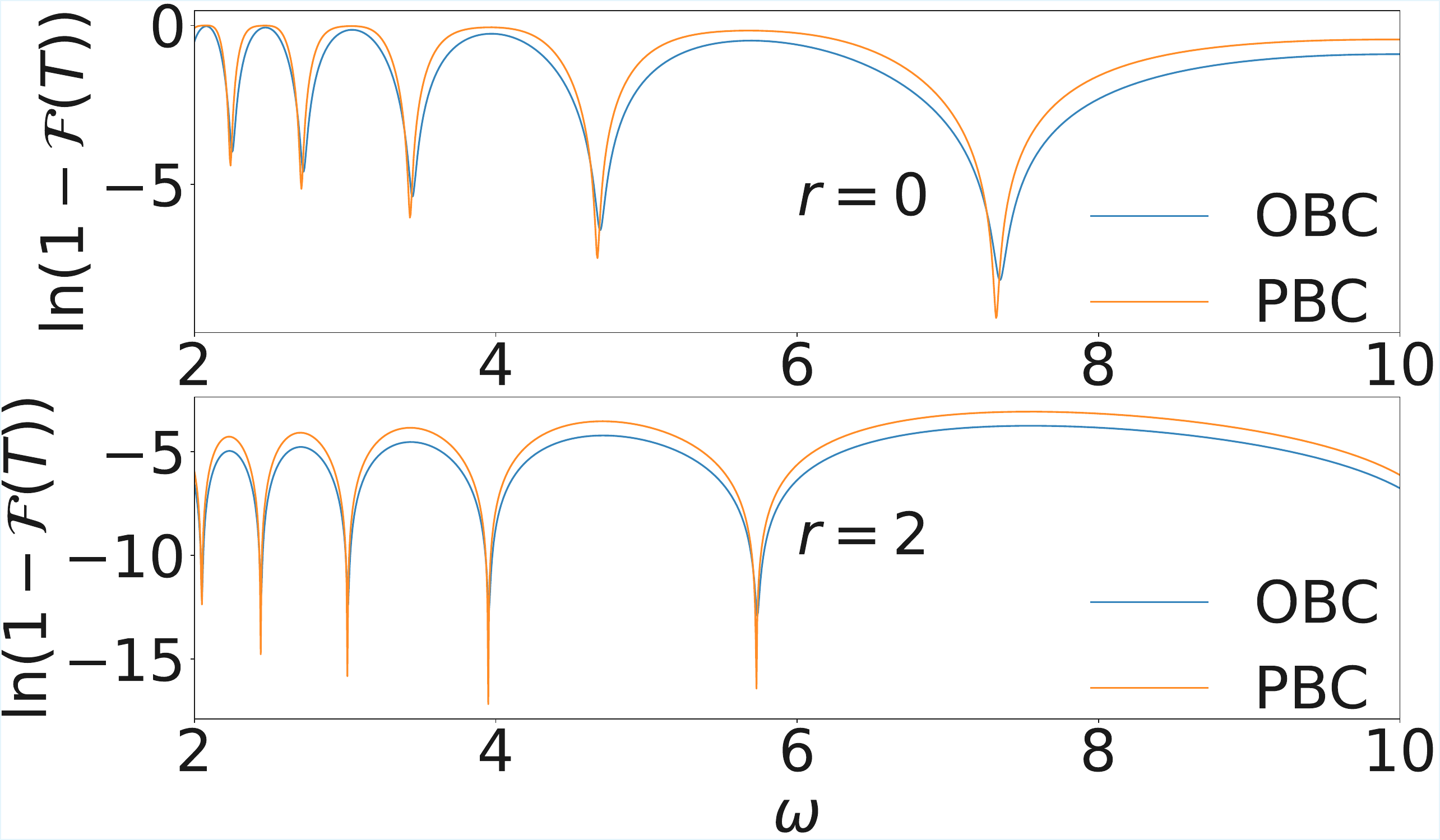}
    \includegraphics[width=0.32\linewidth]{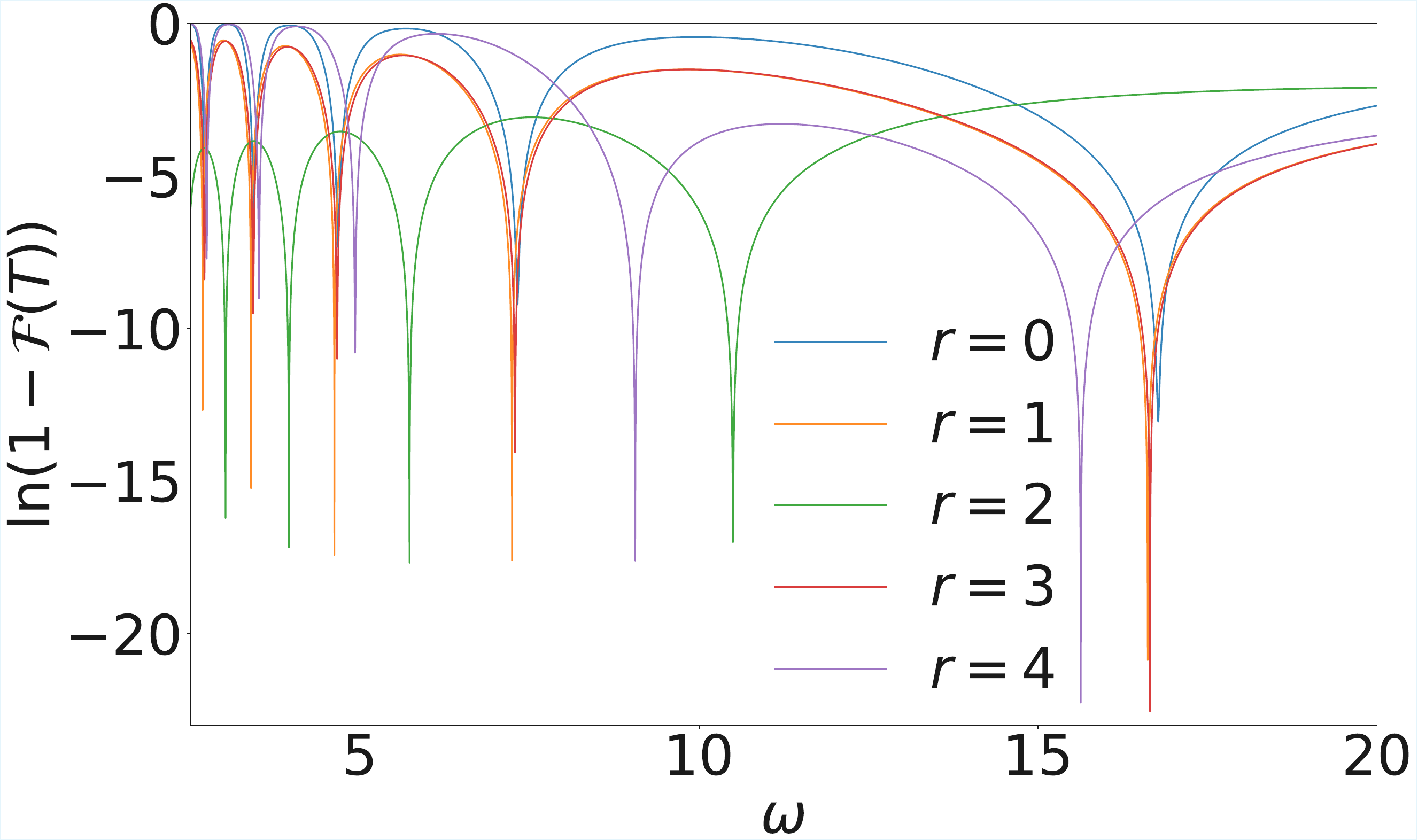}
    \caption{\textbf{System-size, boundary-condition, and harmonic-ratio dependence of dynamical freezing in the PXP model.}
    Left: $\ln[1-\mathcal{F}(T)]$ for $4\times4$ and $6\times6$ square lattices with $\Delta_0=10$, $\Omega_0=2$, and $r=2$. Increasing the system size preserves the freezing minima, with a slight shift towards higher frequencies and a reduction in their depth. Middle: $\ln[1-\mathcal{F}(T)]$ for a $4\times4$ lattice with open (OBC) and periodic (PBC) boundary conditions for $r=0$ (upper panel) and $r=2$ (lower panel), with $\Delta_0=40$. The positions of the freezing minima are largely insensitive to the boundary conditions. Right: $\ln[1-\mathcal{F}(T)]$ for a $4\times4$ lattice with $\Delta_0=40$ for harmonic ratios $r=0,1,2,3,$ and $4$. The $r=2$ protocol exhibits the strongest suppression of $1-\mathcal{F}(T)$ over the frequency range shown. In all cases, the system is initially prepared in the ground state of the Hamiltonian.}
    \label{fig:diff_sizes}
\end{figure}

To access larger systems within exact diagonalization, we exploit the spatial symmetries of the constrained Hilbert space. For the square lattice, translations along both spatial directions, reflections about the $x$ and $y$ axes, and inversion about the diagonal are resolved. This symmetry reduction substantially decreases the effective Hilbert-space dimension. For example, for a $6\times6$ lattice, restricting to the symmetry sector
\begin{equation}
    K_x=K_y=0,\qquad
    I_x=I_y=I_d=+1
\end{equation}
reduces the matrix dimension to $9702$, compared with a full constrained Hilbert-space dimension of $2\,406\,862$. This allows us to investigate system sizes up to $L=36$ in suitable geometries and parameter regimes.

We first examine the dependence of the freezing response on system size. The left panel of Fig.~\ref{fig:diff_sizes} compares $\ln[1-\mathcal{F}(T)]$ for $4\times4$ and $6\times6$ square lattices for $r=2$. The pronounced freezing minima persist upon increasing the system size, demonstrating that the effect is not restricted to the smallest lattice considered. The larger system exhibits a modest shift of the minima towards higher driving frequencies together with a reduction in their depth. Thus, finite-size effects modify the quantitative strength and position of the freezing features, but do not eliminate the underlying interference-induced suppression.

We next investigate the dependence on boundary conditions. The middle panel of Fig.~\ref{fig:diff_sizes} compares open (OBC) and periodic (PBC) boundary conditions for a $4\times4$ lattice, shown for both $r=0$ in the upper panel and $r=2$ in the lower panel. The locations of the principal freezing minima remain largely unchanged between the two boundary conditions, although their depths show some quantitative differences. This indicates that the characteristic frequency dependence of the freezing is not strongly tied to a particular choice of boundary condition. In particular, the persistence of the minima under OBC supports the relevance of these results to the finite arrays realized experimentally.

Finally, we study the dependence of the freezing response on the ratio $r$ of the two driving frequencies. The right panel of Fig.~\ref{fig:diff_sizes} shows $\ln[1-\mathcal{F}(T)]$ for $r=0,1,2,3,$ and $4$. A pronounced dependence on the harmonic content of the drive is observed. Among the values considered, $r=2$ exhibits the strongest suppression of $1-\mathcal{F}(T)$ over a broad frequency range, corresponding to the highest fidelity and hence the most pronounced dynamical freezing. This numerical result is consistent with the theoretical analysis presented in the main text and provides the motivation for using $r=2$ in the bifrequency protocol.

Overall, the PXP simulations demonstrate that the freezing response persists upon increasing the system size and is largely insensitive to the choice of boundary conditions, while its strength depends significantly on the harmonic content of the drive. The enhanced suppression obtained for $r=2$ identifies this frequency ratio as the most effective among those considered and supports its use in the main results.

\section{Resolving interference within individual drive cycles}

\begin{figure}
\hspace{0.05\linewidth}\includegraphics[width=0.9\linewidth,height=0.27\linewidth]{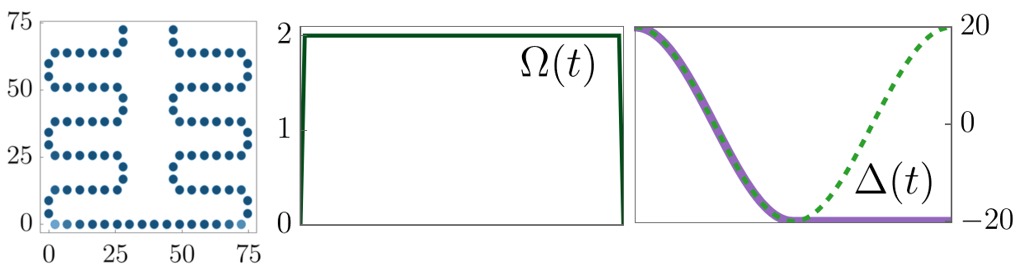}\\
\includegraphics[width=0.48 \linewidth,height=0.3\linewidth]{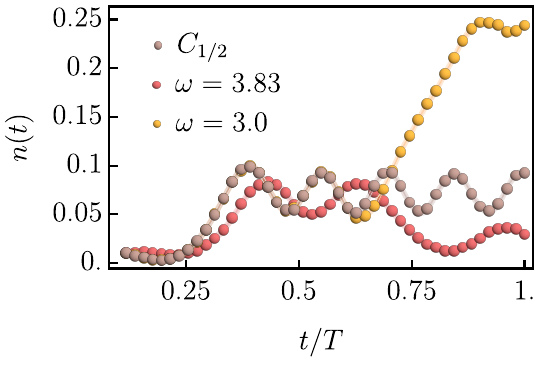}
\includegraphics[width=0.48 \linewidth,height=0.3\linewidth]{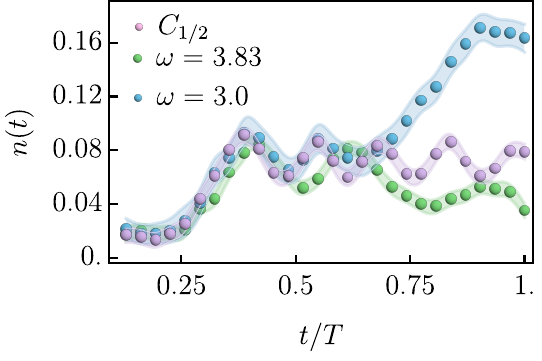}
\caption{
\textbf{Microscopic dynamics within a drive cycle.}
Time-resolved measurements reveal how St\"{u}ckelberg interference produces vacuum-state freezing.
\textbf{Top:} Experimental geometry and drive protocol. The purple line indicates the half-cycle control sequence, where the detuning is held fixed beyond $t=T/2$, $\Delta(t>T/2)=\Delta(T/2)$.
\textbf{Left:} Classical simulations using Bloqade ($L=14$) comparing constructive ($\omega=3.0$, yellow) and destructive ($\omega=3.83$, red) interference. Brown circles denote the half-cycle protocol ($C_{1/2}$), which shows only residual oscillations and no interference signature.
\textbf{Right:} Experimental verification on Aquila ($L=100$) at the same frequencies. The clear contrast between full-cycle (red/green and yellow/blue) and half-cycle (brown/purple) dynamics demonstrates that vacuum freezing arises from coherent amplitude recombination via second passage through the avoided crossing, rather than from an individual passage.
}
\label{fig:timeevo}
\end{figure}
The frequency-dependent vacuum freezing raises a fundamental question: how do quantum phases interfere across repeated passages to produce dynamical stability? To probe this microscopically, we track the time-resolved Rydberg population within a single drive cycle in 100-atom chains on Aquila (Fig.~\ref{fig:timeevo}), benchmarked against Bloqade simulations at $L=14$.

At frequencies yielding destructive interference ($\omega \sim 3.825~\text{rad}/\mu\text{s}$), the excitation probability rises approaching the first avoided crossing but falls during the second, completing the cycle with $n(T)<3\%$—near-perfect vacuum recovery. In contrast, at constructive frequencies ($\omega \sim 3.0~\text{rad}/\mu\text{s}$), excitations accumulate coherently across both traversals, leading to final densities exceeding $15\%$.

To confirm that interference requires completion of the full temporal loop, we implement a truncated protocol halting evolution at half-period ($t=T/2$), keeping $\Delta$ fixed thereafter. In this case, the characteristic contrast vanishes. Excitation grows monotonically and settles into residual oscillations, demonstrating that the observed freezing originates from phase recombination across the full cycle, not from single-pass dynamics.

These time-resolved measurements directly visualize the Stückelberg mechanism at the many-body level. By resolving constructive and destructive recombination within each drive cycle of a $100$-atom ensemble, our results establish temporal interference as a coherent, scalable mechanism underlying Floquet control in interacting quantum systems.

\section{Excitation density distribution}
\begin{figure}[h!]
\includegraphics[width=0.98 \linewidth,height=0.5\linewidth]{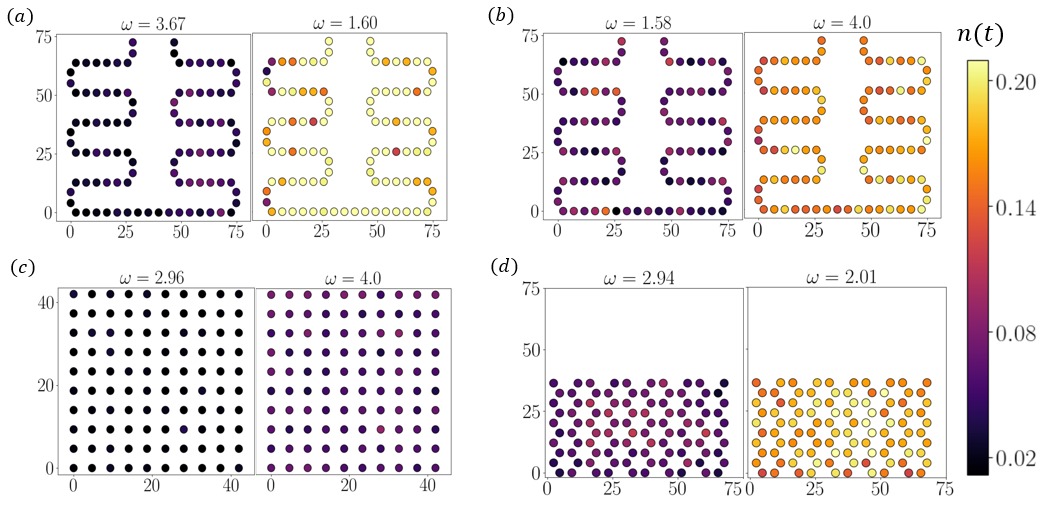}
\caption{\textbf{Local Rydberg excitation probability of each individual atom:} Fig.\ref{fig:1dchaindistribution} (a),(b) corresponds to single- and bi-frequency driving respectively for one-dimensional chains (represented by the snake configuration). (c) Shows the bi-frequency result for two-dimensional square lattice showing that the excitation is suppressed even at the frequencies where we have the peak, and (d) shows the bi-frequency result for two-dimensional hexagonal lattice. For all the figures, Left: Frequencies corresponding to the freezing points, and Right: frequencies at which the total excitation density is maximized (right). The frequencies are mentioned at the top of each plot.}
\label{fig:1dchaindistribution}
\end{figure}

\begin{figure}
\includegraphics[width=0.49 \linewidth,height=0.35\linewidth]{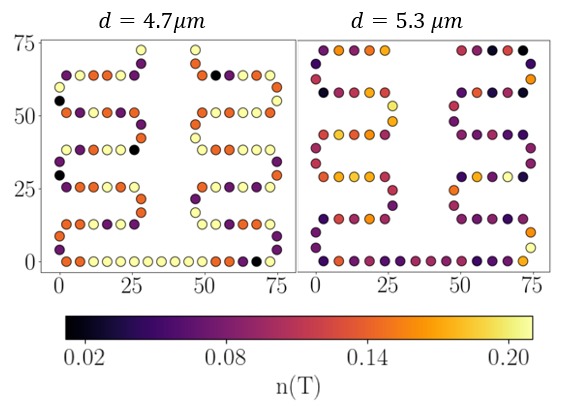}
\caption{\textbf{Local Rydberg excitation probability of each individual atom demonstrating effect of inter-atomic distance:} This plot represents the effect of changing the interatomic distance for the same geometry at $\omega \sim 2.5 ~rad/\mu s$. It demonstrate that at $d=5.3 \mu m$ the excitations are more suppressed compared to that at $d=4.7 \mu m$ for the same frequency.}
\label{fig:1dchaindistribution2}
\end{figure}

In the main text we characterized the system response using the Rydberg excitation density $n(T)$ averaged over all atoms in a particular configuration as a function of the driving frequency. This global observable efficiently captures the emergence of non-equilibrium interference effects, including the appearance of freezing points and enhanced excitation regimes, and allows for a direct comparison across different system geometries and driving protocols. However, such spatial averaging necessarily masks the underlying site-resolved structure of the many-body excitation patterns that develop during the driven dynamics.\\
Here, we therefore complement the averaged analysis by presenting the local Rydberg excitation probability of each individual atom different lattice geometries and driving protocols. Fig.~\ref{fig:1dchaindistribution} (a),(b) corresponds to single- and bi-frequency driving for one-dimensional chains (represented by the snake configuration). The bi-frequency driving for two-dimensional square and hexagonal geometries are shown in Fig.~\ref{fig:1dchaindistribution} (c) and (d) respectively. We choose two representative driving frequencies for Fig.~\ref{fig:1dchaindistribution}(a,b,c,d): (i) frequencies corresponding to the freezing points (left), and (ii) frequencies at which the total excitation density is maximized (right). Fig.~\ref{fig:1dchaindistribution2} represents the effect of changing the interatomic distance for the same geometry at $\omega \sim 2.5 rad/\mu s$. By resolving the excitation distribution at the single-site level, we demonstrate that the observed extrema in the averaged excitation are not the result of localized or edge-dominated effects, but instead arise from a highly coherent and spatially homogeneous excitation pattern across the entire atomic configuration.\\

At both the freezing points and the excitation maxima, the site-resolved excitation probabilities remain approximately uniform across most atoms. Modest deviations are observed in regions of local geometric variation, such as near the bends of the snake-like configuration. These deviations do not result in any abrupt redistribution of the overall excitation density toward higher or lower values.

The observed spatial variations arise from interaction effects and reflect local geometric constraints. Their systematic presence across different geometries and driving protocols indicates that the non-equilibrium features discussed in the main text are collective in origin rather than dominated by local inhomogeneities. The site-resolved data therefore support the interpretation of these features as global many-body effects in driven Rydberg systems.

\end{document}